\documentclass[twocolumn]{aastex61}
\usepackage{graphicx}
\usepackage{float}
\usepackage[boxruled]{algorithm2e}
\usepackage{url}

\newcommand{\Isd}{{I_{sd}}}

\newcommand{\Vsd}{{V_{sd}}}
\newcommand{\Vint}{{V_{int}}}
\newcommand{\wsd}{{w_{sd}}}
\newcommand{\Pnu}{{P_{\nu}}}
\newcommand{\ALGO}{{SDINT }}

\accepted{16 April, 2019}
\submitjournal{The Astronomical Journal}

\shorttitle{\ALGO Reconstruction}
\shortauthors{Rau et al.}

\begin{document}

\title{A joint deconvolution algorithm to combine single dish and
interferometer data for wideband multi-term and mosaic imaging}

\correspondingauthor{Urvashi Rau}
\email{rurvashi@nrao.edu}

\author{Urvashi Rau}
\affiliation{National Radio Astronomy Observatory, 
Socorro, NM 87801, USA}

\author{Nikhil Naik}
\affil{University of Southern California,
Los Angeles, CA 90007, USA }

\author{Timothy Braun}
\affil{University of New Mexico,
Albuquerque, NM 87131, USA }




\begin{abstract}

Imaging in radio astronomy is usually carried out with a single-dish radio telescope doing a raster scan of 
a region of the sky or with an interferometer that samples the visibility function 
of the sky brightness. Mosaic observations are the current standard for imaging 
large fields of view with an interferometer and multi-frequency observations are now routinely 
carried out with both types of telescopes to increase the continuum imaging sensitivity and
 to probe spectral structure.  
This paper describes an algorithm to combine wideband data from these two types of telescopes in
a joint iterative reconstruction scheme that can be applied to spectral cube or 
wideband multi-term imaging both for narrow fields of view as well as mosaics.
Our results demonstrate the ability to prevent instabilities and error that typically arise
when wide-band or joint mosaicing algorithms are presented with spatial and 
spectral structure that is inadequetely sampled by the interferometer alone.
For comparable noise levels in the single dish and interferometer data, 
the numerical behaviour of this algorithm is expected to be similar to the idea of 
generating artificial visibilities from single dish data. 
However, our discussed implementation is simpler and more flexible in terms of applying relative 
data weighting schemes to match noise levels while preserving flux accuracy, fits within
standard iterative image reconstruction frameworks, is fully compatible 
with wide-field and joint mosaicing gridding algorithms that apply corrections
specific to the interferometer data and may be configured to enable spectral cube and wideband multi-term
deconvolution for single-dish data alone.

\end{abstract}




\section{Introduction} \label{sec:intro}

The combination of images and data from single dish telescopes and radio interferometers 
has been a topic of
interest for many years. 
A single-dish radio telescope forms an image by doing a raster scan of a region of the sky
and the observed image can be described as a convolution of the true sky with an 
effective single dish beam pattern. 
An interferometric array samples the visibility function of the sky brightness distribution
over a region of sky covered by the forward gain pattern of each array element and 
forms an image by Fourier
inversion and iterative model reconstruction. A processed interferometric image 
represents a convolution of the true sky
with a Gaussian corresponding to the main lobe of the interferometer impulse response function.
At a given observing frequency, a radio interferometer usually offers excellent 
angular resolution compared to a
single dish telescope, but it suffers from the short-spacing problem where the 
total power and visibility function for 
sources with large angular size and low surface brightness are often not measured 
at all or well enough for an accurate reconstruction. This is especially relevant for 
mosaic observations of large fields of view containing
spatial structure that extends beyond the field of view offered by each individual pointing. 

With the widespread adoption of wideband recievers on both types of radio 
telescopes, wideband imaging techniques are
now used to combine data  across a large range of observing frequencies. 
The main purpose is to increase imaging
sensitivity and fidelity as well as to reconstruct both the spatial and spectral 
structure of the sky brightness 
distribution. Using joint reconstruction algorithms such as 
MTMFS \citep{MSMFS, MFCLEAN}, reconstructions of both
intensity and spectral structure can be done at an angular resolution offered 
by the joint spatial frequency 
coverage and not limited to the angular resolution offered by the lowest frequency. 
So far, this technique has been applied only to wideband interferometer data
 although there is nothing conceptual
preventing it from being used on wideband single dish data as well. 

With wideband interferometer data alone, a reconstruction of spectral structure at 
the largest 
spatial scales is completely unconstrained by the data. A subtle difference 
with intensity-only imaging is that
the problem exists even for situations where they may be 
enough data for an adequate reconstruction of large scale intensity. 
As shown in an example in \cite{MSMFS}, there may be no apparent
imaging artifacts as the model adequately fits the measured data, 
but the spectral reconstruction
will still be wrong. 



Over the years, several approaches have been tried and used, with varying degrees of success. 
They range from combining fully processed images from single dish and interferometry
 data, to using the single dish image
as a starting model for the interferometer reconstruction and finally to various schemes
 for joint reconstructions of 
the sky model using constraints from both datasets at once.   
It has been shown that best results are obtained when the iteratively reconstructed model
represents the entire structure on the sky and not just some spatial scales.

This paper formulates a joint reconstruction algorithm that always uses constraints from both sets of 
data and follows the standard major/minor cycle approach to iterative image reconstruction in radio interferometry.
The algorithm can be configured for spectral cube imaging as well as wide-band multi-term imaging and 
is fully compatible with wide-field interferometric imaging schemes such as AW-Projection \citep{AWProjection} and 
Joint Mosaicing \citep{SaultMosaic1996}. It can also be configured to provide spectral cube and wideband multi-term
deconvolution of the single dish data alone and these deconvolved single dish models can in turn be used
in traditional feathering and startmodel approaches. 



Sections \ref{Sec:SDimaging} and \ref{Sec:INTimaging} summarize wideband imaging with a
single dish and interferometer respectively and section \ref{Sec:Methods} summarizes existing
combination approaches. Section \ref{Sec:NewAlgo} describes our algorithm with imaging results
presented in section \ref{Sec:Results}.

\section{Single Dish imaging}\label{Sec:SDimaging}
A single dish radio telescope makes an image of the sky by doing a raster scan across the region of interest,
a process often refered to as basket-weaving. 
This is mathematically equivalent to a convolution of the 
true sky image $I^{sky}$ with the antenna power pattern.
An image is then constructed by resampling the measurements 
onto an image pixel grid using an explicit 
gridding kernel to compute a weighted average of all data points within the support radius around each pixel center. 

The observed image $I^{obs}_{SD,\nu}$ can be represented as 
\begin{equation}
I^{obs}_{SD,\nu} = I^{sky}_{\nu} \star I^{psf}_{SD,\nu}
\label{Eq:sd_obs}
\end{equation}
where $I^{psf}_{SD,\nu} = I^{PB}_{SD,\nu} \star I^{grid}_{SD}$ is an aggregate convolution function constructed
from the antenna power pattern and the image gridding kernel. 
The angular resolution of the observed image is given by
$\theta_{SD,\nu} = \sqrt{ \left[ \theta_{PB,\nu}^2 + \theta_{grid}^2 \right]}$
where $\theta_{PB,\nu} = \frac{\lambda}{D}$ is the half-power beam width of the SD primary 
and $\theta_{grid}$ is the width of the image gridding kernel. 
In the limit of very fine sampling on the sky as well as on the image grid, the sky and pixel 
sampling patterns may be ignored in this analysis. 
At an observing frequency $\lambda$, a single dish telescope is sensitive to spatial frequencies
ranging from zero to $\theta_{PB,\nu} = \frac{\lambda}{D}$ where $D$ is the effective diameter 
of the aperture and $\lambda$ is the observing wavelength. 
Wide band single dish observations will result in images at angular resolutions 
ranging from $\frac{\lambda_{max}}{D}$ to $\frac{\lambda_{min}}{D}$. 

\citet{Mangum_OTF_2007} formally describe the process of constructing a single dish image and choosing
appropriate image gridding kernels and this is currently the approach used for ALMA total power data.
Within the CASA\footnote{Common Astronomy Software Applications (https://casa.nrao.edu)} 
software, wideband single dish data are usually imaged in
spectral line mode, using a fixed image gridding kernel for all frequencies. 
For continuum imaging, data from multiple observing frequencies are combined onto a single 
output image grid, also using the same gridding kernel. 

\section{Interferometric imaging}\label{Sec:INTimaging}
%
An interferometer constructs an image by partially sampling the visibility function of the 
target sky brightness distribution. Measured visibilities are resampled onto a 2D spatial 
frequency grid by a process of convolutional resampling (called gridding) and then 
Fourier transformed to form the observed image. 


The observed image $I^{obs}_{INT,\nu}$ can be approximately represented as 
\begin{equation}
I^{obs}_{INT,\nu} = \left[ I^{sky}_{\nu} \cdot \Pnu \right] \star I^{psf}_{INT,\nu}
\label{Eq:int_obs}
\end{equation}
where $I^{psf}_{INT,\nu} = [F]^{-1} S_{\nu}$ is the point spread function given by the inverse 
Fourier transform of the 
weighted 2D spatial frequency sampling functions at the observing frequency $\nu$ and $\Pnu$ is the primary beam
(baseline power pattern) of each array element. 
The angular resolution of an interferometer (and typically also the image reconstructed from it) 
is therefore given by $\theta_{INT,\nu} = \lambda/B_{max}$.

A model of the true sky $I^{sky}_{model}$ is computed via an iterative image reconstruction scheme that uses
a-priori information about the types of structures being imaged to estimate the visibility function in regions
of the spatial frequency plane that the interferometer has not made measurements. Algorithms such as classic
CLEAN \citep{CLEAN,CLARK_CLEAN} model the sky as a series of delta functions, MS-CLEAN\citep{MSCLEAN} 
uses a basis of inverted truncated
paraboloids of specific widths, MT-MFS\citep{MSMFS} uses a Taylor polynomial model to 
describe the frequency dependent amplitude per multi-scale flux component. 
Wide-field baseline-based effects due to array non-coplanarity (the W-term) and variable antenna
primary beams are handled via the W-Projection \citep{W_Projection_IEEE} and 
A-Projection \citep{AWProjection} that use careful choices of gridding convolution functions during
interferometric imaging. Joint mosaic imaging \citep{SaultMosaic1996} is performed (also often
during gridding) as an appropriately weighted sum of data from a series of shifted pointings.


At a given observing frequency $\lambda$, an interferometer is sensitive to a finite set of spatial scales
within the limits of $\lambda/B_{min}$ and $\lambda/B_{max}$ where $B_{min}$ and $B_{max}$ are the shortest and 
longest baselines that are possible for a given array configuration. 
The angular resolution of the reconstructed image is therefore given by $\lambda/B_{max}$. 
The maximum spatial scale that can be measured is $\theta_{INT} = \lambda/B_{min}$ where $B_{min} > D$
and it is left to the reconstruction process to estimate the visibility function at these and larger scales.
The result therefore depends strongly on the accuracy and flexibility of the models used and 
this uncertainty is known as the short spacing problem. 
A negative bowl in the reconstructed image
implies that large scales are missing from the reconstructed model.
%
Other large scale artifacts could result when the reconstruction attempts 
to construct a model that spans the unmeasured large scales but fails to produce 
something physically plausible. Finally, with the use of 
multi-scale algorithms it is sometimes possible to produce images with 
artifact-free large-scale reconstructions but without external or {\it a-priori} 
information, one can never be sure that 
it is flux-accurate. 

Interferometric mosaicing is often employed to cover fields of view much larger than 
the primary beam of each array element. Here, the short spacing problem becomes especially
relevant when the sky emission has structure on scales larger than the single pointing field of view. 

With wide-band observations, the frequency dependence of the instrument and the sky 
must also be taken into 
account during the reconstruction process.
One option is to make a spectral cube, with varying angular resolution and sensitivity 
to large spatial scales.
Another is the MTMFS algorithm and related approaches that solve for joint wideband 
models to produce 
maps of intensity as well as spectral structure at a resolution given by the combined beam. 
Additionally, for wide-field and mosaic imaging, frequency dependent antenna primary beams 
introduce an artificial spectral index into the wideband sky flux model.
Post-deconvolution wideband beam correction eliminates it from the wideband model but 
methods also exist to remove it prior to the iterative deconvolution and model generation on
a per antenna or timestep basis \citep{WBAWP}.


The short spacing problem for wideband imaging is slightly different from that of narrow-band imaging. 
The extent of the central unsampled region on the spatial frequency plane increases with
 observing frequency. There is a range of scales that are measured adequately enough for an accurate reconstruction 
at lower frequencies but which are completely missing at the higher end of the band, thus 
causing spectral
structure to be completely unconstrained by the data. Most plausible wideband sky models 
will not be able to
distinguish between this missing data and (for example) a genuine steep spectrum 
extended source.

\begin{figure}
\centering
\includegraphics[width=0.5\textwidth]{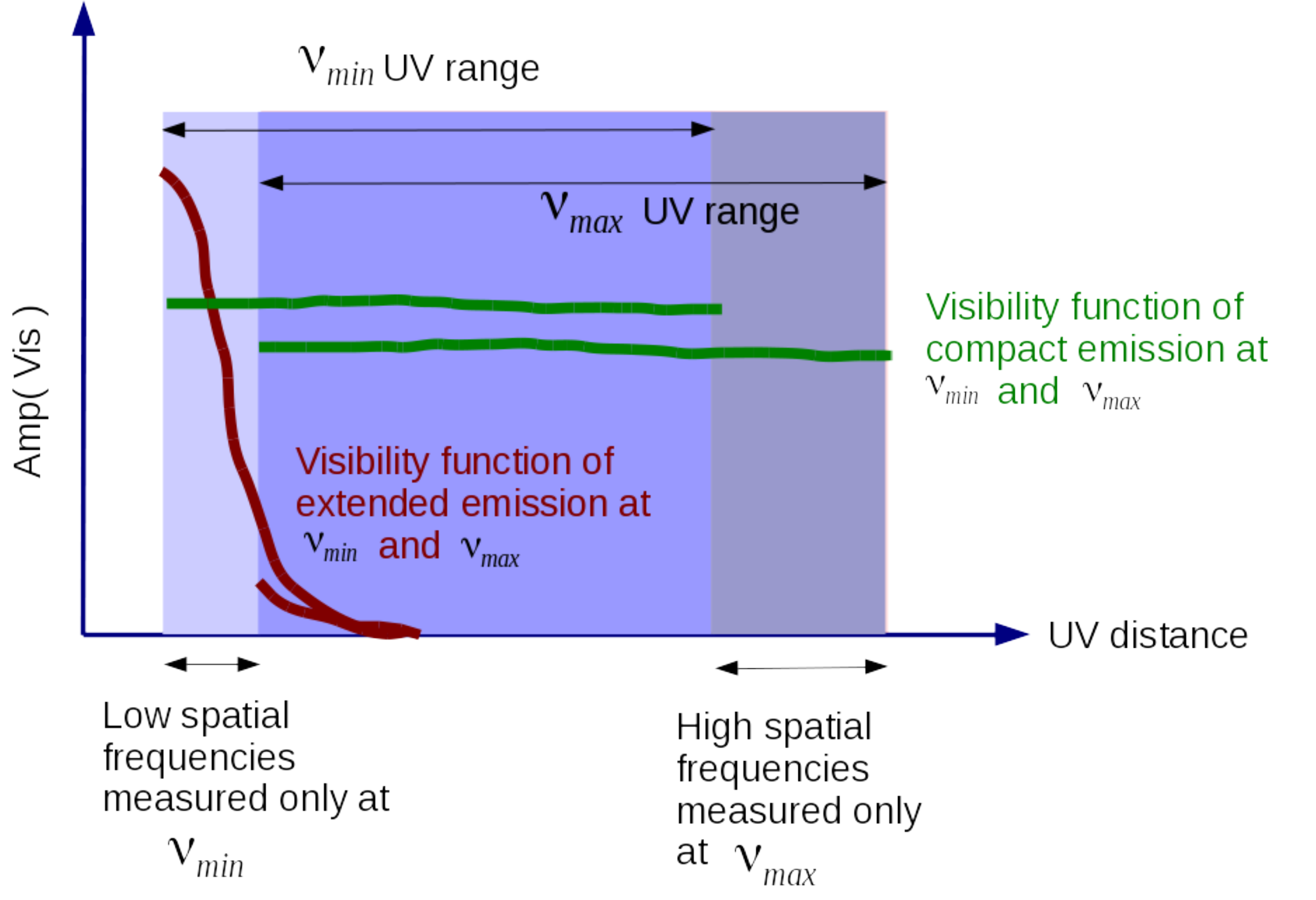}
\caption{This diagram illustrates the spatial scales measured by an interferometer 
at different observing frequencies and its impact on the visibility function of a 
 compact source
as well as extended structure. The shaded rectangles 
represent the uv-range at the low and high frequency ends of an observing band. The green 
curve represents the measured visibility function of a point source with a negative spectral
index, as measured at the low and high frequencies. 
The red curves represent the measured visibility function of an extended source, also with
a negative spectral index. This figure illustrates the source of reconstruction uncertainty
for extended sources, especially for spectral structure. }
\label{fig:example_uvscale}
\end{figure}

As seen in the illustration in Fig.\ref{fig:example_uvscale}, compact sources 
have a signature all
across the uv-plane and therefore their spectra are well constrained, even at the full
angular resolution. The spectral structure of extended emission, however, is particularly
poorly sampled and therefore extremely error prone, even in situations where some frequencies
do measure the visibility function adequately enough to reconstruct an accurate average
intensity.  Symptoms of this effect may include the typical negative bowl in the intensity 
image but it
is also possible to obtain a perfectly legitimate-looking intensity pattern 
with an overly steep spectral reconstruction. 

\section{Combining single dish and interferometer data and images}\label{Sec:Methods}

Sections \ref{Sec:feathering},\ref{Sec:startmodel} and \ref{Sec:jointrecon} summarize
several techniques that have been in use for narrow-band imaging, 
discuss their strengths and weaknesses and explore their compatibility with wideband imaging.
Section \ref{Sec:PBs} lists the general approach followed by all the methods when dealing with interferometer
primary beams for full-beam or mosaic imaging. 

\subsection{Feathering} \label{Sec:feathering}
Feathering is the technique most often used to combine images from single-dish instruments
and interferometers. A composite image is obtained by computing a 
weighted sum of the observed single dish image and the reconstructed interferometer
image, calculated in the spatial frequency domain.
There are many variants of this approach, depending on weighting function choice and scale factors
(for example, CASA {\tt feather}\citep{CASAFeather}, IMMERGE\citep{MIRIADimmerge},IMERG \citep{AIPSimerg},
ObitFeather\citep{Feather_Cotton}).

CASA {\tt feather} implements the following.
\begin{equation}
\mathcal{F}(\textbf{I}^{feathered}) = \wsd \Vsd + [1 - \wsd] \Vint
\end{equation}
where $\wsd \Vsd$ is the Fourier transform of the observed single-dish image given 
by Eqn.~\ref{Eq:sd_obs} and
$\Vint$ is the Fourier transform of the reconstructed interferometer image 
defined at a resolution given by the restoring beam.
The single dish image is first scaled by the 
volume ratio of the interferometer restoring beam to 
the single dish antenna pattern.
Then the weighting function is computed as 
\begin{equation}
\wsd = \mathcal{F} \Isd^{psf}  ~~ \mathrm{scaled~to} ~~ max(\wsd) = 1.0
\label{Eq:sd_wt}
\end{equation}
The weight given to the Fourier transform of the restored interferometer image
is then $[1-\wsd]$. Such a combination has the effect of using only well trusted
regions of the spatial frequency plane from both images and selecting only
those scales that each instrument measures best.
Generally, the single-dish image is also pre-scaled by an additional gain parameter
meant to correct for residual flux calibration errors. This is typically chosen by
manually matching the flux in the overlapping range of spatial frequency. 
The IMMERGE implementation \citep{MIRIADimmerge} solves for this automatically by matching 
visibility function shapes within a user-defined annulus in the uv-domain. 

Feathering is most appropriate when 
there are no reconstruction artifacts at scales
other than the largest ones in the interferometer-only reconstruction 
and noise levels between the two images are 
comparable such that the addition of the single dish image does not degrade
the interferometric image. 
Feathering is a single step process whose results 
depend strongly on the single-dish gain parameter as well
as the specific shape of the weighting function.
When the noise levels between the single dish and interferometer images differ
significantly, one must choose between preserving flux correctness and
keeping the noise level in the combined image close to that of 
the interferometric image. A significant amount of art is often involved in 
carefully constructing non-standard weighting functions in order to achieve
this (i.e. by changing the form of Eqn.\ref{Eq:sd_wt}).
Although this might suffice for specific source structures and known
noise characteristics, it is not a viable generic approach.
Additionally, when structure is present on intermediate scales, 
an interferometer-only wideband reconstruction can perform unconstrained 
extrapolations which burn in errors in
the model that are beyond the spatial frequency range that will be replaced
by the single dish image. 
In such cases, a post-deconvolution combination may not be
able to achieve flux correctness on all scales.


In situations where feathering can work well, it can be directly extended 
to wideband image reconstructions as well.
The output of the MT-MFS algorithm is a set of Taylor coefficients that
represent the spectral structure of the sky brightness distribution. 
A similar set of coefficient images must first be constructed from the
single dish data, and the images feathered together term by term before
recomputing derived quantities such as spectral index and curvature.

\subsection{Single Dish image as a starting model or prior}\label{Sec:startmodel}
A single dish observed image can be used to derive a 
starting model for the interferometric reconstruction.
This approach is effective only when there is considerable overlap 
in measured spatial frequencies between the single dish and interferometer
data such that the input starting model already accounts for much of 
the shorter spacing interferometer data. 
When there isn't much overlap it is unlikely to be any different from 
feathering as the single dish image will provide no extra constraints for 
the interferometer reconstructions.
Techniques such as MEM \citep{MEM} make use of a-priori information and a single dish
image can be used as a prior such that the reconstruction is biased towards
it.


In situations where this startmodel approach is effective, wideband single dish 
data (or image cubes) can also be deconvolved separately and then converted to a set of 
Taylor coefficient model images to be supplied as a starting model for a 
wideband interferometer reconstruction. 
\newline

\subsection{Joint reconstructions}\label{Sec:jointrecon}

The most successful and robust approach is a joint reconstruction by which a single
sky model is constructed using data and constraints from both data
sets at once. Several methods are summarized below. 

\subsubsection{Joint image-domain constraints}\label{Sec:joint1}
Image solvers that directly implement optimization algorithms may modify their
objective functions to include constraints from single dish as well as interferometric data. 
For example, the MOSMEM \citep{MIRIADimmerge} task in MIRIAD employs 
an image-domain chi-square constraint with separate
terms for the interferometer dirty image as well as the single dish image, and 
implements this within a Maximum Entropy reconstruction algorithm \citep{MEM}. This implementation
includes the option of pre-scaling the single dish data, along with the ability to
optionally auto-scale it to match visibility functions in an annulus on the uv-plane
that matched the overlap region. 
\cite{SDINT_Wright_2012} analyses the joint imaging problem for ALMA+ACA+TP and the
tests demonstrate that such a joint reconstruction is clearly superior to the previous two methods. 
For wideband imaging, the MEM sky model would have to be augmented to support a
wideband flux model, a detail currently not included in existing literature or software. 
Algorithms such as RESOLVE \cite{RESOLVE_2016} that have a Bayesian formulation along with a
built-in wideband sky model may be an effective alternative algorithm within which to include
similar single dish constraints.

\citet{SDINT_SS_1999} describes another approach in which dirty images and point
spread functions from both the interferometer and single dish are 
combined to form a new pair of images to feed into a deconvolution algorithm. 
In this particular application, a single scaling multiplier is optionally applied during 
an image-domain combination step.
\citet{SDINT_SS_1999} make no mention of how to handle primary beams or to do wideband imaging, but
this method can easily be extended on those axes. 

Note that both of these methods \citep{MIRIADimmerge,SDINT_SS_1999} have been described 
as purely image-domain
deconvolution approaches without any feedback loop to the raw data themselves. 
However, it is trivial to place these methods within the standard major/minor cycle
imaging framework \citep{Schwab_Cotton_Clean, IMAGING_THEORY_IEEE} and treat them as minor 
cycle image-domain
modeling algorithms. The only extra required step would be to implement suitable
model-to-data tranforms for the single dish data. Our algorithm described in
section \ref{Sec:NewAlgo} describes such a framework.

\subsubsection{Creating visibilities from single dish data}\label{Sec:joint2}
\citet{SDINT_KODA_2011} and \citet{SDINT_KURONO_2009} perform joint 
reconstructions by creating
aritificial visibilities from single dish data and then including them in interferometer
reconstructions.
The single dish image is first linearly deconvolved using an estimate of the
aggregate single dish beam. A randomised visibility sampling distribution is generated to
match the shape of the Fourier transform of the single dish 
beam (similar to Eqn.~\ref{Eq:sd_wt}) so as to preserve 
information about the intrinsic resolution of the single dish observation. 
New visibilities are then generated by a de-gridding step. 
For a mosaic observation, \citet{SDINT_KODA_2011} generate single dish visibilities for
each interferometer pointing so as to be consistent with pointing table
metadata already present in the interferometer dataset and which standard 
algorithms for joint mosaicing require. 
{\it tp2vis}\footnote{https://github.com/keflavich/tp2vis} is a recently developed 
solution targeted towards narrow-band (spectral cube) joint reconstructions for ALMA.

\citet{SDINT_KODA_2011} describe this algorithm for narrow-band (or spectral cube) 
imaging but it is a generic approach that can easily be extended to wideband
image reconstruction.
Qualitatively, this approach achieves the same results as the 
joint reconstructions mentioned above, with the added advantage of being naturally
usable in existing software that implements iterative image reconstruction with 
major and minor cycles.  Two practical caveats are that any software implementation
would require very carefully crafted 
meta-data to ensure that the artificially generated visibilities are treated 
appropriately when used with interferometry-specific imaging options
such as W-Projection \citep{W_Projection_IEEE} or A-Projection \citep{AWProjection}
as they use instrument-specific gridding convolution functions.
Finally, the relative scaling or weighting of single-dish and interferometer data 
has to be done external to the reconstruction process.

\subsection{Handling primary beams}\label{Sec:PBs}
From the forms of  Eqns.~\ref{Eq:sd_obs} and \ref{Eq:int_obs}, it is clear 
that the interferometer primary beam $P_{\nu}$ must be taken into account carefully 
for any field of view that goes beyond the inner central region of
the beam where it can be approximated as unity. Only in that inner region
can both observed images be described as convolutions of the sky brightness with
a point spread function, thus allowing a simple linear combination to be usable
within standard deconvolution routines.

A reconstructed wide-field interferometric image typically represents $I_{\nu}P_{\nu}$
and this is known as a flat-noise normalization. 
Interferometric image reconstruction is usually done with a flat-noise normalization
and so any combination with single dish images prior to or during the reconstruction
will require that the single dish image be modified to mimic the field-of-view as
seen by the interferometer.  This can be done by first performing a linear deconvolution
of the single dish image using a model of the single dish beam, multiplying the resulting
image by the interferometer primary beam (single field or mosaic) and then re-convolving
the result by the single dish beam.
The resulting single dish image now also represents a convolution of $I^{sky} \cdot P_{nu}$ with a point spread function, similar to Eqn.\ref{Eq:int_obs}. 
\citet{SDINT_KODA_2011} and \citet{SDINT_KURONO_2009} also describe this as steps 
done prior to calculating artificial single dish visibilities. 

Note that for the startmodel approach (Sec.\ref{Sec:startmodel}) the last 
re-convolution step must be
left out as the sky model is usually devoid of any point spread function. 
Alternatively, a flat-sky combination may also be performed. 
by first dividing out a primary beam model
from the (flat-noise) interferometer image before feathering it with the single dish image.
Similarly, a flat-sky startmodel approach would only require the single dish data to be
linearly deconvolved prior to its use within the reconstruction. 
%

\section{\ALGO  reconstruction}\label{Sec:NewAlgo}
In this paper we present a generic joint reconstruction algorithm for single dish and 
interferometer data.
It combines and extends several of the
ideas summarized above, and demonstrates a solution to the wideband short-spacing 
problem described in section \ref{Sec:INTimaging}. 

\begin{figure}[h]
\centering
\includegraphics[width=0.5\textwidth]{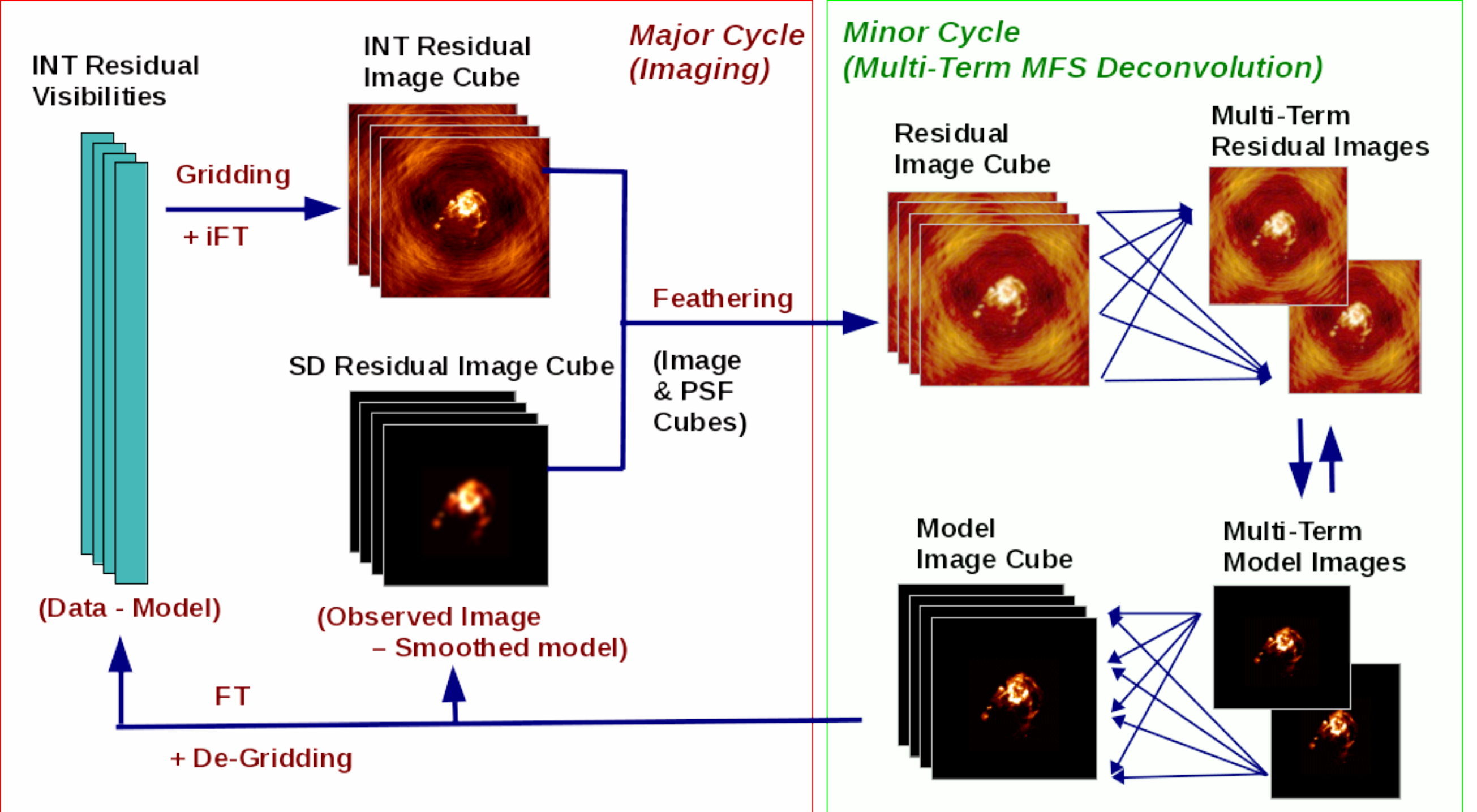}
\caption{Joint reconstruction algorithm for wideband single dish and interferometer data.
PSF and Residual image cubes from the interferometer and single dish data are combined
prior to the minor cycle deconvolution step. This illustration emphasizes the ability to
configure the algorithm to operate in spectral line mode as well as multi-term wideband
continuum mode with either a point-source or multiscale sky model.
The single dish data are represented as images but it is possible to
include a data-image transform there as well. }
\label{algofig}
\end{figure}

A graphical representation of the basic algorithm is presented in Figure.\ref{algofig}.
Interferometer visibilities are gridded and imaged as a spectral cube 
and then combined with the single dish image cube. A similar combination is done
with the interferometer point spread function and the representative single dish 
kernel ($I^{psf}_{SD,\nu}$  in Eq.\ref{Eq:sd_obs}). The resulting pair of observed and
PSF cubes are then
sent to the minor cycle for image-plane deconvolution.  Spectral cube deconvolution is
performed directly with this new set of images and PSFs. Wideband multi-term 
deconvolution is performed by first constructing Taylor weighted averages across
frequency to construct the set of inputs required by the MT-MFS algorithm.
After a series of image-domain deconvolution iterations, the current model images
are converted to model cubes (in the case of multi-term imaging) or used as is for
spectral cube model prediction. Residuals are then computed for both the interferometer and the
single dish data. For the interferometer data, it is the standard major cycle and for the
single dish data a smoothed model is subtracted per channel from the observed single dish
image cube.  The new residuals are feathered together and the process repeats until 
an exit criterion is satisfied. 

Pseudo-code listings based on the CASA PySynthesisImager tool \citep{CASAImager} are 
provided in Algorithm.\ref{algolist_1} for joint spectral cube imaging and 
in Algorithm.\ref{algolist_2} for joint wideband multi-term imaging. The lines in black 
are the basic algorithm that follows Fig.\ref{algofig} and the lines in blue
are steps required to include primary beams into the basic algorithm so that it may be
applied to mosaic interferometer data.  The recipes presented here assume a flat noise
normalization of the interferometric gridder and minor cycle algorithm. Therefore, 
primary beams are divided out of the observed flat-noise image prior to feathering
with the single dish data, and the result is re-multiplied by a primary beam
to present a flat-noise input image to the minor cycle.  For the wideband multi-term
algorithm, an average or common primary beam is used in this second step in order
to conform to the WBAWP algorithm (\cite{WBAWP}) that eliminates the primary 
beam frequency dependence prior to the multi-term minor cycle.

In this framework, spectral cube and multi-term deconvolution of the single dish data 
alone may also be done simply leaving out interferometer data and the 
feathering step. Examples from this algorithm configuration are included in the next section.

\begin{algorithm}[ht!]
\label{algolist_1}
  \DontPrintSemicolon
  \tcc*[h]{Initialize imagers and deconvolver}\\
  IntCube = Imager(gridder='interferometer')\\
  SDCube = Imager(gridder='singledish')\\
  JointCube = Imager(deconvolver='multiscale')\\
  \vspace{0.1cm}
  \tcc*[h]{Make PSFs}\\
  IntCube.MakePSF()\\
  SDCube.MakePSF()\\
  JointCube.psf=Feather(IntCube.psf, SDCube.psf)\\
  \vspace{0.1cm}
  \tcc*[h]{Make initial images}\\
  IntCube.MakeRES()\\
  SDCube.MakeRES()\\
  {\color{blue}IntCube.res = IntCube.res $\div$ IntCube.PB}\\
  JointCube.res=Feather(IntCube.res, SDCube.res)\\
  {\color{blue}JointCube.res = JointCube.res $\times$ IntCube.PB}\\
  \vspace{0.1cm}
  \Repeat {Convergence criteria are satisfied} 
  {
  \vspace{0.1cm}
    \tcc*[h]{Deconvolve per channel PSF}\\
    JointCube.deconvolve()   \\
    \vspace{0.1cm}
    \tcc*[h]{Update residual images}\\
    IntCube.MakeRES(JointCube.mod)\\
    SDCube.MakeRES(JointCube.mod{\color{blue}/IntCube.PB})\\
  {\color{blue}IntCube.res = IntCube.res $\div$ IntCube.PB}\\
  JointCube.res=Feather(IntCube.res, SDCube.res)\\
  {\color{blue}JointCube.res = JointCube.res $\times$ IntCube.PB}\\
  \vspace{0.2cm} 
  }
  \vspace{0.1cm} 
  \tcc*[h]{Restore Model and PB-correct}\\
  JointCube.restore()\\
  {\color{blue}JointCube.image = JointCube.image $\div$ IntCube.PB}\\
  \caption[\ALGO Algorithm with Mosaicing]
         {Pseudo-code for the \ALGO joint reconstruction algorithm for Spectral Cube Imaging. Primary Beam manipulations (shown in Blue) represent the steps required with an interferometric imager operating with flat-noise normalization. }
\end{algorithm}

\begin{algorithm}[ht!]
\label{algolist_2}
  \DontPrintSemicolon
  \tcc*[h]{Initialize imagers and deconvolver}\\
  IntCube = Imager(gridder='interferometer')\\
  SDCube = Imager(gridder='singledish')\\
  JointMT = Imager(deconvolver='mtmfs')\\
  \vspace{0.1cm}
  \tcc*[h]{Make PSFs}\\
  IntCube.MakePSF()\\
  SDCube.MakePSF()\\
  JointCube.psf=Feather(IntCube.psf, SDCube.psf)\\
  JointMT.psf = CubeToTaylor(JointCubePSF)\\
  \vspace{0.1cm}
  \tcc*[h]{Make initial images}\\
  IntCube.MakeRES()\\
  SDCube.MakeRES()\\
  {\color{blue}IntCube.res = IntCube.res $\div$ IntCube.PB}\\
  JointCube.res=Feather(IntCube.res, SDCube.res)\\
  {\color{blue}JointCube.res = JointCube.res $\times$ AvgPB}\\
  JointMT.res = CubeToTaylor(JointCubeRes)\\
  \vspace{0.1cm}
  \Repeat{Convergence criteria are satisfied} 
  {
  \vspace{0.1cm}
    \tcc*[h]{Deconvolve Multi-Term PSFs}\\
    JointMT.deconvolve()  \\
    JointCube.mod = TaylorToCube(JointMT.mod)\\
    \vspace{0.1cm}
    \tcc*[h]{Update residual images}\\

    {\color{blue}JointCube.mod = JointCube.mod / avgPB}\\
    IntCube.MakeRES(JointCube.mod{\color{blue}$\times$IntCube.PB})\\
    SDCube.MakeRES(JointCube.mod)\\
  {\color{blue}IntCube.res = IntCube.res $\div$ IntCube.PB}\\
  JointCube.res=Feather(IntCube.res, SDCube.res)\\
  {\color{blue}JointCube.res = JointCube.res $\times$ AvgPB}\\
  JointMT.res = CubeToTaylor(JointCubeRes)\\
  \vspace{0.2cm} 
  }
  \vspace{0.1cm} 
  \tcc*[h]{Restore Model and PB-correct}\\
  JointMT.restore()\\
  {\color{blue}JointMT.image = JointMT.image $\div$ avgPB}\\
  \caption[\ALGO Algorithm with Mosaicing]
         {Pseudo-code for the \ALGO joint reconstruction algorithm for Wideband Multi-term Imaging. Primary Beam manipulations (shown in Blue) represent the steps required with an interferometric imager operating with flat-noise normalization. The use of common avgPB prior to multi-term deconvolution is based on the Wideband AProjection algorithm that eliminates the primary beam frequency dependence prior to the multi-term minor cycle.}
\end{algorithm}


\subsection{Algorithm Flexibility }
In addition to being easily configured to operate in spectral cube or wideband
multi-term imaging modes, with and without primary beam support
for joint mosaics, and using either interferometer data only or single dish data
only or both together, 
two key modules of the basic algorithm may also be replaced easily for 
even greater flexibility.
\paragraph{The feather step :} As currently implemented, 
the single dish and interferometer residual and PSF cubes are combined using 
the CASA Feather task (sec \ref{Sec:feathering}). 
Custom weighting schemes may be implemented as 
stand-alone modules that apply identical operations to the residual cubes as well as
the PSFs. The frequency dependence of the chosen weighting scheme may also be
varied and common preconditioners such as uniform or robust or tapered weighting schemes
may additionally be applied after combination.
. 
\paragraph{Single dish major cycle :} 
Single dish data may be used either as input image cubes or as
calibrated single dish data. For either option, residual single dish images
must be constructed from the current model, in every major cycle iteration. 
\begin{enumerate}
\item Images : When working with SD image cubes, the simplest option is 
to smooth the model image cube
by the effective single dish PSF per channel
before subtracting it from the original single dish image cube.
\item Data : A more complete single-dish major cycle would require the prediction of model 
single dish data, the calculation of residuals against the calibrated single dish
data, and the re-imaging of these residuals using the single dish image domain gridding 
kernel. This approach would have a stabilizing effect similar to that seen in 
interferometric imaging that employs periodic major cycles. 
\end{enumerate}

\subsection{Scaling vs Weighting : }\label{Sec:ScaleWeight}
The choice of relative scale factors and weighting functions during the
feather step applies to both the PSF as well
as the dirty image and therefore can be interpreted simply as a data weighting 
scheme similar to those routinely used in interferometric imaging.
The situation is analogous to a joint reconstruction using multi-configuration
data from the Very Large Array. 
%
In both situations, the exact choice of relative scale factor
during combination in the spatial frequency domain
will bias the reconstruction but will not be as embedded in the final 
product as a purely post-deconvolution 
single step feathering scheme.
  For a given pair of datasets, there will be a
viable range of scale factors that the algorithm is robust to.
The maximum that the single-dish data may be weighted relative to the interferometer
data can by identified by the value above which the width of the joint PSF begins to 
increase beyond the angular resolution offered by the interferometer. If driven beyond this limit, the
reconstruction will be at an angular resolution closer to the single dish data.
The lower limit is harder to establish in a sky-independent way, but may be derived
empirically using simulations for a given observing setup.
It can be defined as the value below which the reconstruction begins to show signs
of error typical of the interferometery-only reconstruction. In the situation of
wideband multi-term imaging, the symptoms would first manifest on the large
scale spectral index.

The presence of a viable range of relative weighting scale factors 
is particularly helpful in the context of widely different noise levels
between the single dish and interferometer data, especially when the flux scale of the
single dish data is also inaccurate. 
In traditional single step feathering, one has to navigate a trade-off 
between flux accuracy and the resulting image noise level. However, by considering scaling 
for flux accuracy as distinct from weighting to control noise levels, one
can achieve flux correctness in a pre-scaling step and then match noise levels 
via weighting so as to not degrade the interferometer image during 
the combined reconstruction process. 
This scheme can be effective as long as the relative scale factor 
required to suppress noise from the single dish data is within the viable range discussed
above.
%
%

\subsection{Comparison with other methods : }
The \ALGO  approach uses feathering in between the major and minor cycles
to construct joint point spread functions and residual images that are
then fed into an image domain deconvolution algorithm. 
In this respect, it is similar to the method described in \citet{SDINT_SS_1999}
but formulates it more generally such that it employs a uv-dependent
relative weighting during combination (via feathering), applies to narrow-band as well as 
wide-band applications and is a part of the iterative major/minor cycle loops
so that the reconstruction can benefit
from the stabilizing effect of periodically returning to the data (interferometric as 
well as single dish).

From the general point of view of using constraints from both datasets during the 
reconstruction of a sky model by constructing joint dirty images and point spread functions,
the \ALGO algorithm will produce results similar to the
approach of generating artificial visibilities \citep{SDINT_KODA_2011}.
The primary difference arises from the realization that there is no fundamental need to 
construct pseudo-visibilities from the single dish data 
and that the same results can be achieved by a more 
flexible (and therefore tunable) combination scheme implemented as part of the
standard iterative reconstruction process. 
Relative weights (for example) between the single dish and interferometer data need not
be burnt into the single dish visibilities (as they are according to the prescription
in \citet{SDINT_KODA_2011}).
\citet{SDINT_KODA_2011} also discuss some specific relative weighting schemes for use with ALMA
data and the same prescriptions can be employed within our feather step. 

Finally, by joining the data just prior to the
minor cycle step, major cycles can remain functionally 
separate for the interferometer and single dish
data and therefore be customized via different gridding schemes that allow
the algorithm to iterate directly with the data from both instruments.
This eliminates the need to 
carefully construct meta-data for artificial visibilities 
to make the single dish data match what standard
interferometric imaging algorithms expect. Algorithms such as joint mosaics and 
A-Projection that employ very specific baseline-based gridding convolution kernels 
to the interferometer data will therefore be usable in our framework without 
any extra programming complexity regarding the correct interpretation and
handling of artificial single dish visibilities.
A single dish major cycle may also be implemented independent of the 
interferometer data.

\newpage
\section{Tests and Results}\label{Sec:Results}

The \ALGO algorithm was tested on a simulated interferometer dataset representing
the Jansky Very Large Array (JVLA) D-configuration and a simulated single dish observation 
with the Green Bank Telescope (GBT), both spanning a frequency range
of 1.0 to 2.0 GHz (i.e. L-Band) with three channels at 1.0, 1.5 and 2.0GHz.


\begin{figure}
\centering
\includegraphics[width=0.23\textwidth]{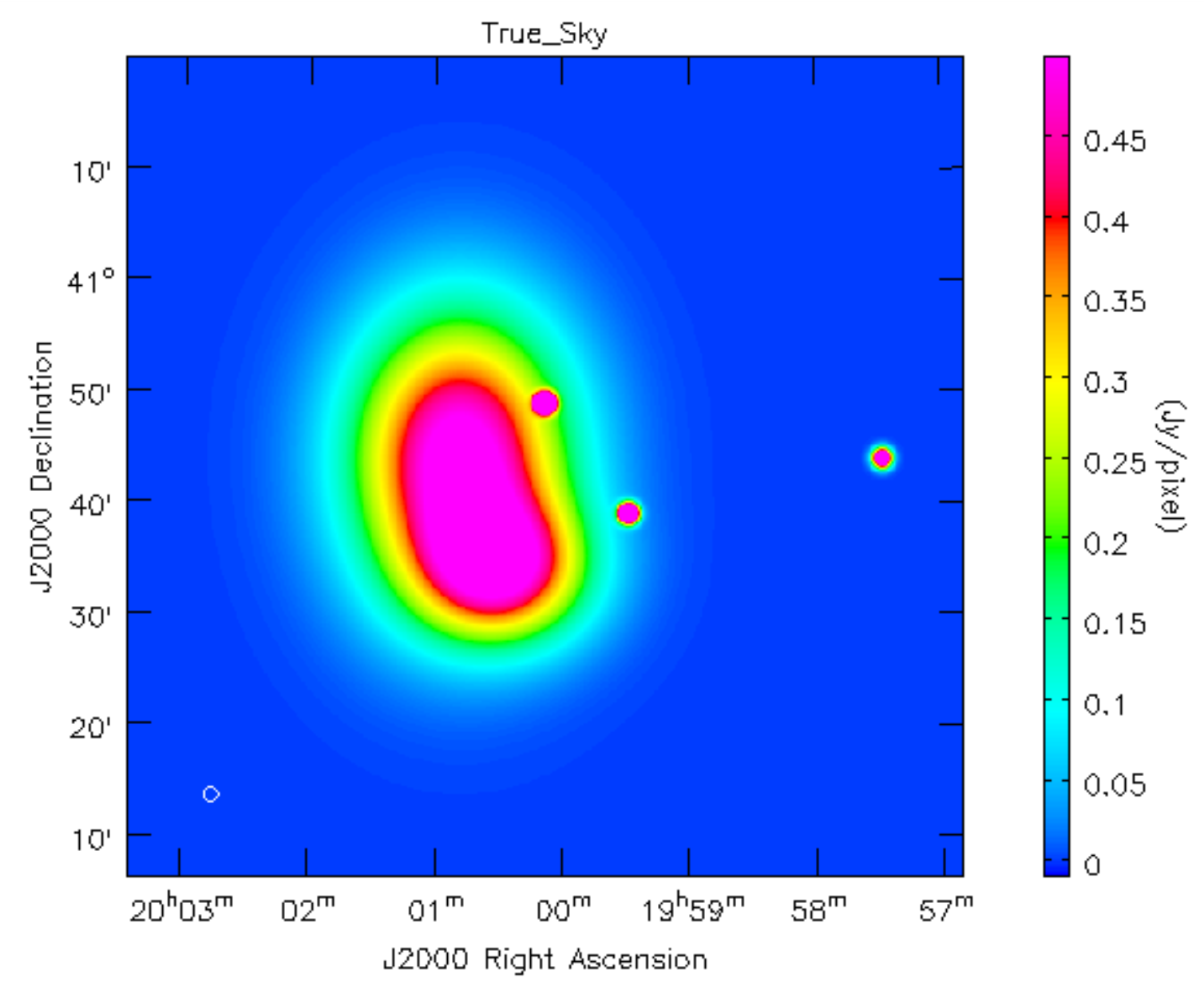}
\includegraphics[width=0.23\textwidth]{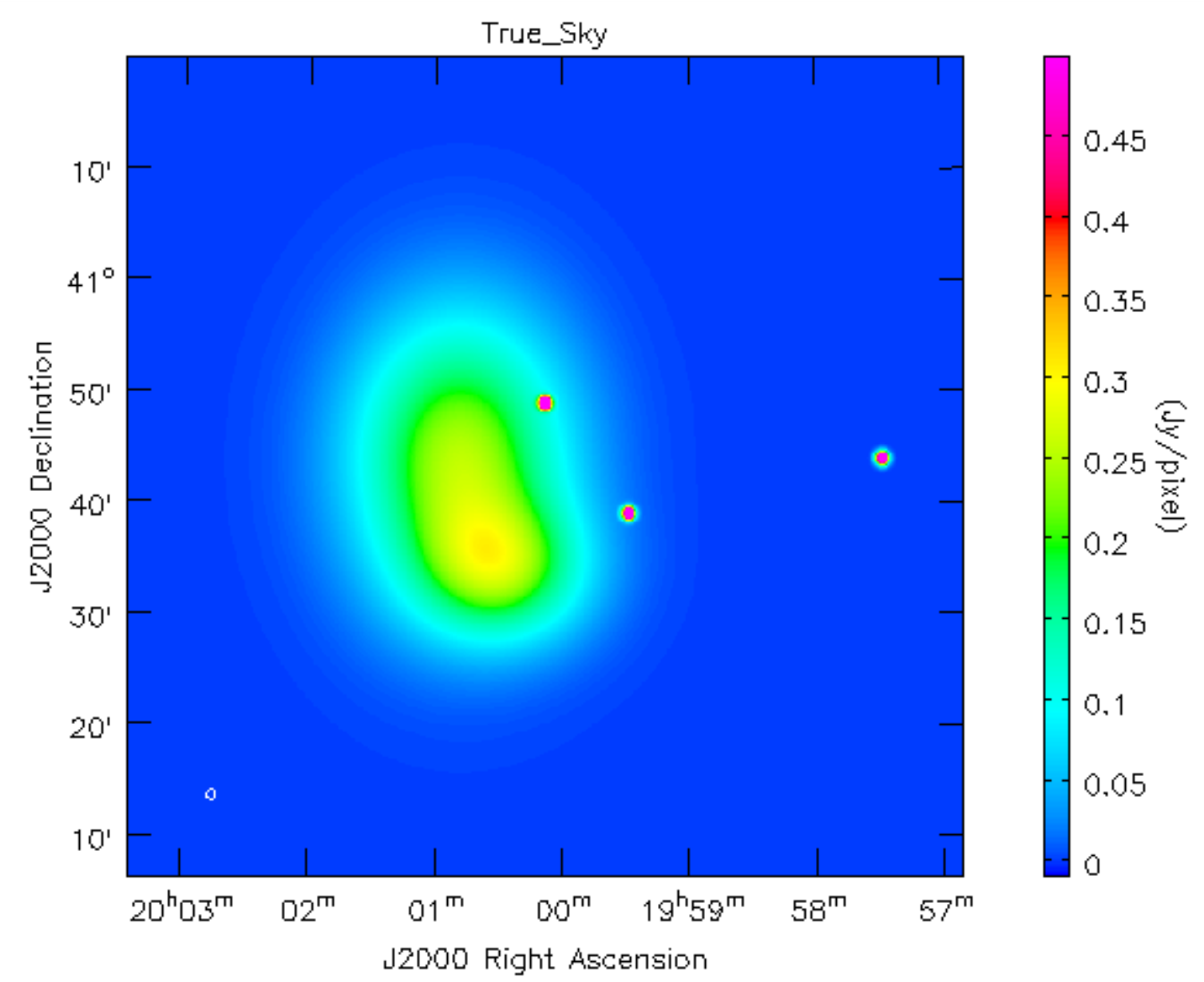}
\includegraphics[width=0.23\textwidth]{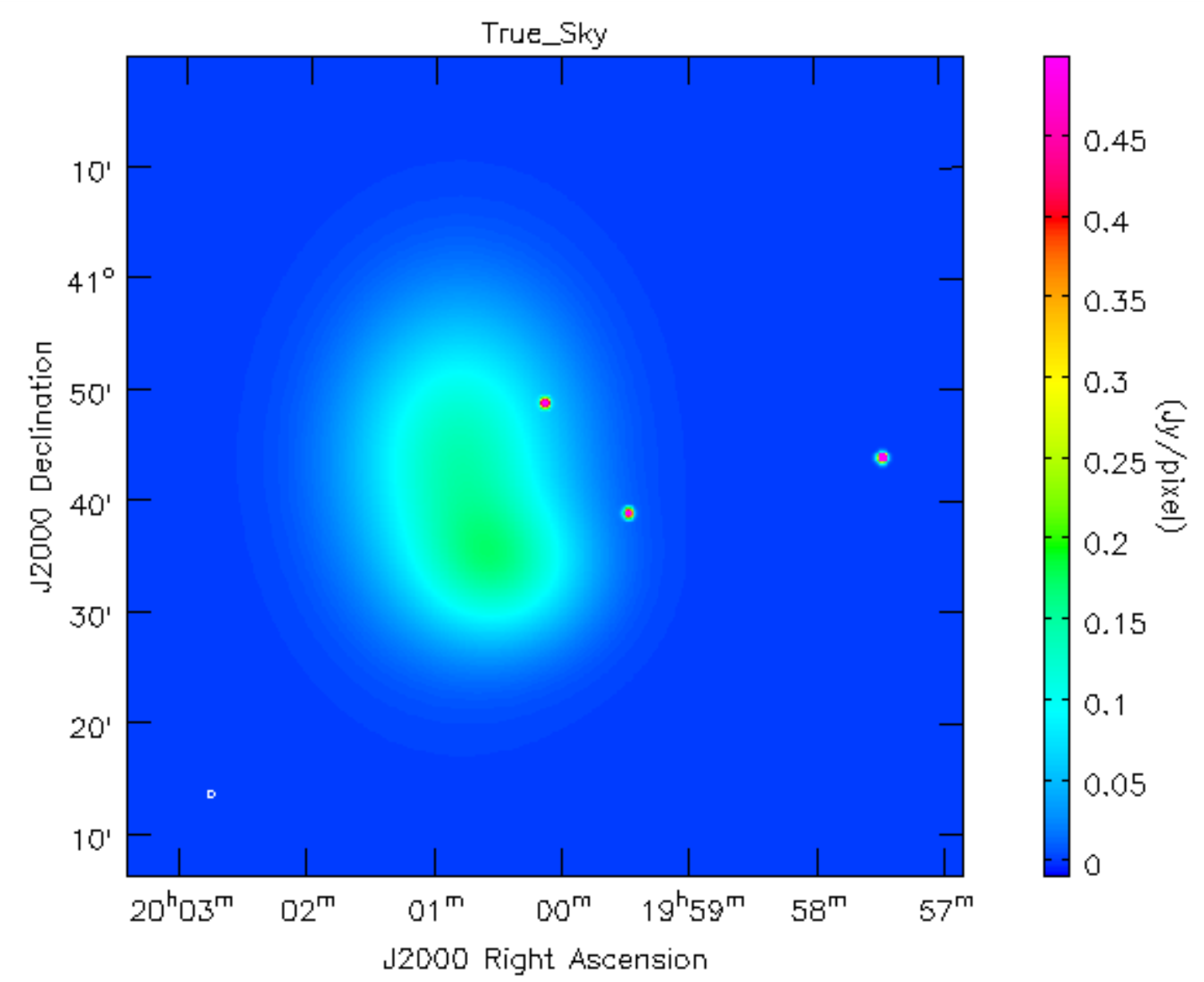}
\caption{Simulated true sky intensity distribution, smoothed to match the 
interferometer resolution
at 1.0 GHz (TOP LEFT), 1.5 GHz (TOP RIGHT) and 2.0 GHz (BOTTOM).
Spectral indices are -1.0, -1.0 and 0.0 for the three point sources (from left to right) and 0.0 
for the extended components on the left. The image at the reference frequency of 
1.5 GHz (TOP LEFT) is also the true wideband intensity image.}
\label{Fig:true_intensity}
\end{figure}

Fig.\ref{Fig:true_intensity} shows the true intensity distribution smoothed to the
angular resolution per channel offered by the interferometer. 
The extended source consists of two Gaussian components, one of size 
$15\times20$ arcmin (bottom half) and the other of size $10\times12$ arcmin (top half). 
The extended components were given a flat spectrum (spectral index = 0.0) and
the point sources from left to right had spectral indices of -1.0, -1.0 and 0.0. 
The scales were chosen such that the top half of the extended source
is unsampled by the interferometer at
most observed frequencies and the bottom half is only partially sampled.
The resulting structure therefore has significant flux in
the region between the interferometer and single dish measurements. This was chosen
to test the algorithm's robustness in a situation where operations 
such as single-step feathering and startmodel approaches are expected to have problems 
due the inadequacy of interferometer-only constraints during the modeling of the large scale
spectrum. 
Table \ref{Tab:specs} lists the angular resolutions and maximum measured 
scales for both instruments betwee 1.0 and 2.0 GHz. 


\begin{table}
\begin{tabular}{|c|c|c|c|p{1.2cm}|}
\hline
Frequency & 1.0GHz & 1.5 GHz & 2.0 GHz & Min Spacing \\
\hline
INT (resolution) &1.0' & 0.67' &0.5' & 1030m\\
INT (max scale) &30.0' &19.6' & 14.7' & 35m\\
SD (resolution) &10.3' &6.8' &5.1' &100m \\
\hline
\end{tabular}
\caption{Angular resolutions and max scales for different instruments and 
frequencies, For comparison,
the large scale structures of the simulated source were Gaussians with 
widths of 20.0' x 15.0' and 15.0'x12.0' } 
\label{Tab:specs}
\end{table}

Two types of interferometer datasets were simulated. The first was a single pointing
observation, with no primary beams. The second was a 25-pointing mosaic covering
the same region of sky. 
Spectral cube as well as wideband multi-term imaging was carried out using
SD-only, INT-only and joint \ALGO reconstructions. 
Section \ref{Sec:compare_algo} compares these approaches as well as the 
traditional feathering and startmodel approaches.
Interferometer primary beams are included in the simulations discussed in
section \ref{Sec:withbeams} within the context of mosaic spectral line and wideband imaging. 

A consistent set of imaging parameters were used in all the tests (imsize=$800\times800$
for the basic simulation and $1500\times1500$ for the mosaic, 
cellsize=$9$arcsec, scales = [0,12,20,40,60,80,100], niter=1000 per plane, 
cycleniter=200, threshold = 0.0 
as supported by the CASA {\it tclean} task \citep{CASAImager}). 

With these settings, it was found that deconvolution masks were required for 
all the INT-only reconstructions without which they very easily diverged. 
It is important to note that the joint SD+INT reconstructions did not need masks
although for consistency all our tests used the same mask.
A separate test was performed without masks in which all but the \ALGO algorithm failed
and the resulting \ALGO images had no qualtitative difference with 
the ones in which masks were used.
%
Additionally, all the INT-only spectral cube reconstructions diverged easily even with a
mask and it was necessary to stabilize it via shallower minor cycles (cycleniter=20). 
It is important to note that this modification
was not required for wideband multi-term imaging that used the combined 
interferometric spatial frequency coverage. It was also not required for spectral cube
joint reconstructions where the single dish data were included. 

These simulations therefore probe situations where the spatial frequency coverage
of the entire combined data (multi-frequency for the interferometer plus single dish) 
is critical to the reconstruction algorithm, without which the data simply do not 
adequately constrain the sky structure being modeled. 

\subsection{Algorithm comparison - Basic Imaging}\label{Sec:compare_algo}

\begin{figure}
\centering
\includegraphics[width=0.23\textwidth]{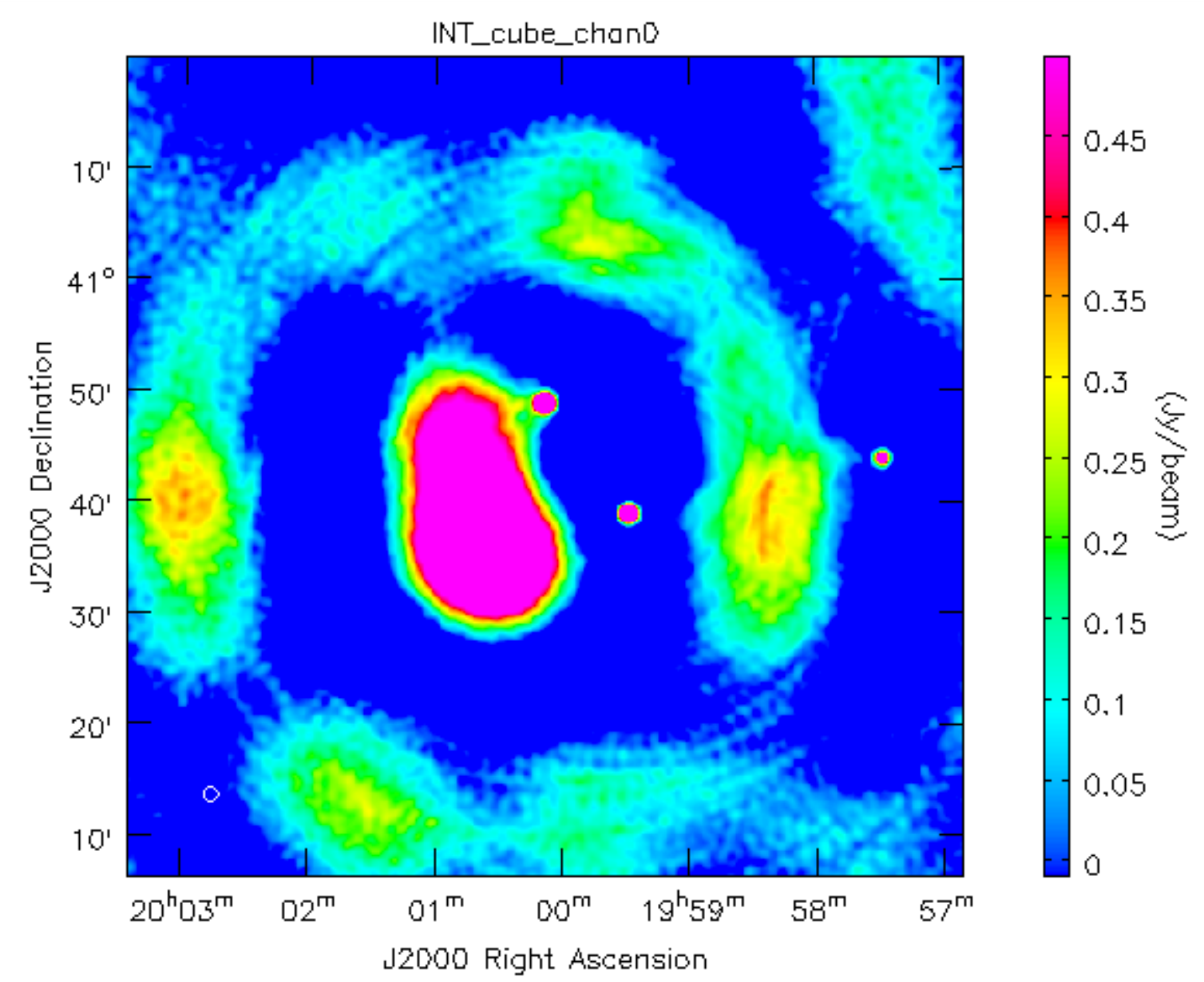}
\includegraphics[width=0.23\textwidth]{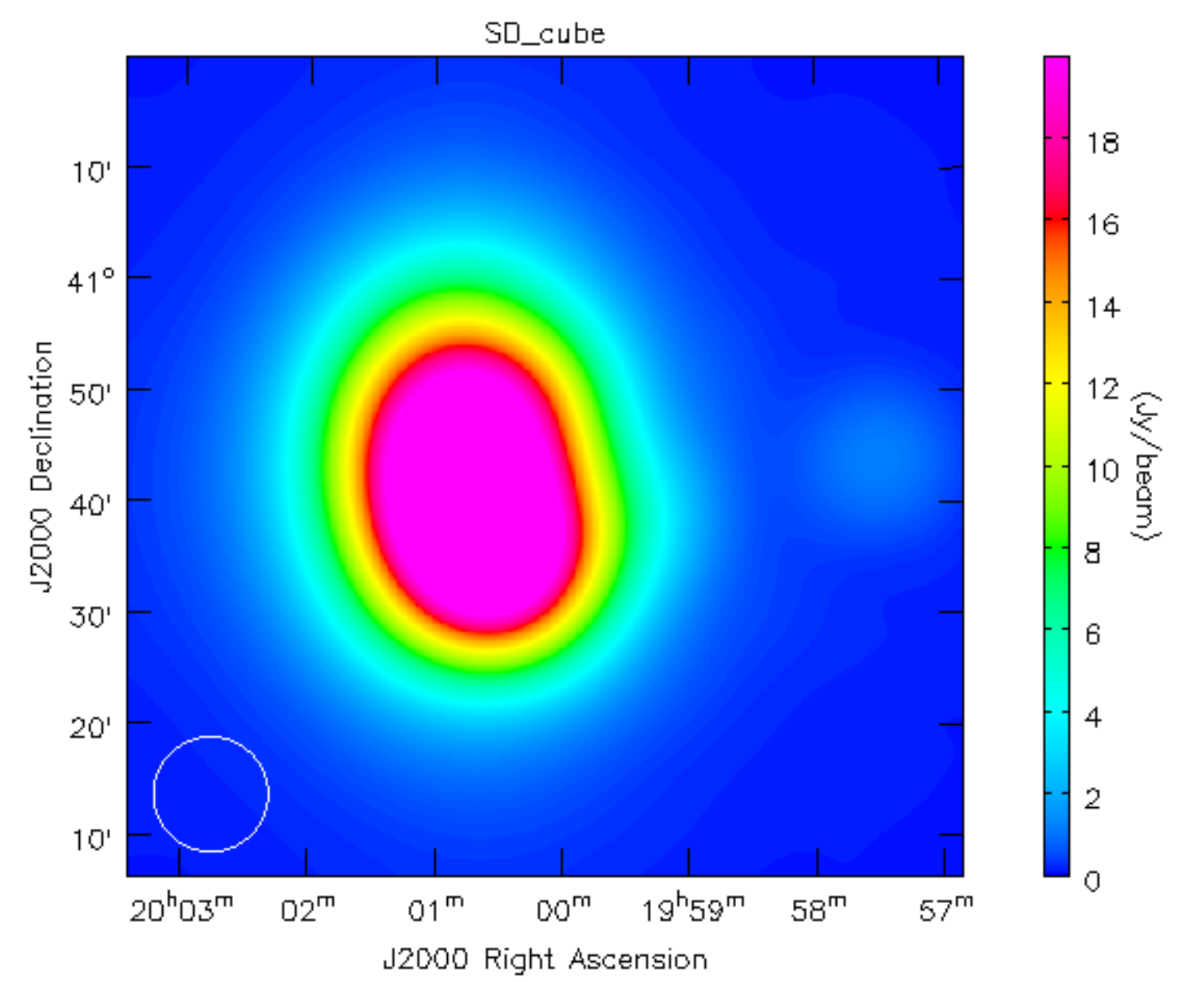}
\includegraphics[width=0.23\textwidth]{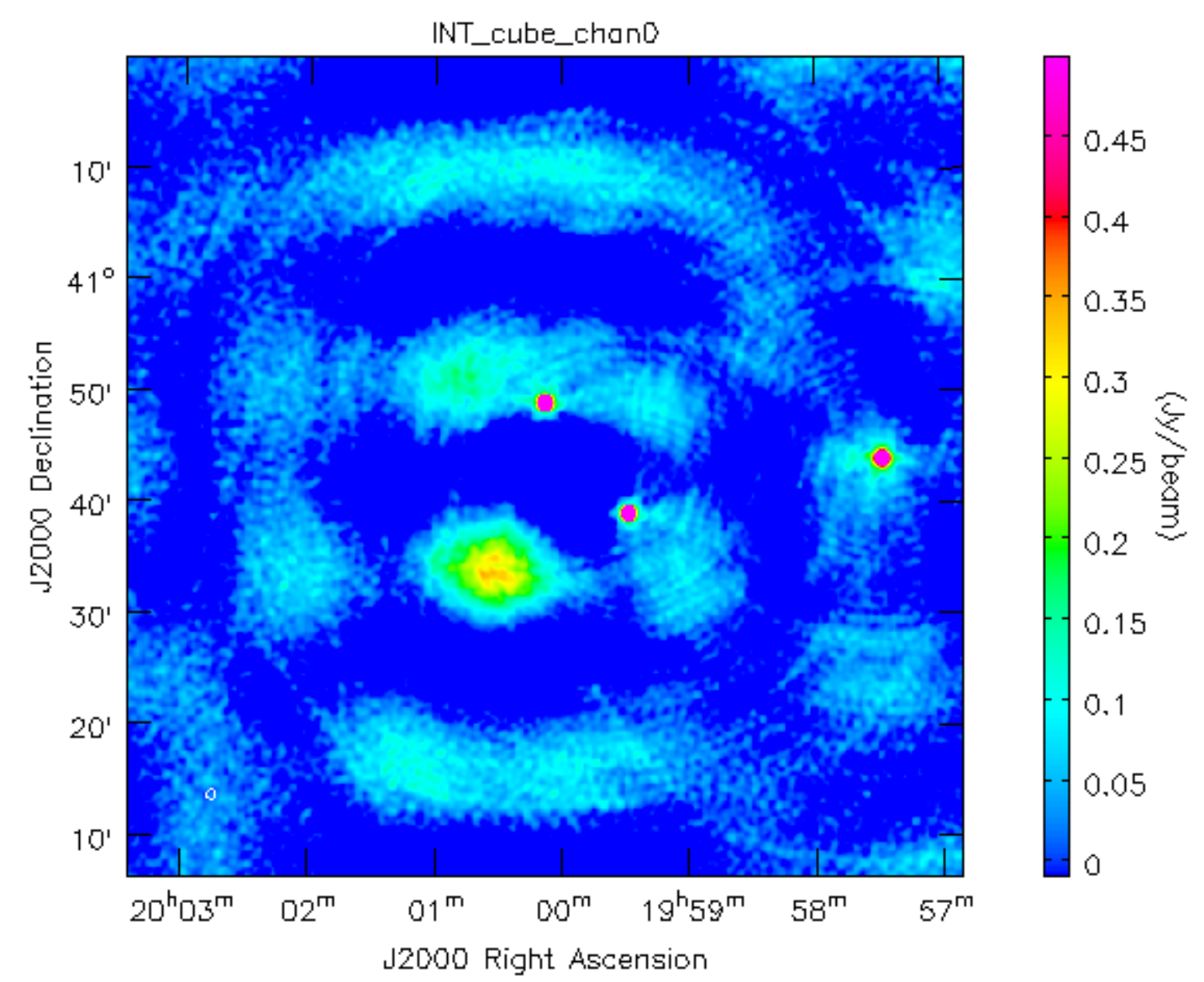}
\includegraphics[width=0.23\textwidth]{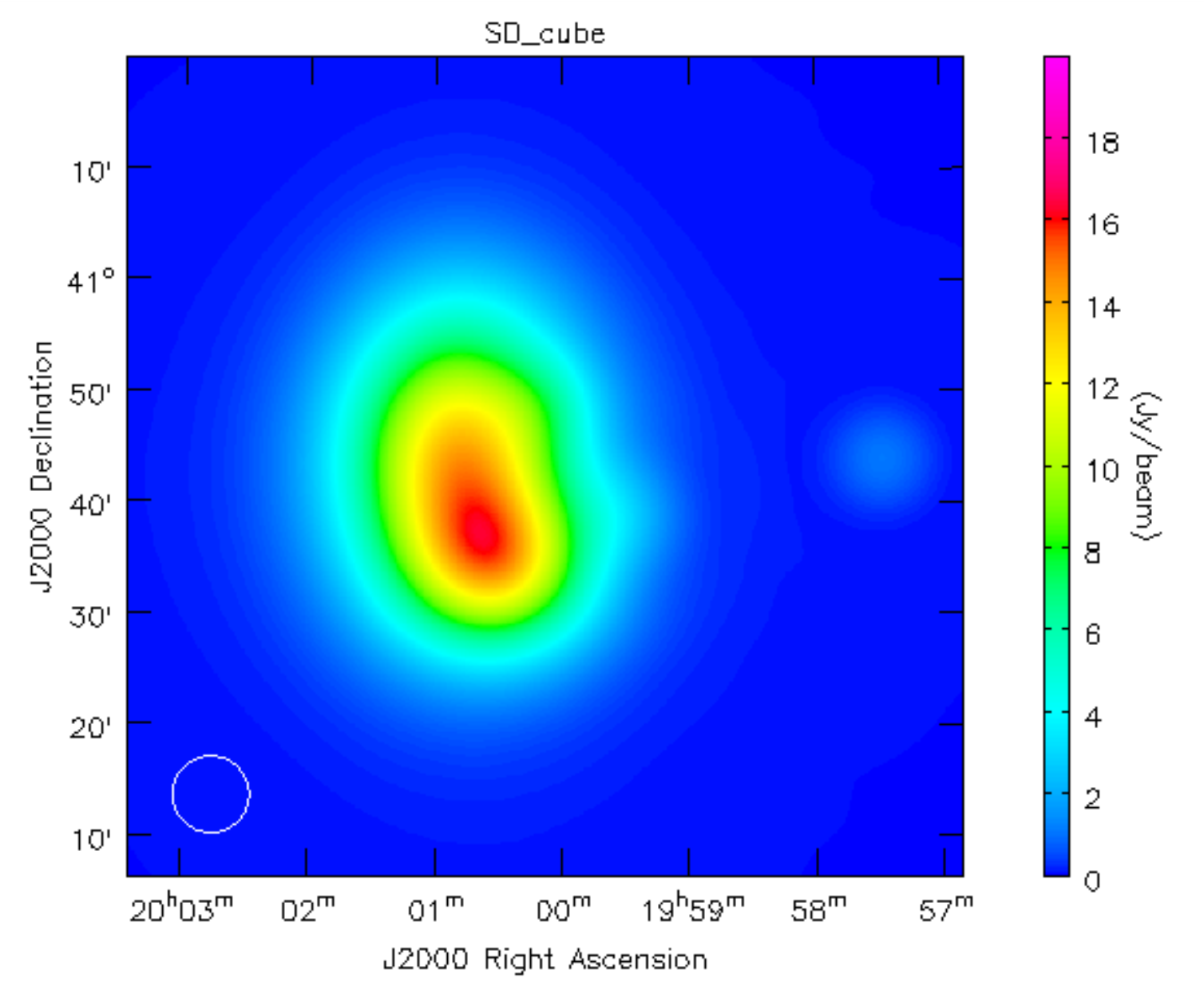}
\includegraphics[width=0.23\textwidth]{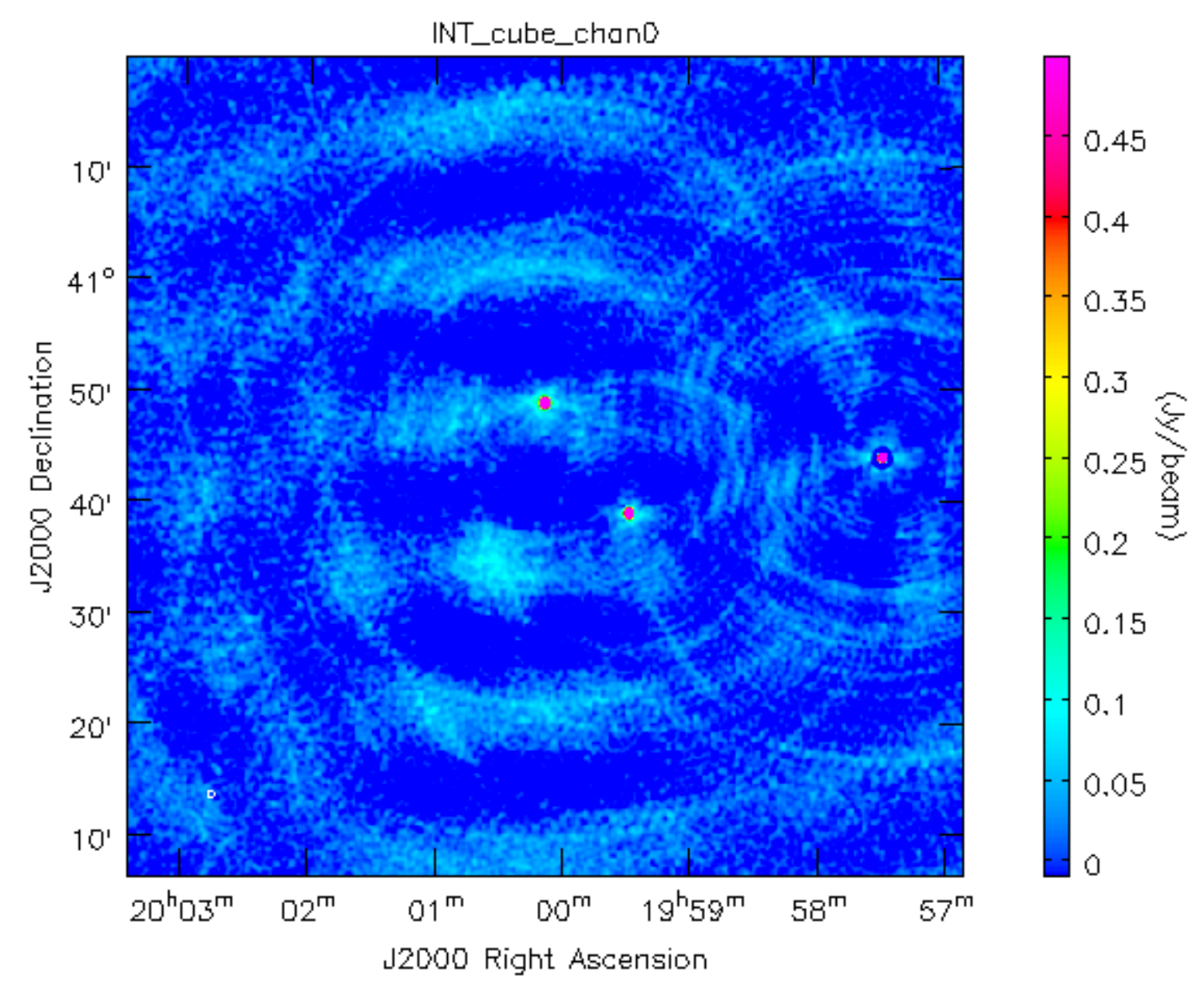}
\includegraphics[width=0.23\textwidth]{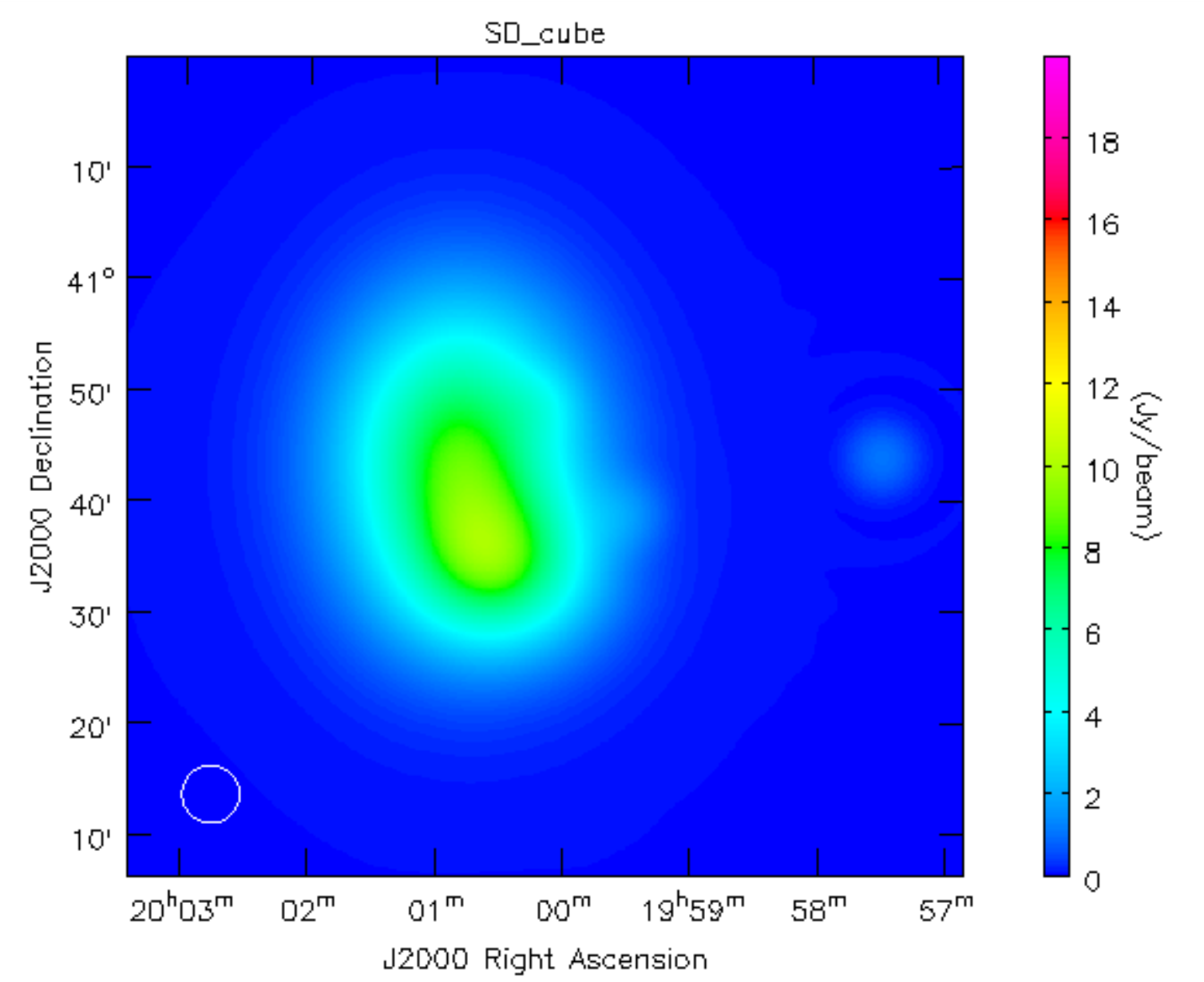}
\caption{Deconvolved INT-only (LEFT) and SD-only (RIGHT) Spectral Cubes (1.0, 1.5, 2.0 GHz). The INT-only reconstructions of the extended emission are clearly underconstrained.}
\label{Fig:cube_intonly_sdonly}
\end{figure}

Figure \ref{Fig:cube_intonly_sdonly} shows deconvolved images made at 1.0, 1.5 and 2.0 GHz,
without any combination. The left column shows deconvolved interferometer images and the 
right column shows deconvolved single dish images, all at their native angular resolutions.
The interferometer images show the typical negative bowls and progressive disappearance of
the large scale emission as observing frequency increases. 

\begin{figure}
\centering
\includegraphics[width=0.23\textwidth]{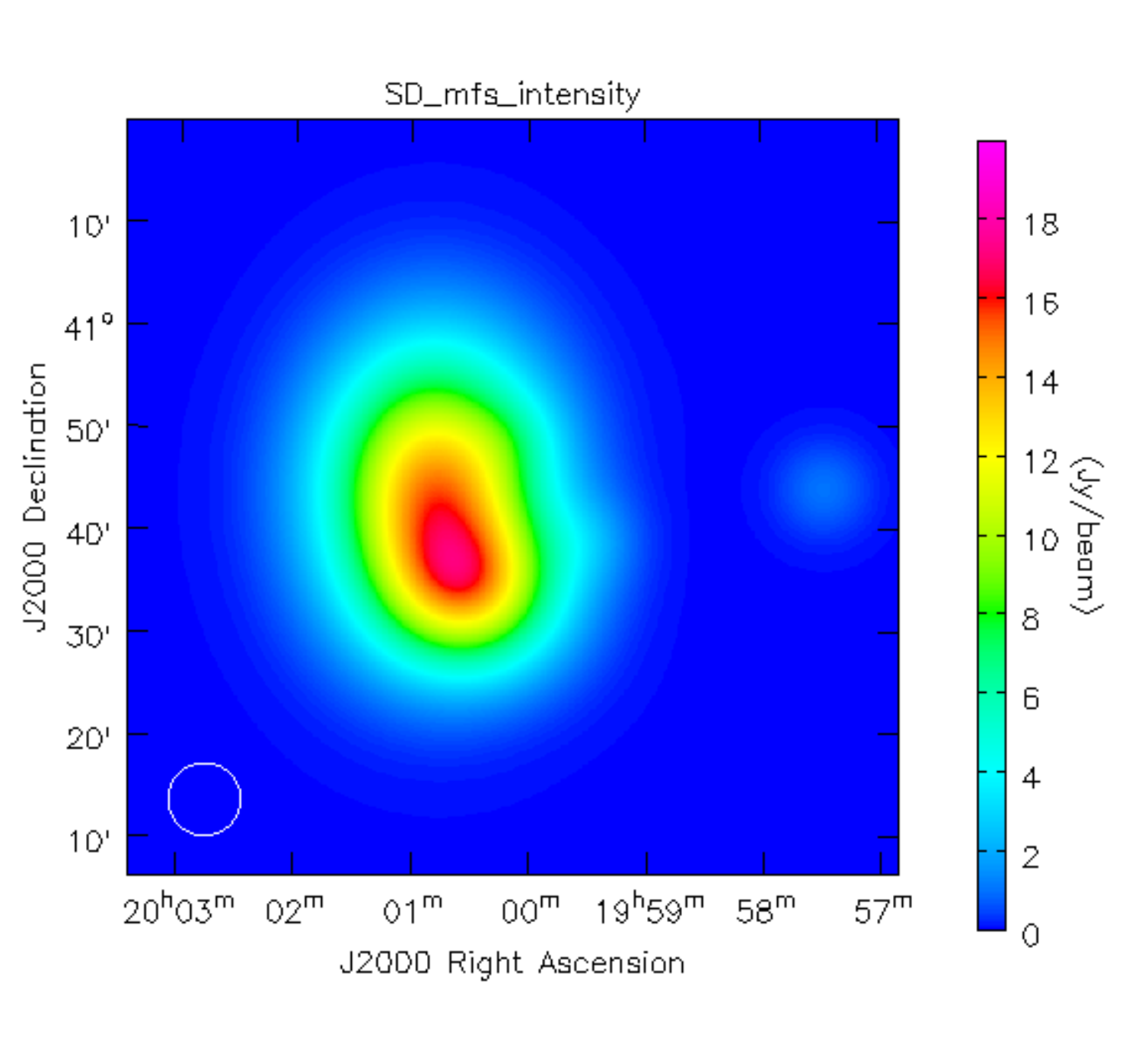}
\includegraphics[width=0.23\textwidth]{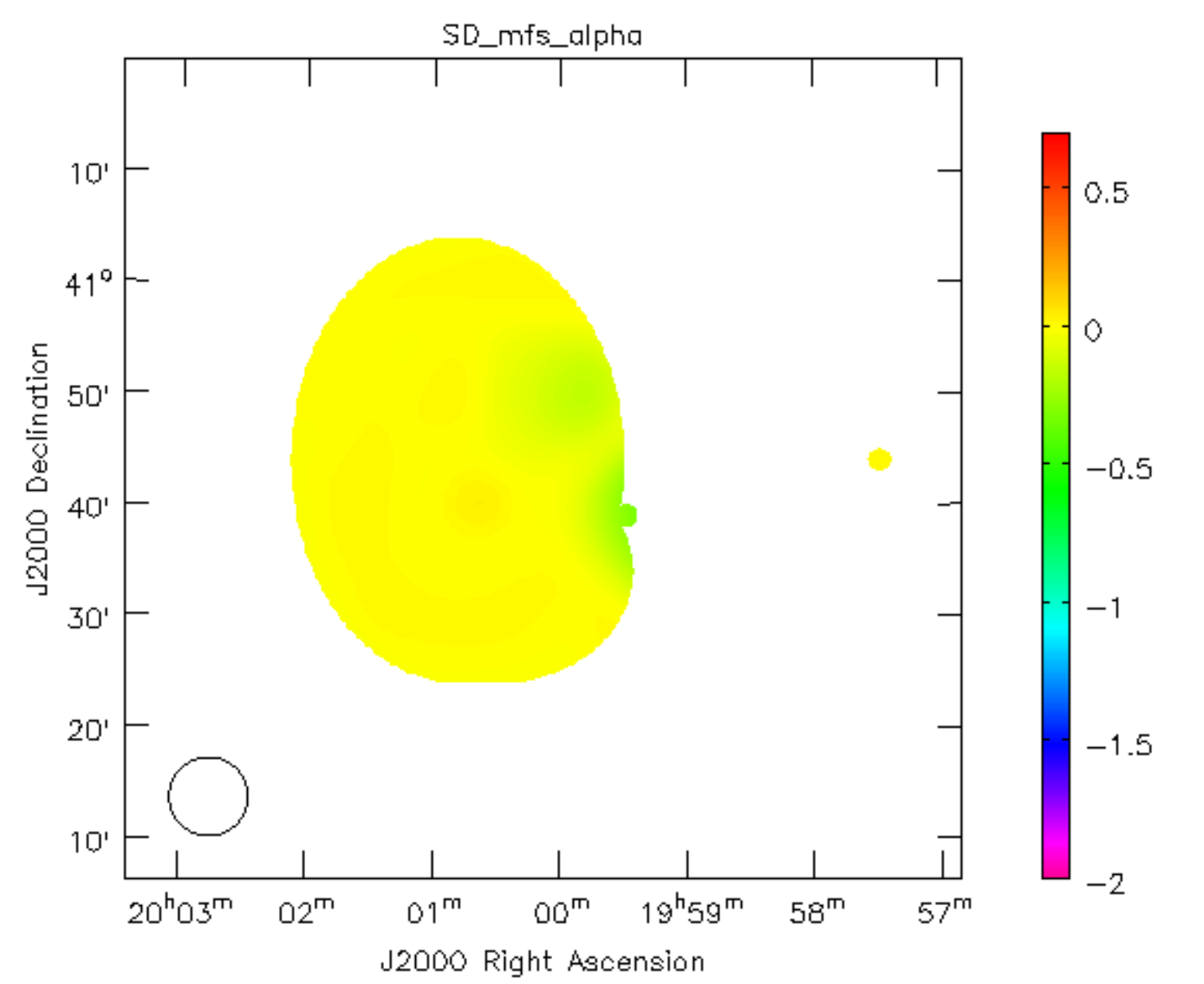}
\caption{Wideband MT-MFS Intensity and Spectral index from SD data only }
\label{Fig:mtmfs_sdonly}
\includegraphics[width=0.23\textwidth]{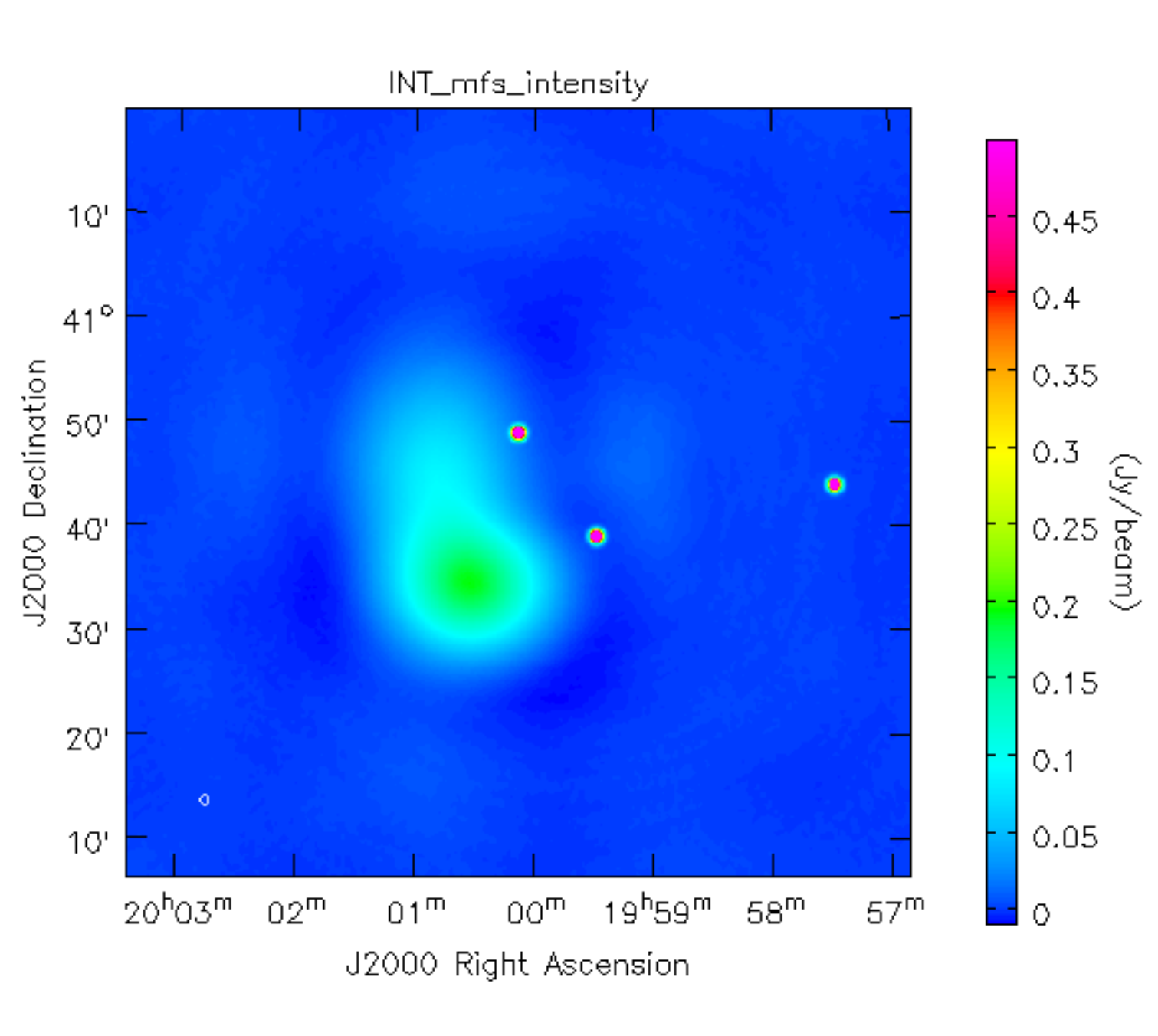}
\includegraphics[width=0.23\textwidth]{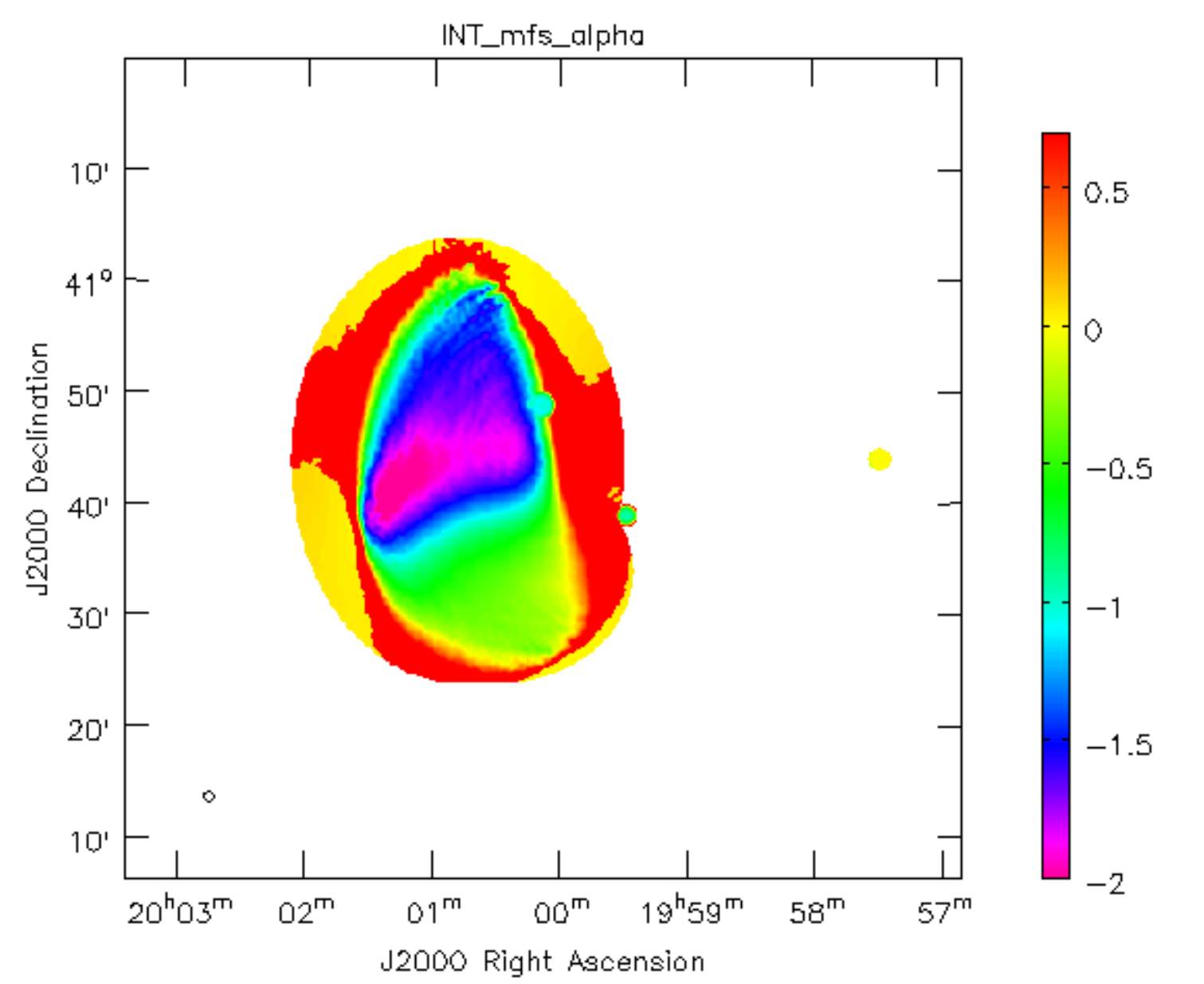}
\caption{Wideband MT-MFS Intensity and Spectral index from INT data only}
\label{Fig:mtmfs_intonly}
%
\includegraphics[width=0.23\textwidth]{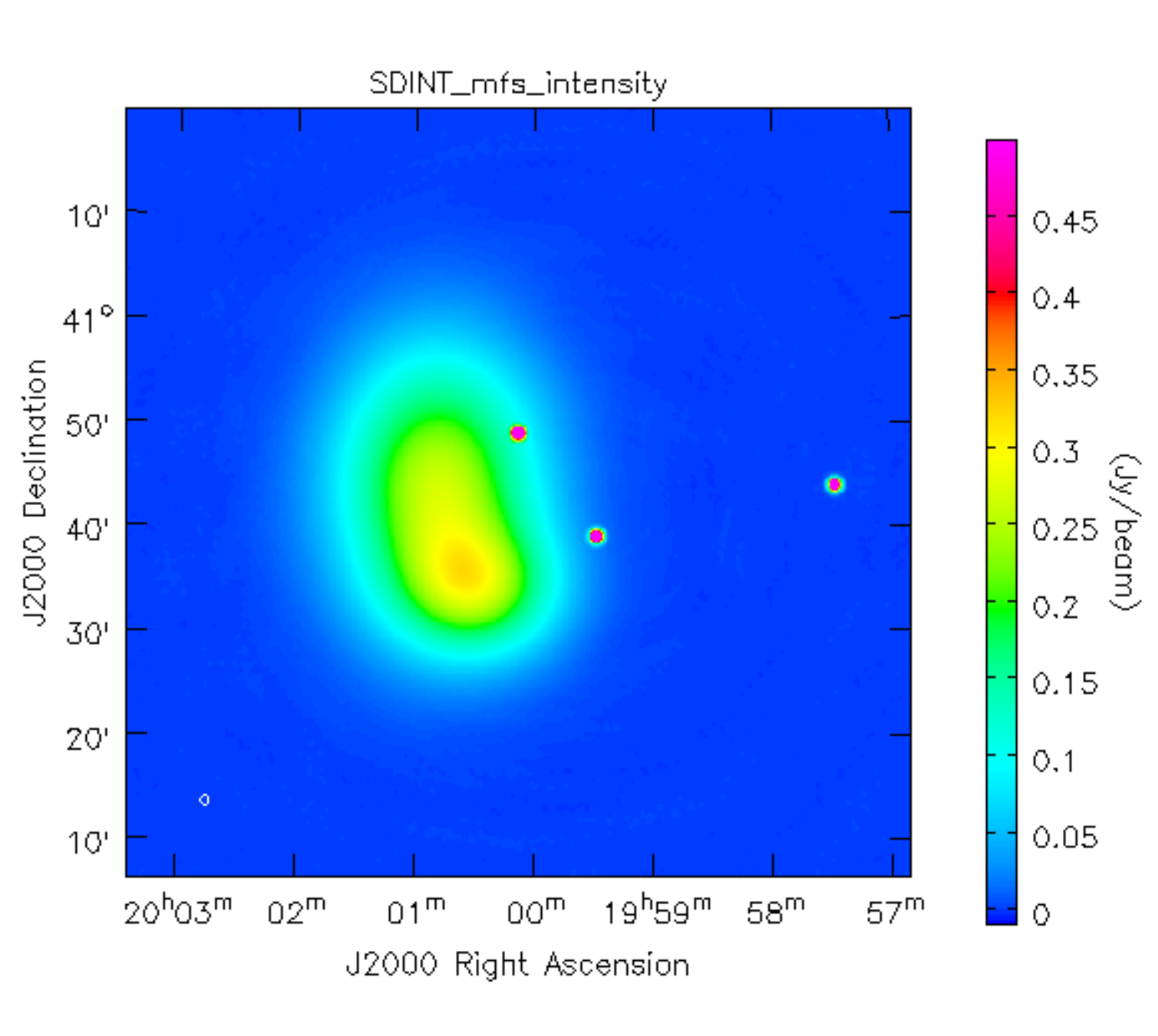}
\includegraphics[width=0.23\textwidth]{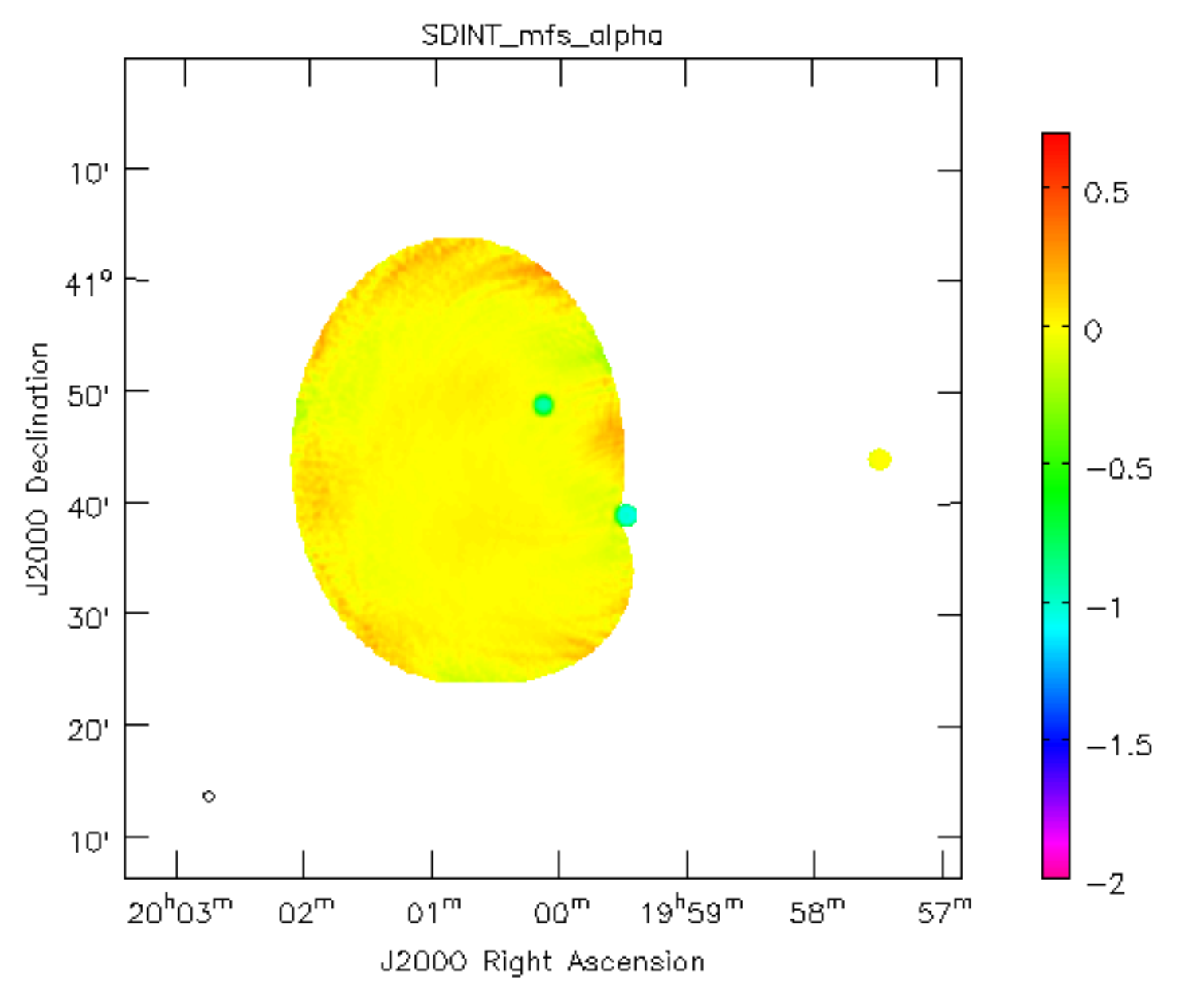}
\caption{Wideband MT-MFS Intensity and Spectral index from Joint SDINT reconstruction}
\label{Fig:mtmfs_sdint}
\includegraphics[width=0.23\textwidth]{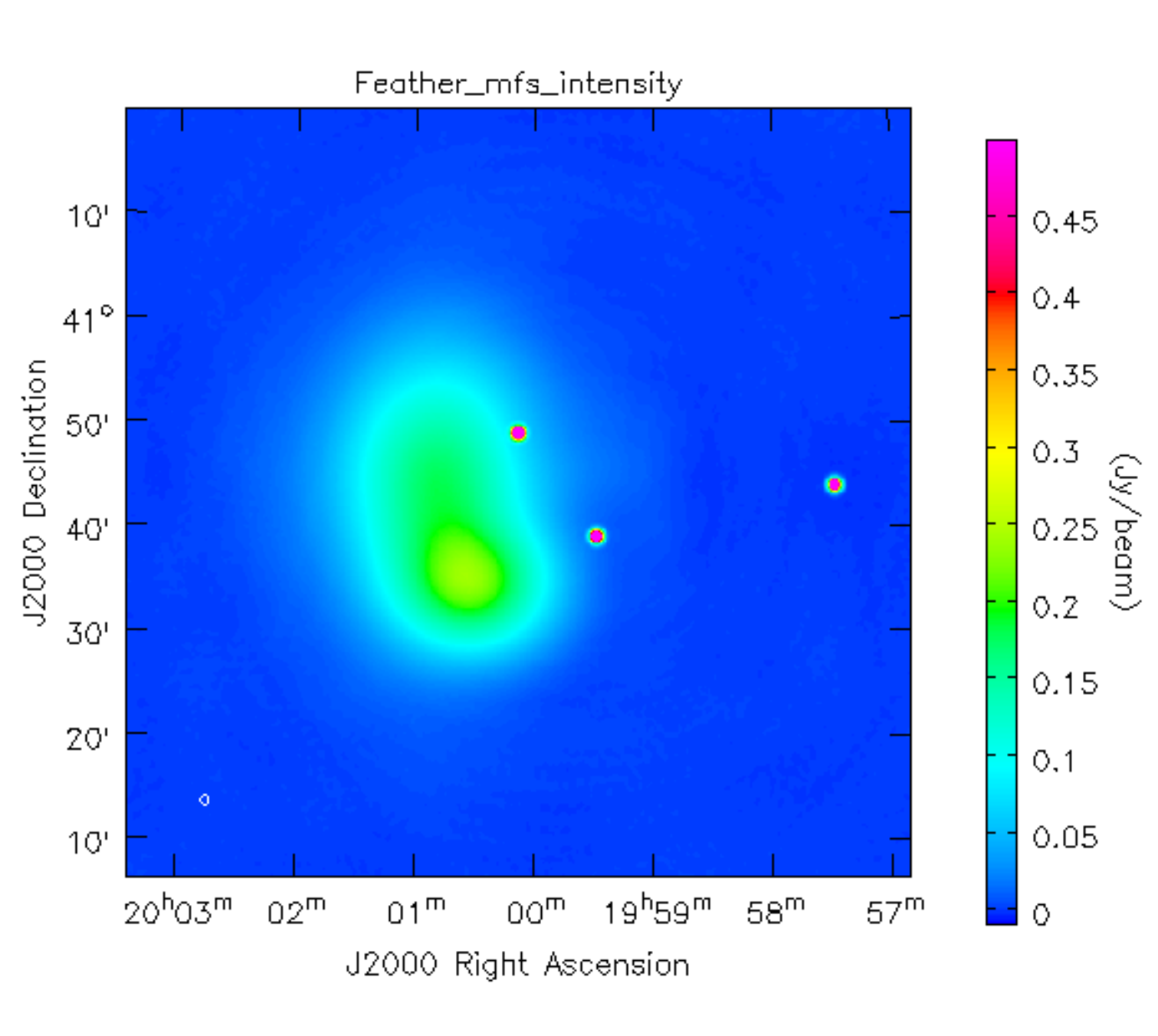}
\includegraphics[width=0.23\textwidth]{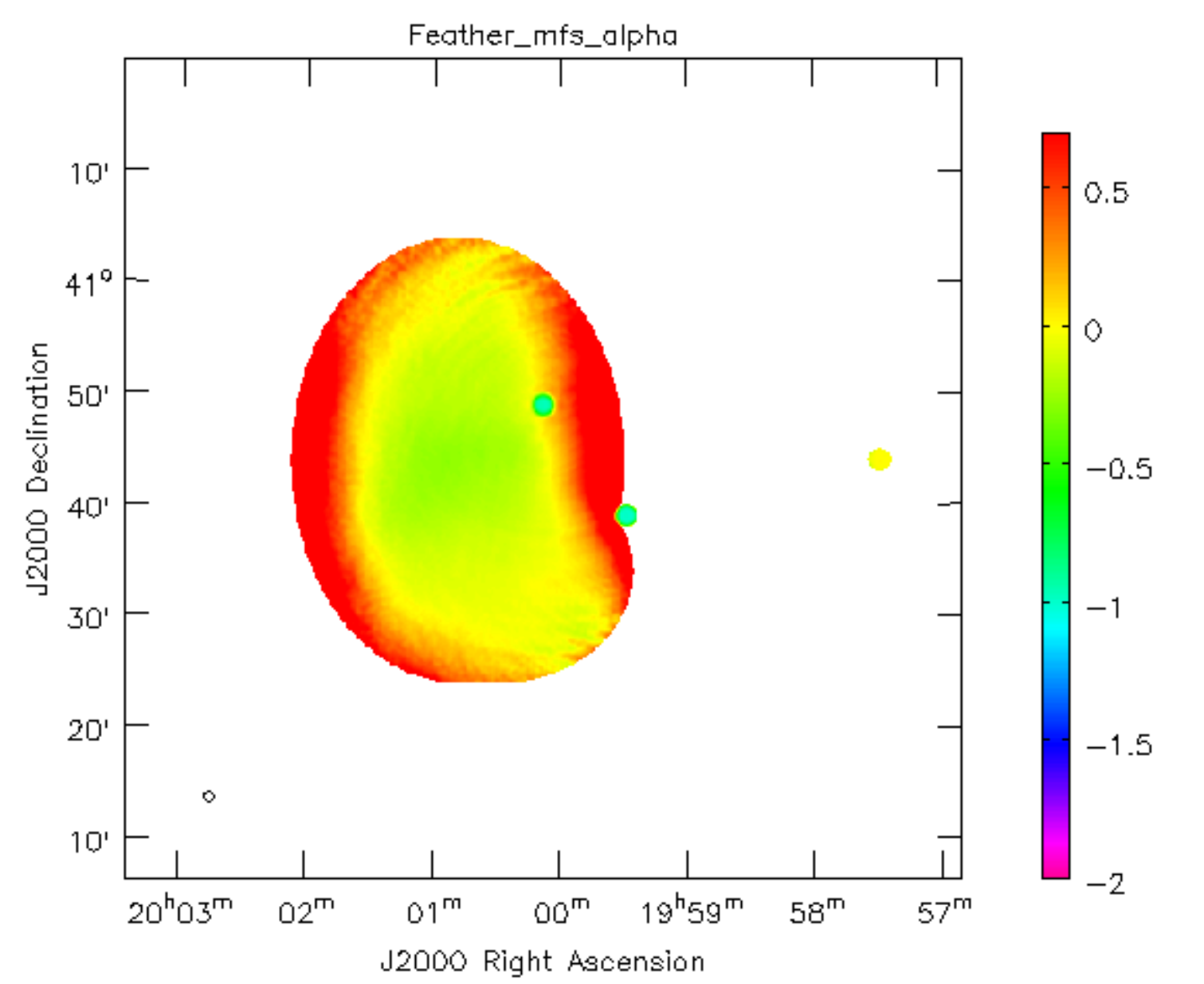}
\caption{Wideband MT-MFS Intensity and Spectral index from Feathering of Taylor coefficients}
\label{Fig:mtmfs_feather}
\includegraphics[width=0.23\textwidth]{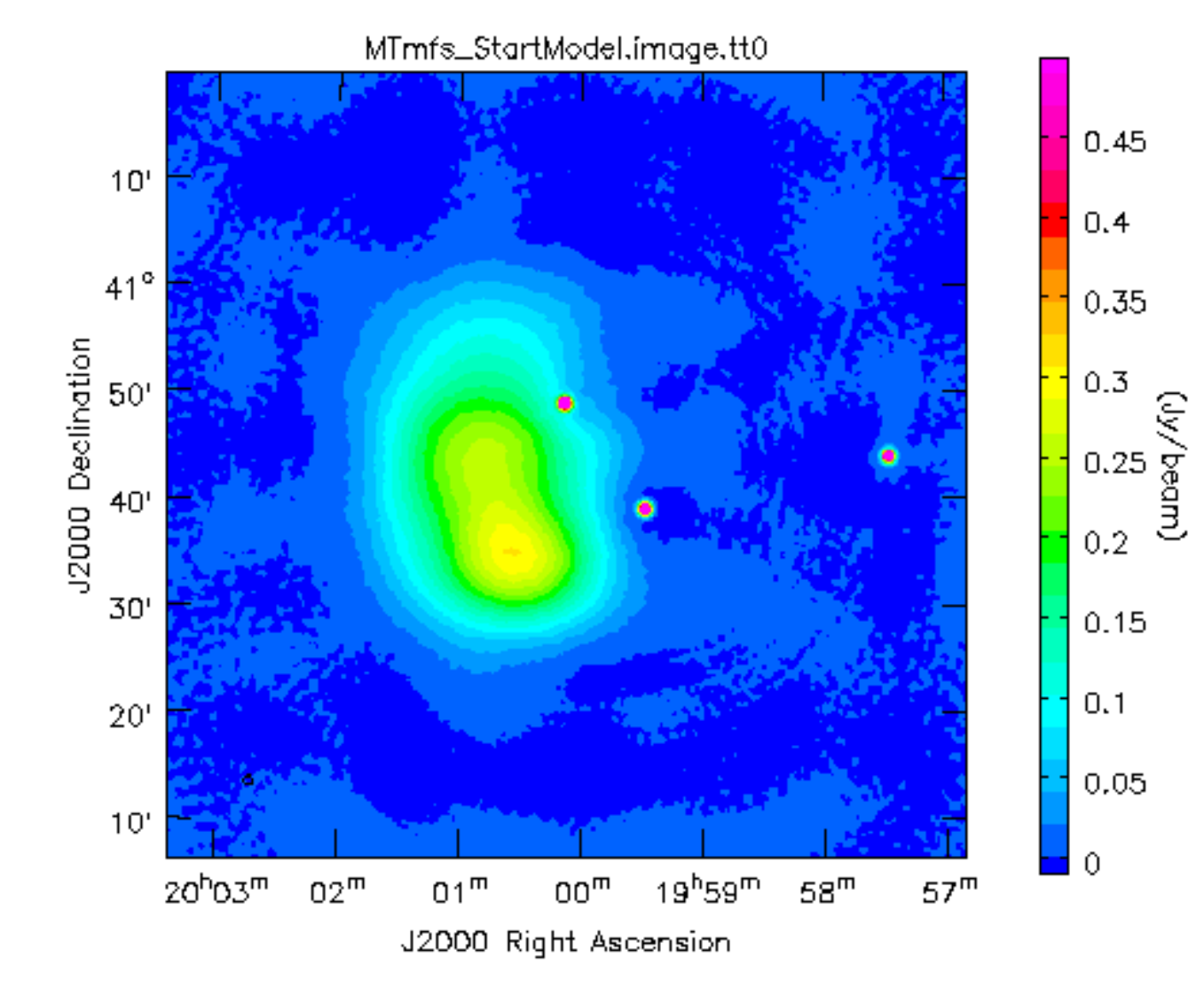}
\includegraphics[width=0.23\textwidth]{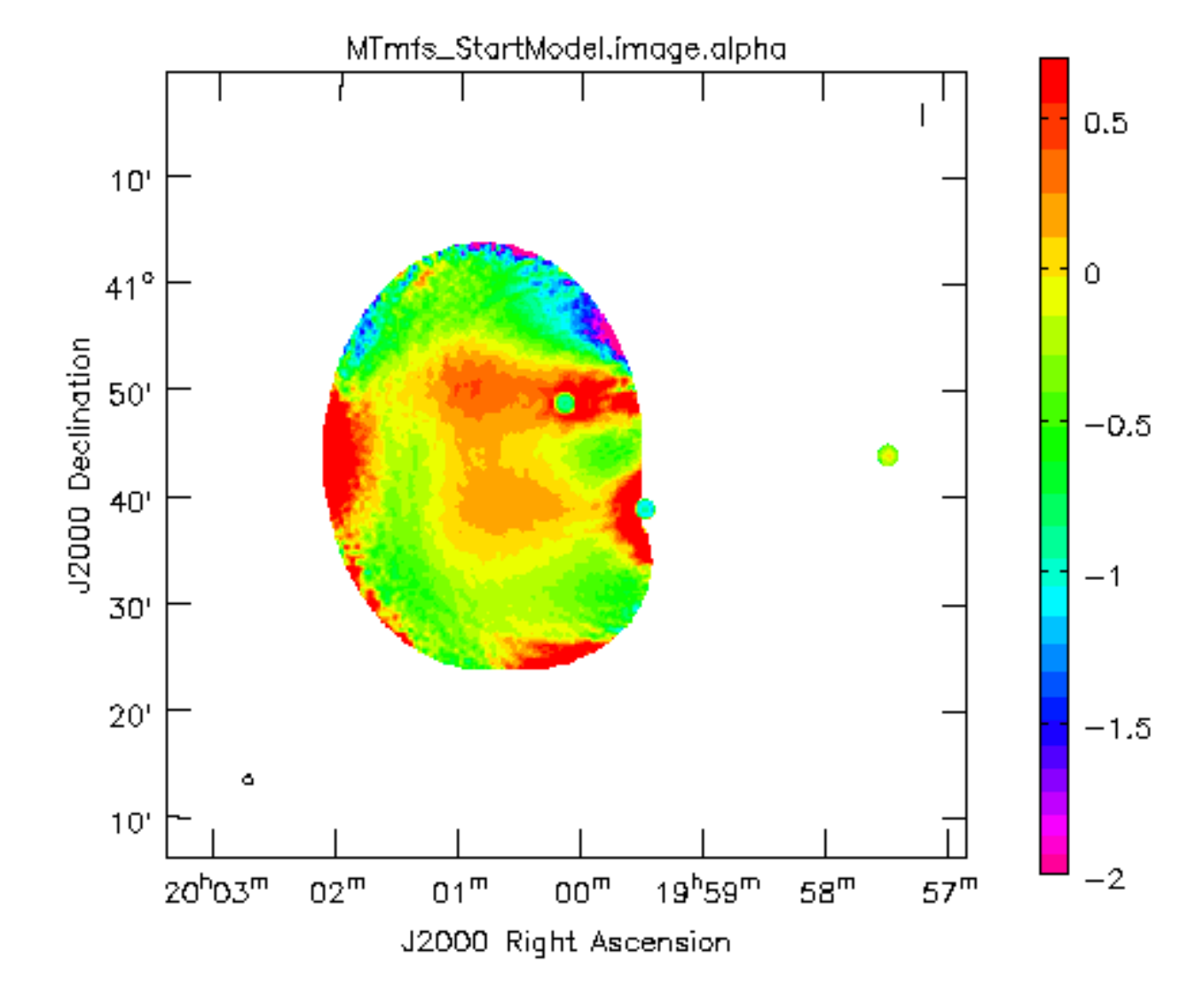}
\caption{Wideband MT-MFS Intensity and Spectral index from the Startmodel approach} 
\label{Fig:mtmfs_startmodel}
\end{figure}

Figures \ref{Fig:mtmfs_sdonly} through \ref{Fig:mtmfs_startmodel} show intensity
and spectral index maps for different algorithms and data combinations. 
Intensity maps may be compared with the true intensity at 1.5 GHz from 
Fig.\ref{Fig:true_intensity}. The ideal spectral index of the entire extended 
component is 0.0 and mapped to yellow in the chosen color scheme. The
three point sources have spectral indices of -1.0, -1.0 and 0.0, mapping to
green, green and yellow respectively. 
Figure \ref{Fig:mtmfs_sdonly} shows wideband multi-term deconvolution results from the 
single dish data alone, showing accurate spectral index recovery but at the
low single dish angular resolution.
Figure \ref{Fig:mtmfs_intonly} shows results from the
interferometer data alone, showing missing flux, large scale sidelobes and 
overly-steep spectral structure with the largest spatial structure feature
having the most error. 

Figure \ref{Fig:mtmfs_sdint} shows results from the
\ALGO algorithm which produced the most accurate representation of the true
multi-scale wideband sky at high angular resolution.  
Figure \ref{Fig:mtmfs_feather} shows the result of feathering the output
Taylor-coefficient images from the SD and INT algorithms 
(i.e.from Figs.\ref{Fig:mtmfs_sdonly} and \ref{Fig:mtmfs_intonly}). 
A post-deconvolution wideband primary beam correction was applied
to the INT-only Taylor coefficient images prior to feathering. 
Feathering certainly produced an improvement over INT-only imaging
but was not as accurate as the \ALGO approach.
Figure \ref{Fig:mtmfs_startmodel} shows results from the startmodel approach in which a
multi-term imaging run was performed using single dish Taylor coefficient model images
(i.e. model images Fig.\ref{Fig:mtmfs_sdonly} as 
starting models per term. In this example, the constraints from the single dish model 
definitely helped, but were insufficient to produce an accurate spectral index
reconstruction. 

\begin{figure}
\includegraphics[width=0.23\textwidth]{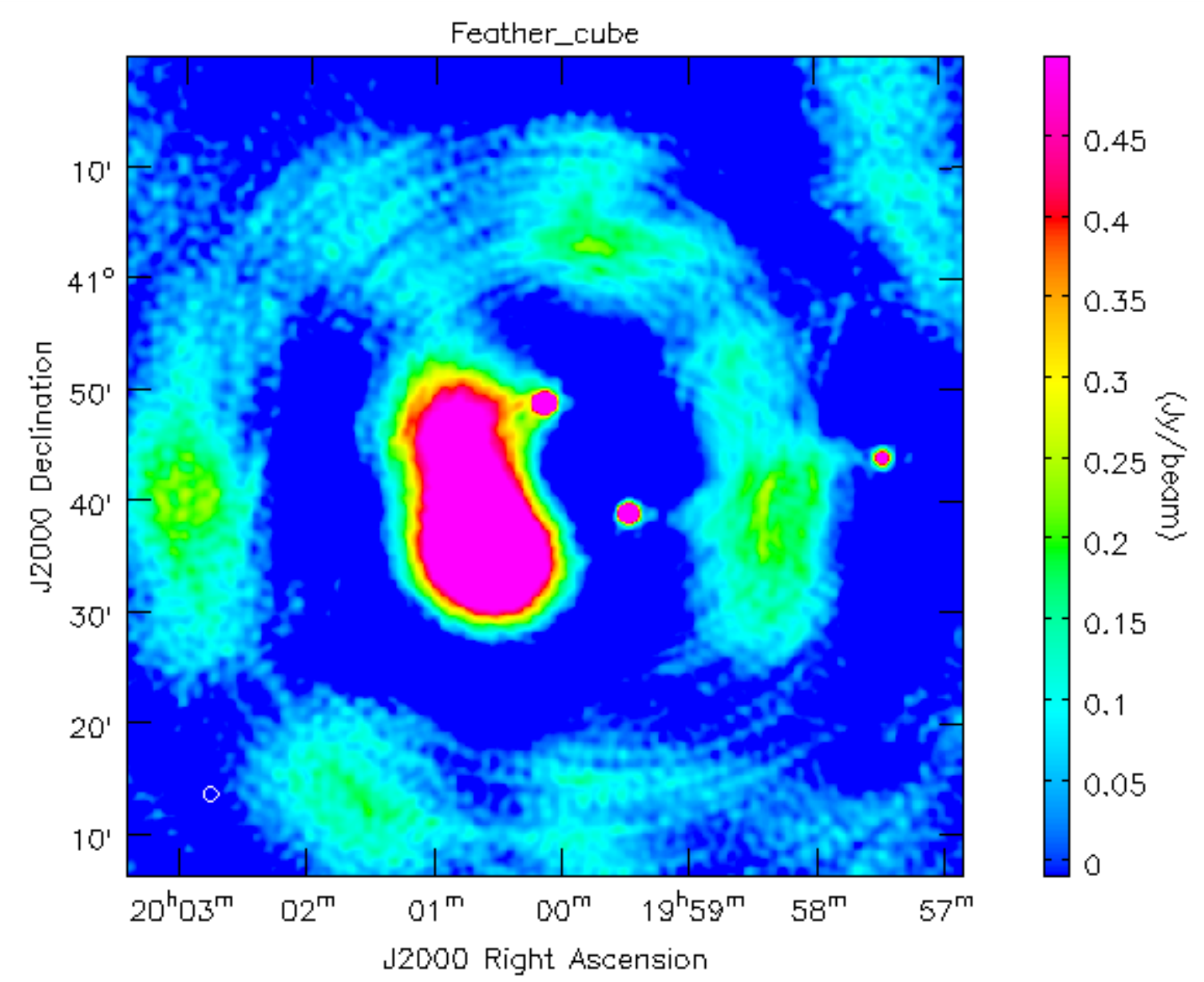}
\includegraphics[width=0.23\textwidth]{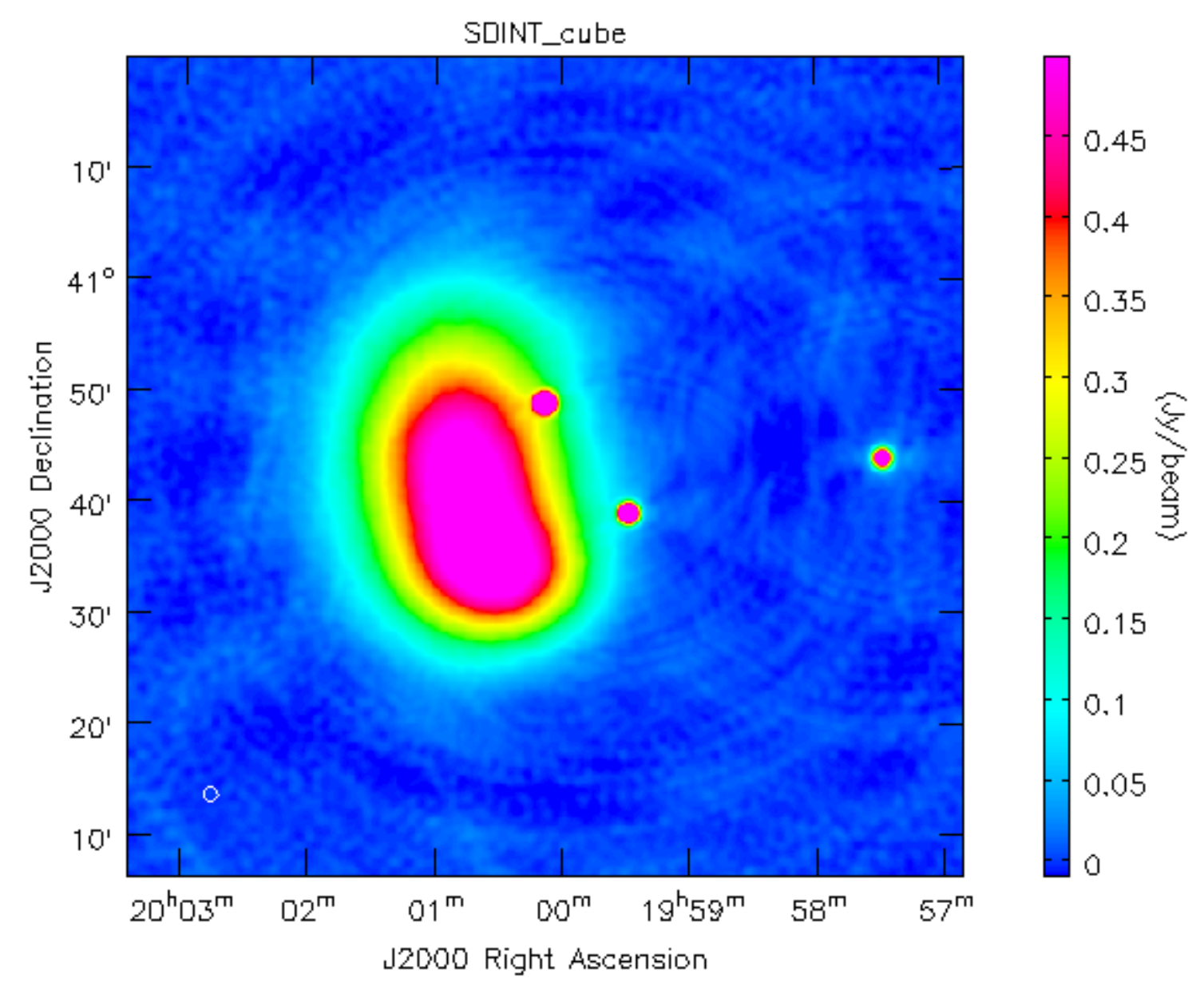}
\includegraphics[width=0.23\textwidth]{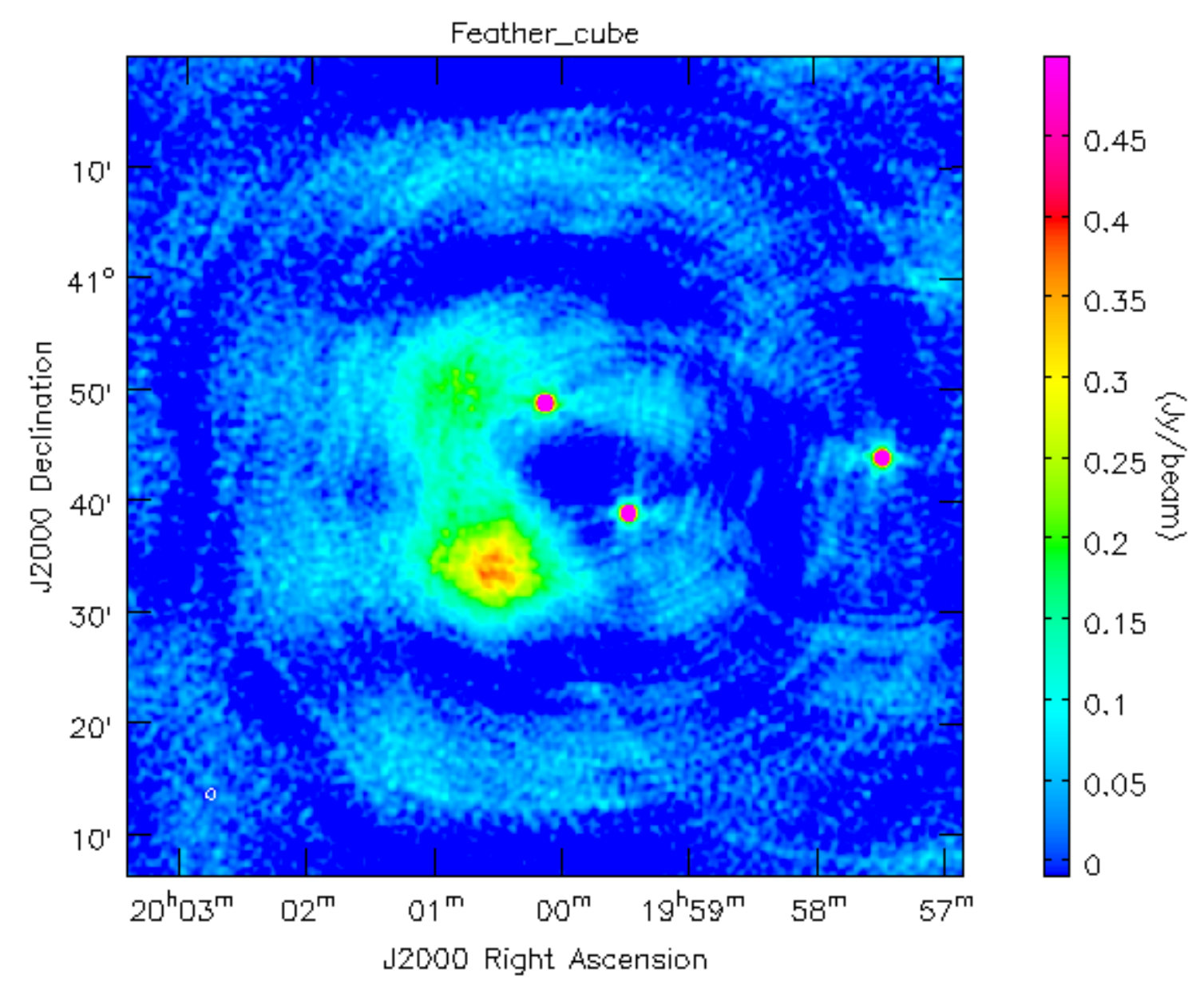}
\includegraphics[width=0.23\textwidth]{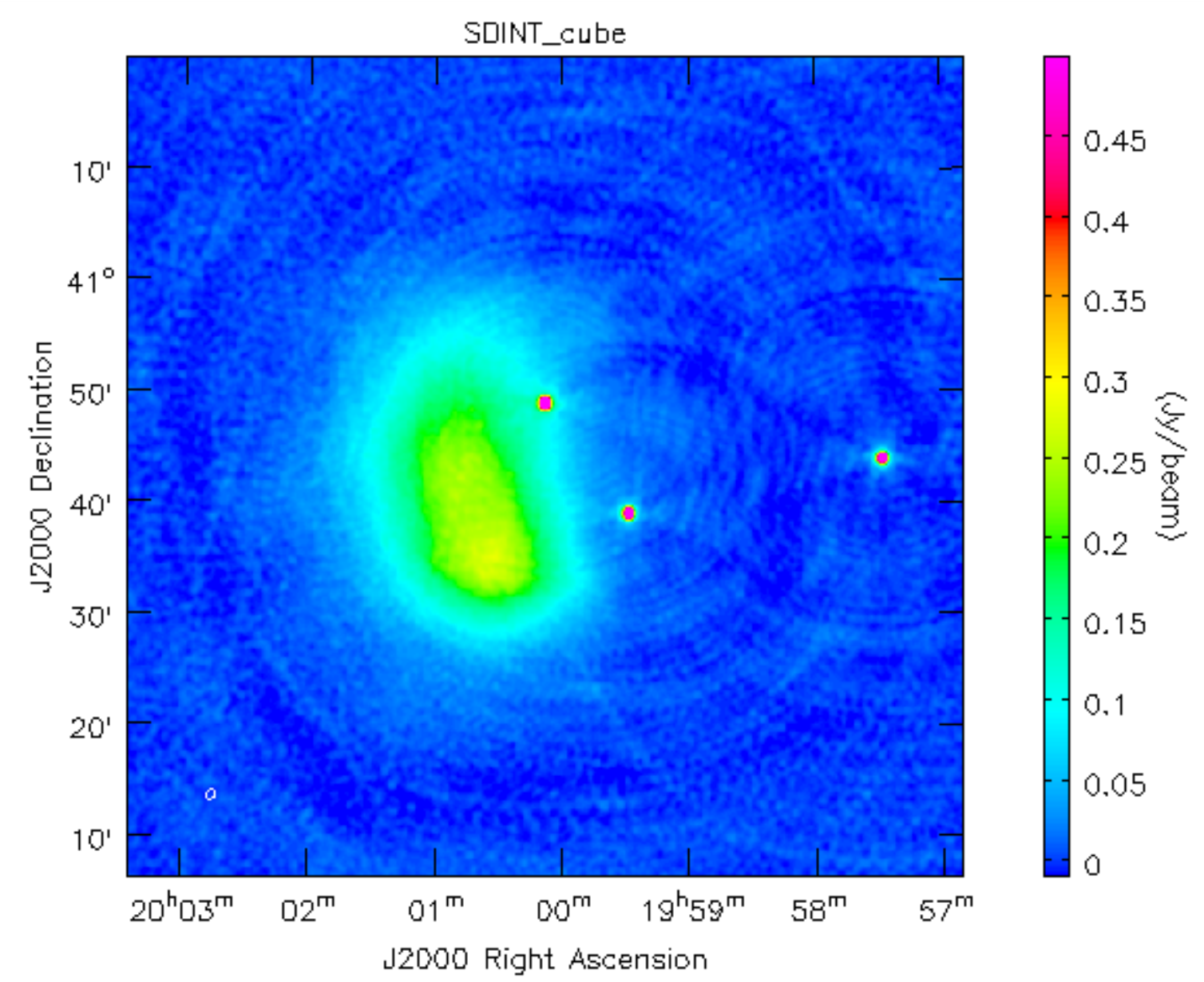}
\includegraphics[width=0.23\textwidth]{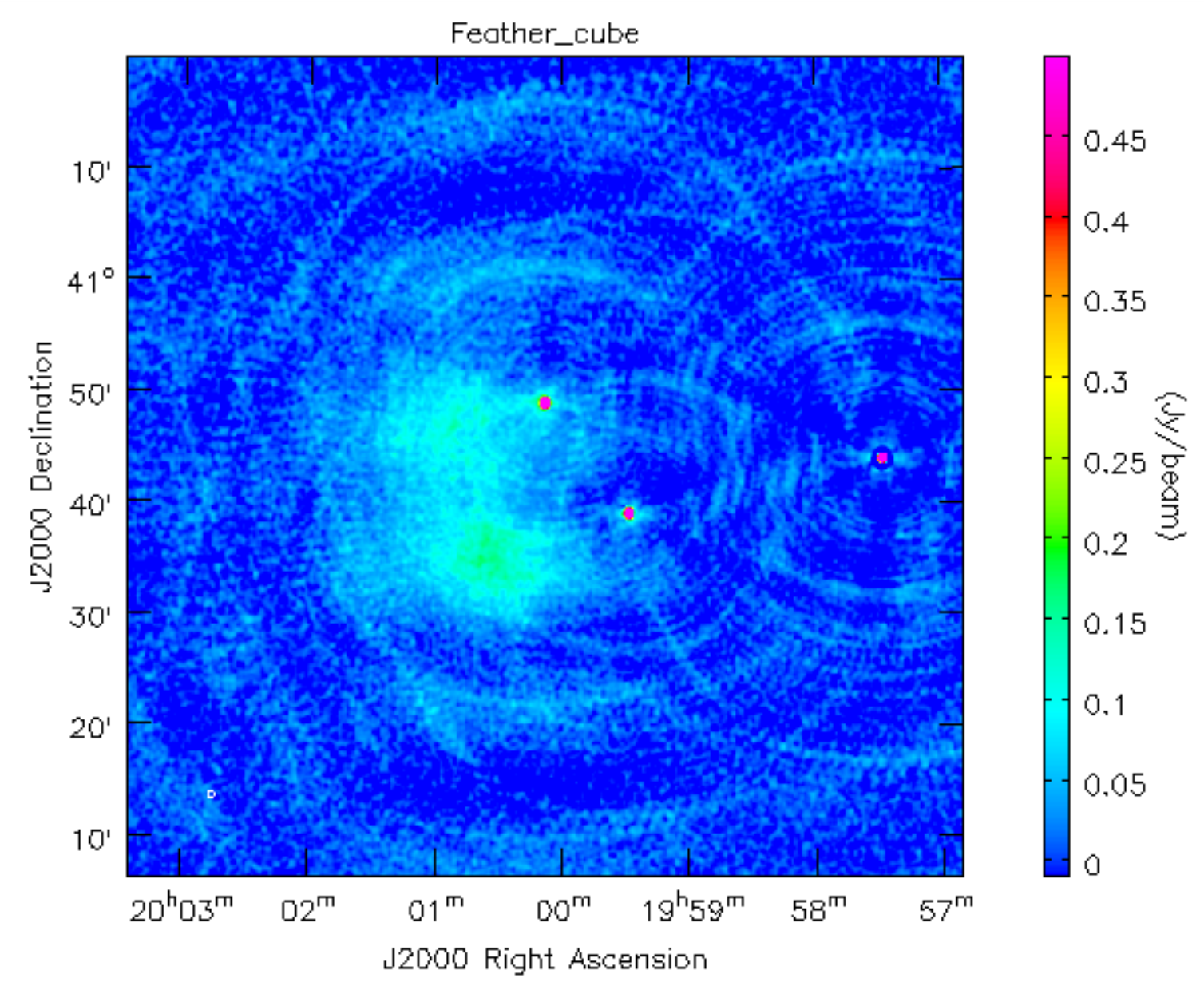}
\includegraphics[width=0.23\textwidth]{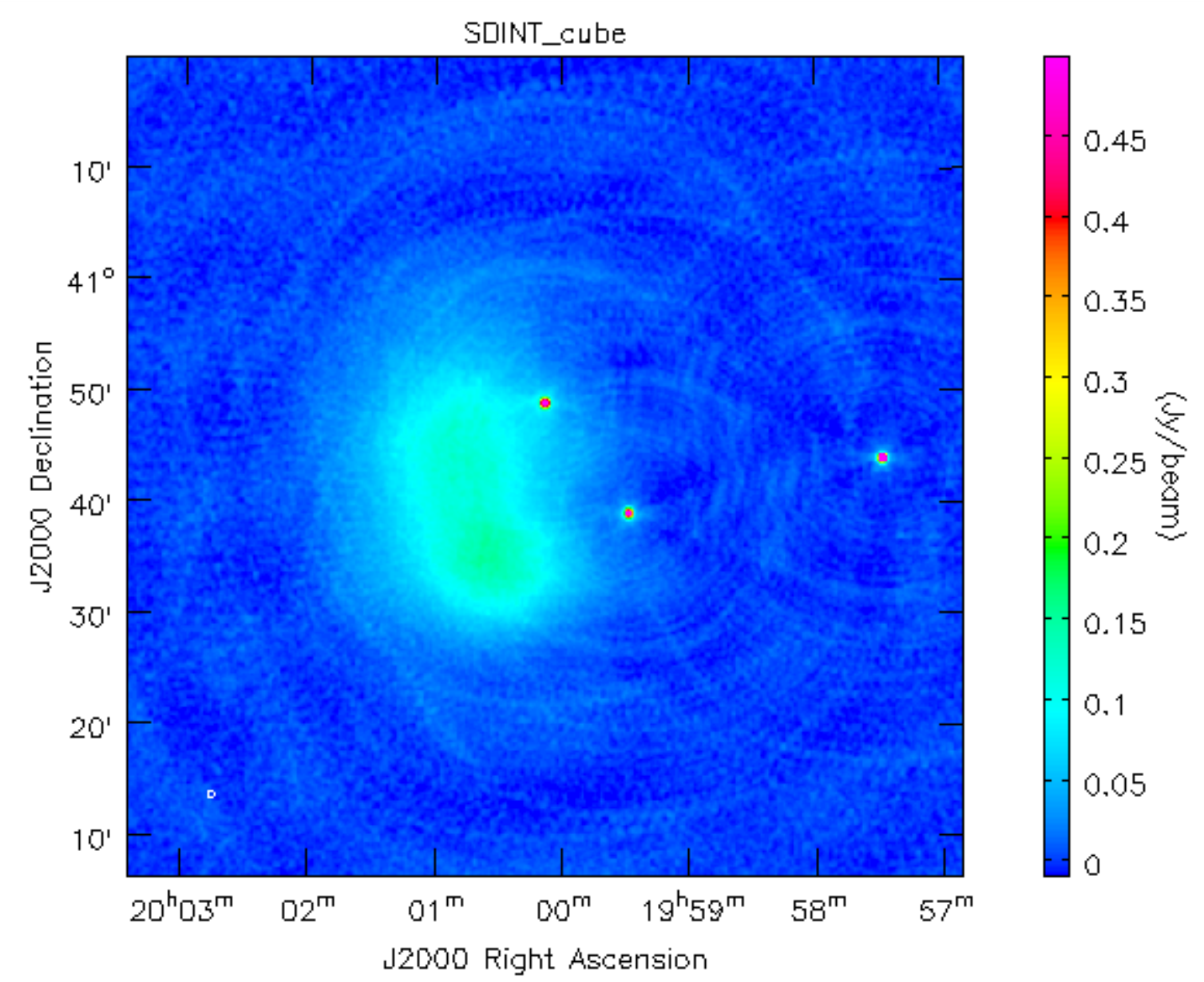}
\caption{Feathered (LEFT) and Joint SDINT (RIGHT) Spectral Cube (1.0, 1.5, 2.0 GHz)}
\label{Fig:cube_feather_sdint}
\end{figure}

Figure \ref{Fig:cube_feather_sdint} shows results from spectral cube imaging with
feathering on the left column and joint \ALGO deconvolution on the right. 
With the relatively higher PSF sidelobes for single-channel imaging, the fixed
number of iterations chosen for all imaging runs was insuffiicent for perfect
spectral cube reconstructions but are sufficient to illustrate the reconstruction
quality at the larger spatial scales.


\subsection{Algorithm Comparison - Mosaic Imaging}\label{Sec:withbeams}

\begin{figure}
\centering
\includegraphics[width=0.23\textwidth]{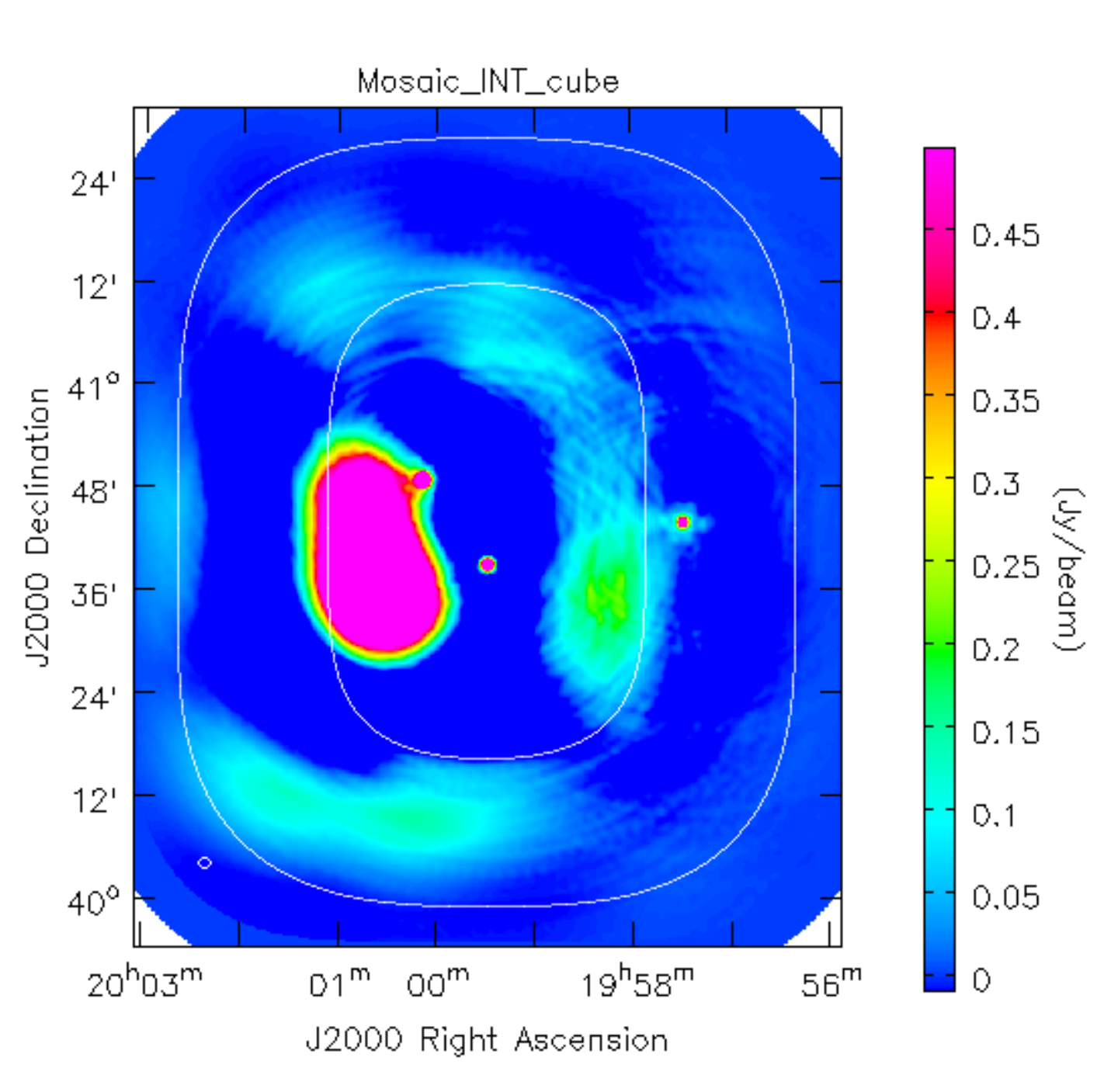}
\includegraphics[width=0.23\textwidth]{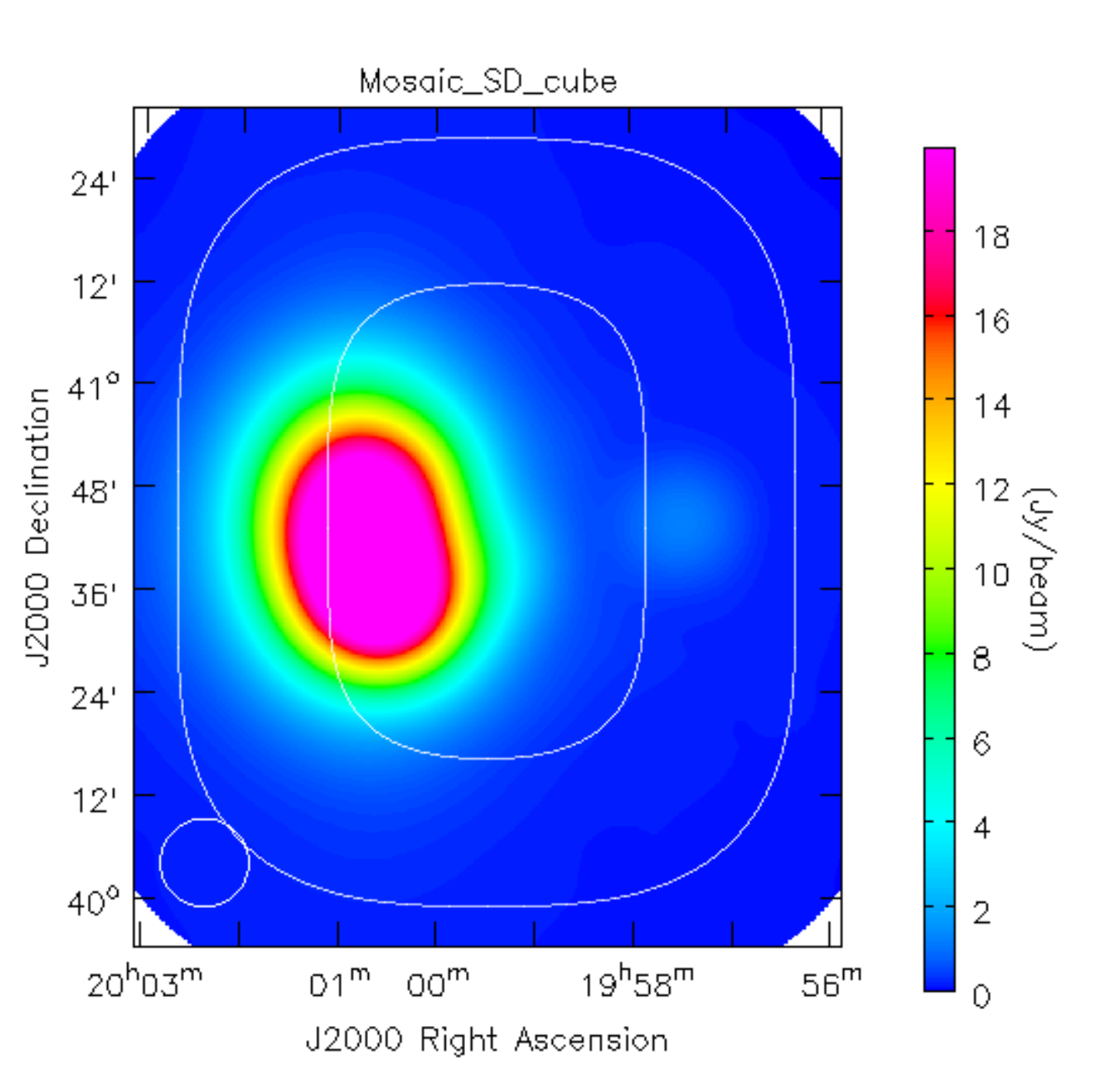}
\includegraphics[width=0.23\textwidth]{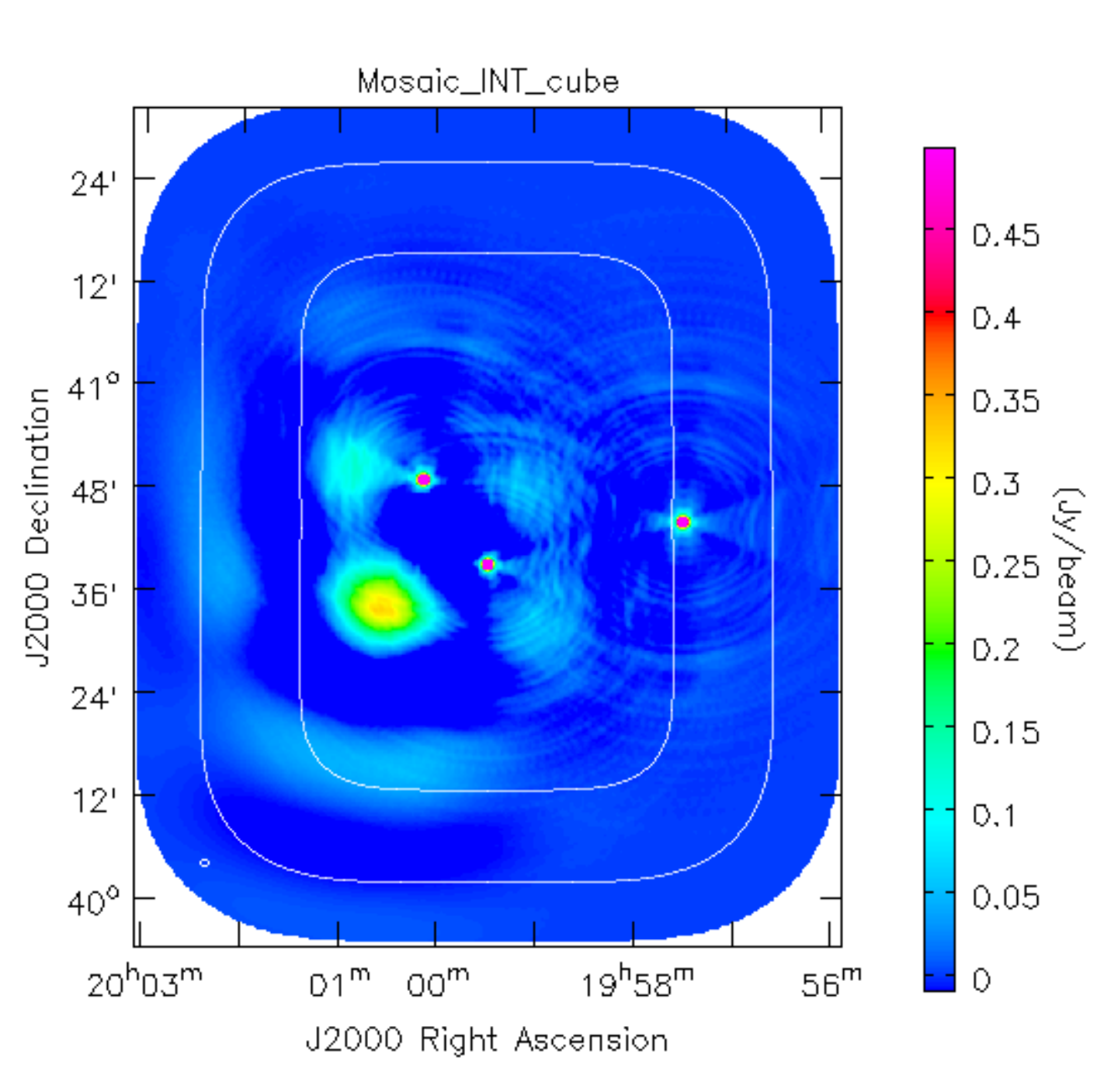}
\includegraphics[width=0.23\textwidth]{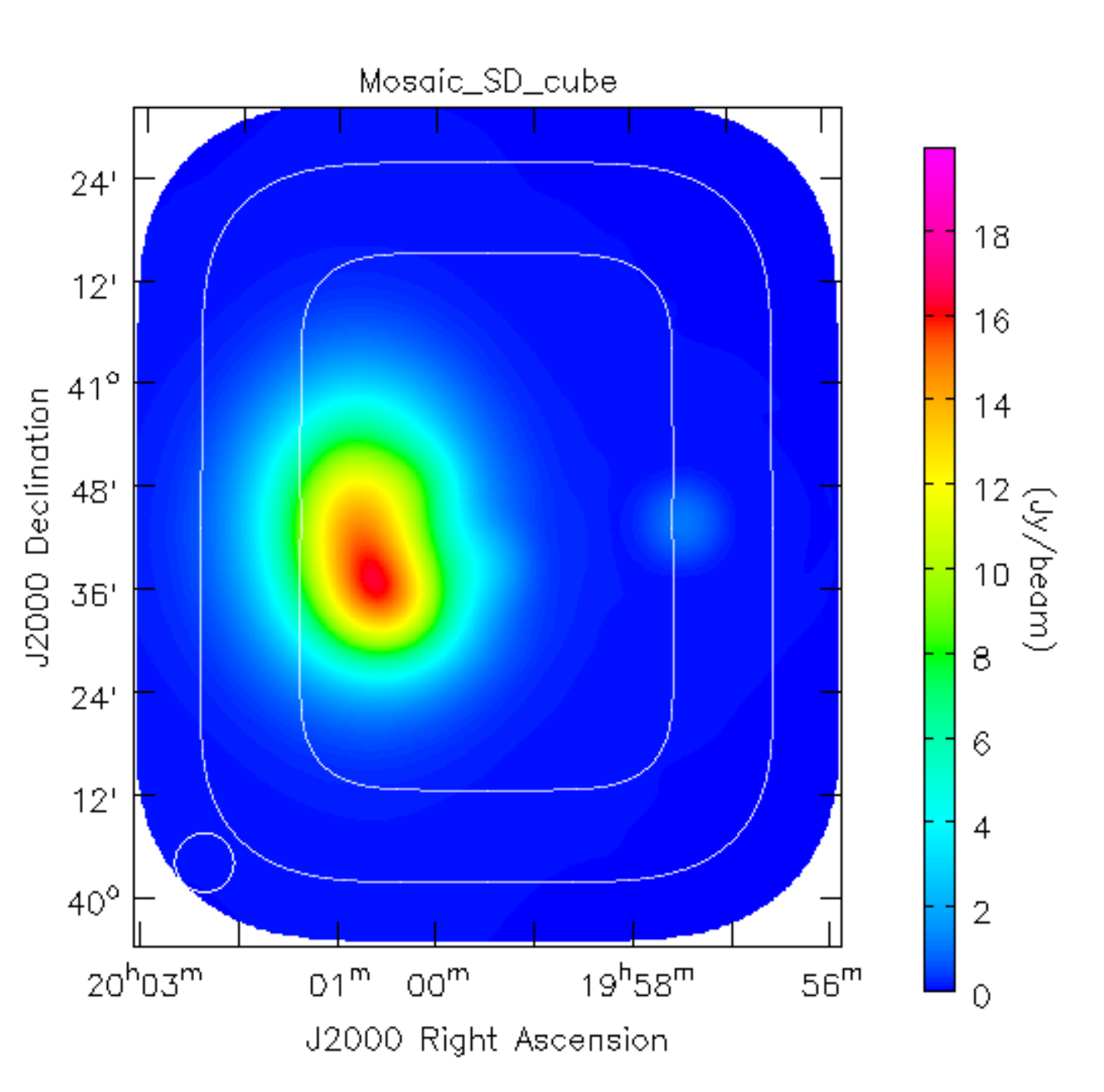}
\includegraphics[width=0.23\textwidth]{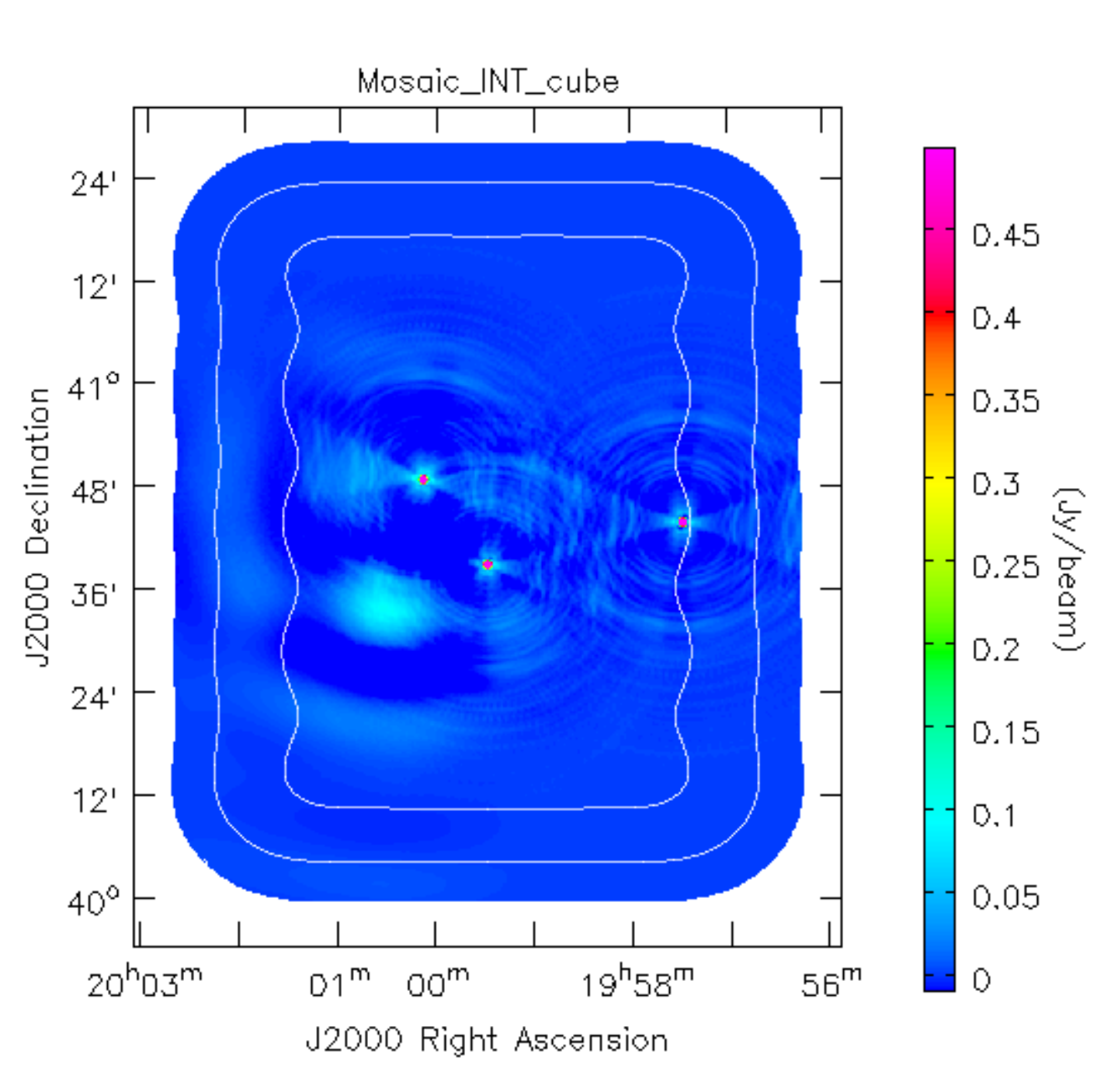}
\includegraphics[width=0.23\textwidth]{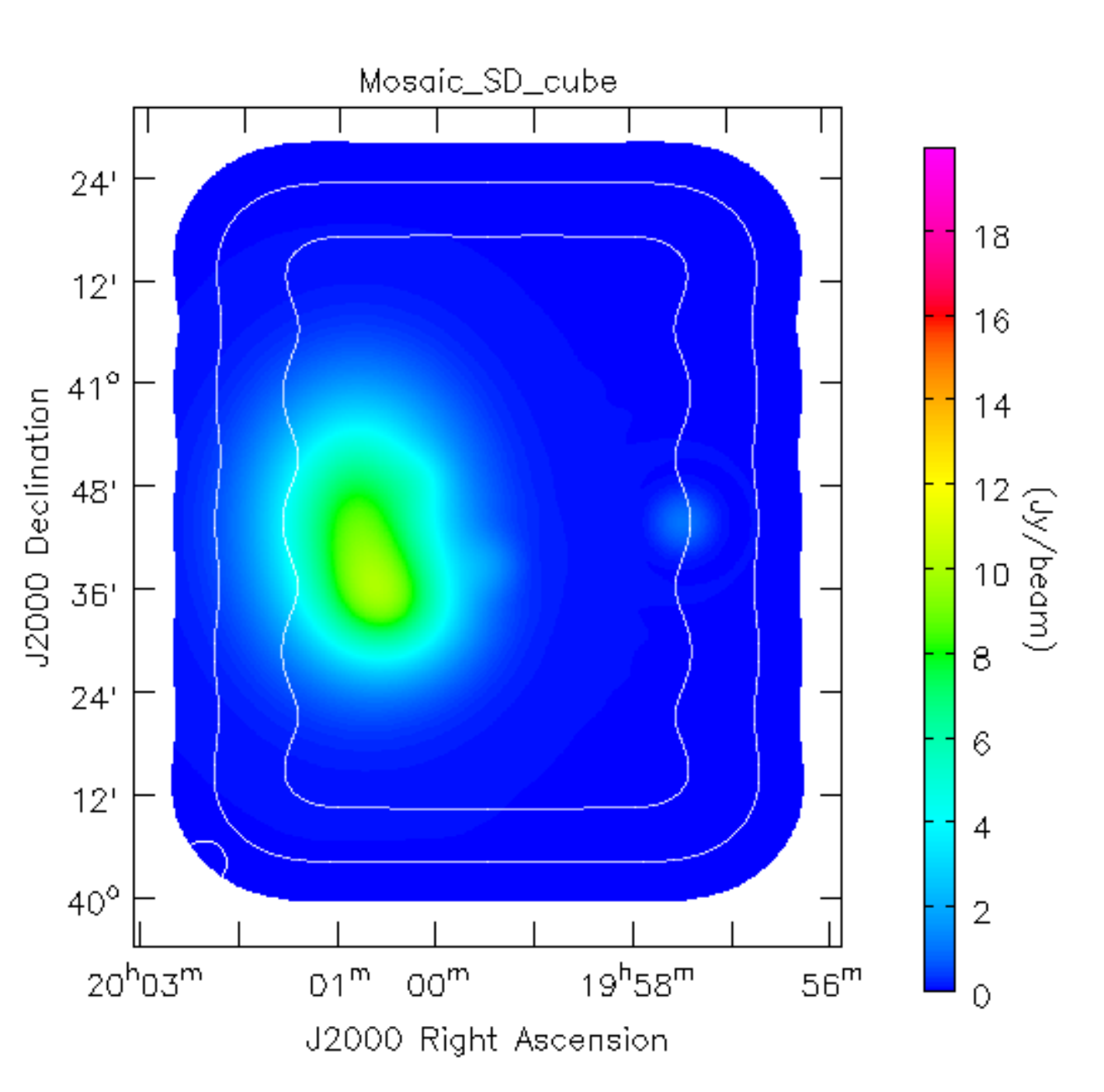}
\caption{Mosaic Spectral Cube from INT data only (LEFT) and SD data only (RIGHT) 
at 1.0GHz, 1.5GHz and 2.0GHz (top to bottom).}
\label{Fig:mos_cube_intonly_sdonly}
\end{figure}

\begin{figure}
\includegraphics[width=0.23\textwidth]{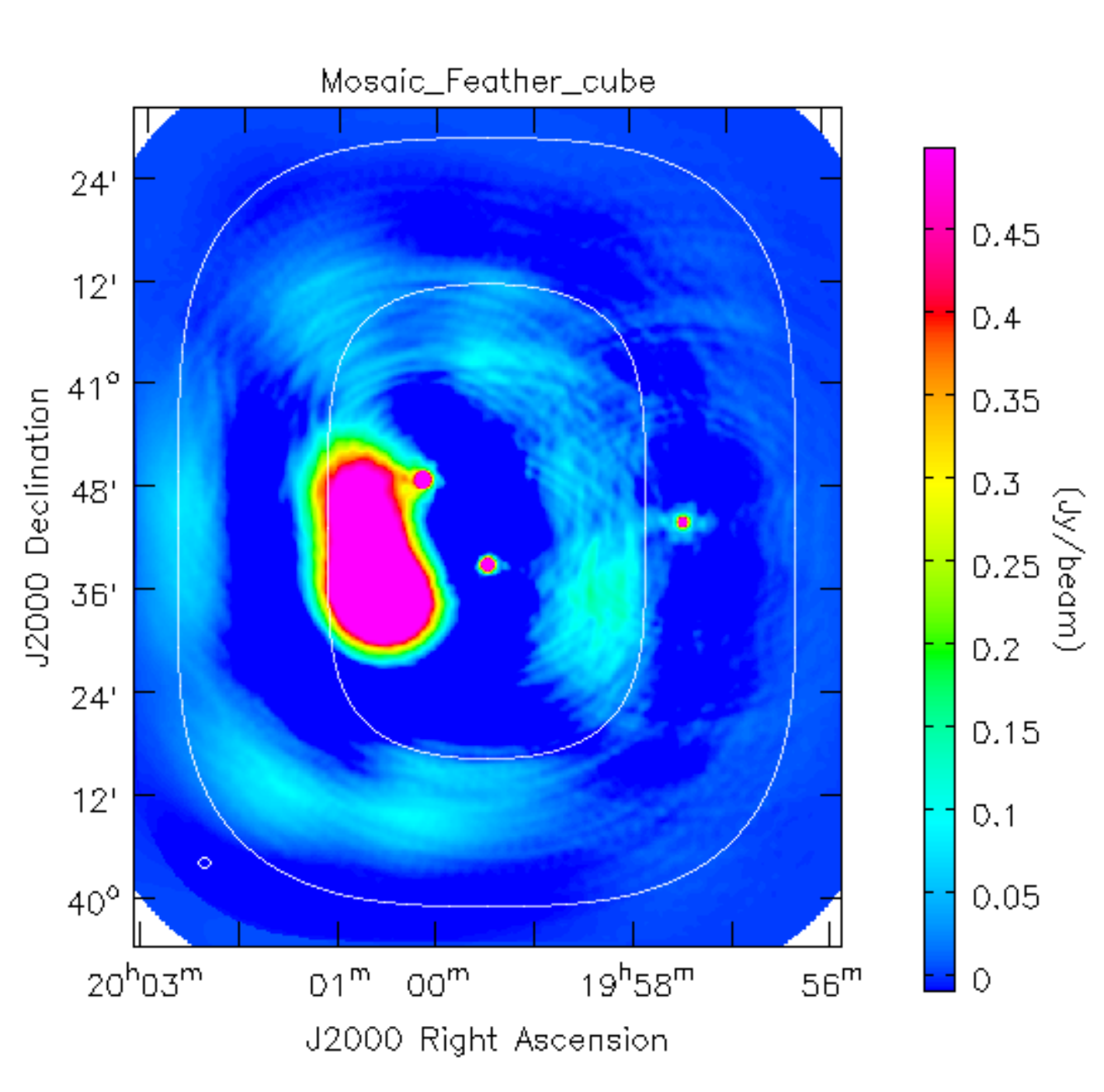}
\includegraphics[width=0.23\textwidth]{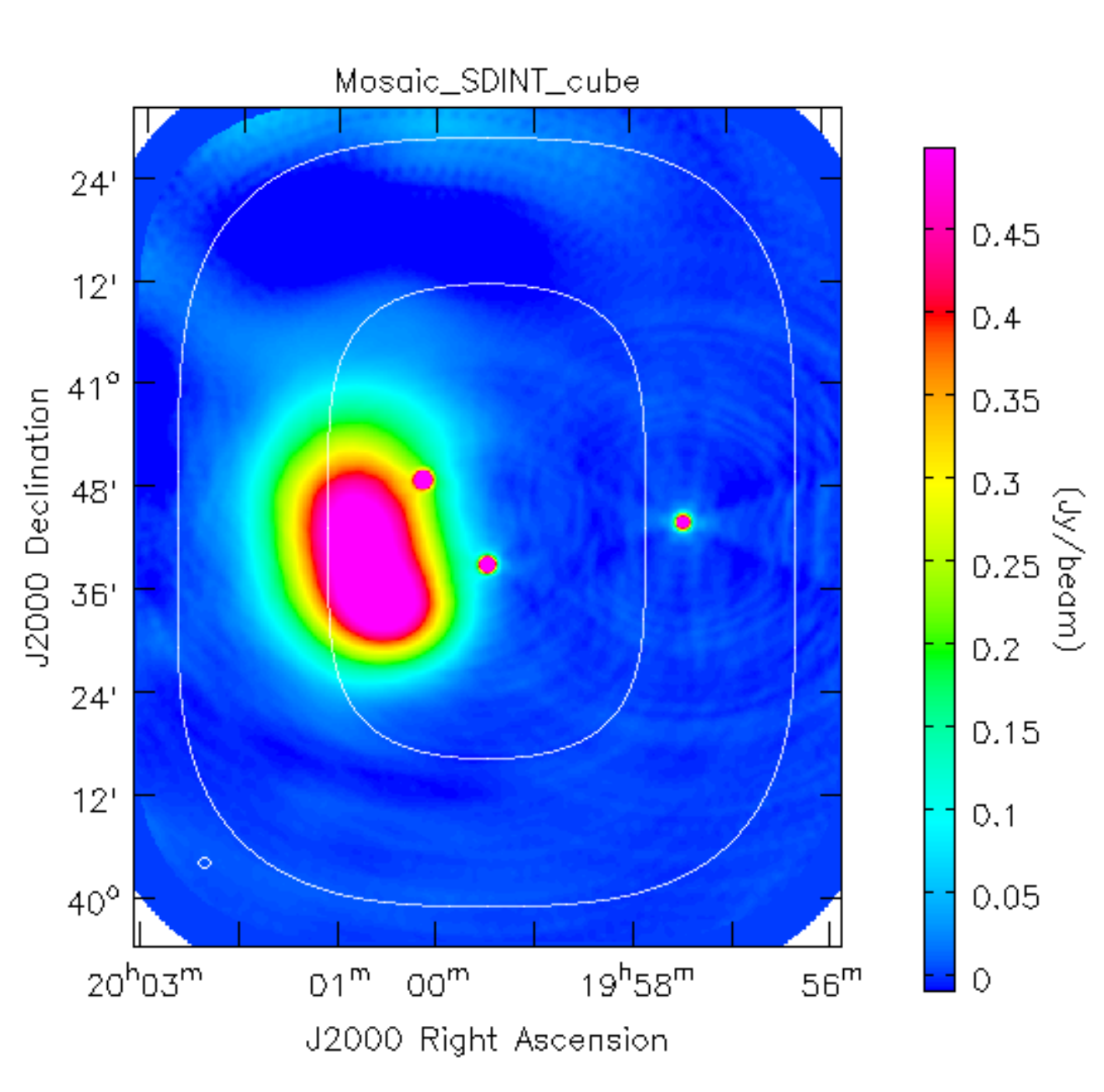}
\includegraphics[width=0.23\textwidth]{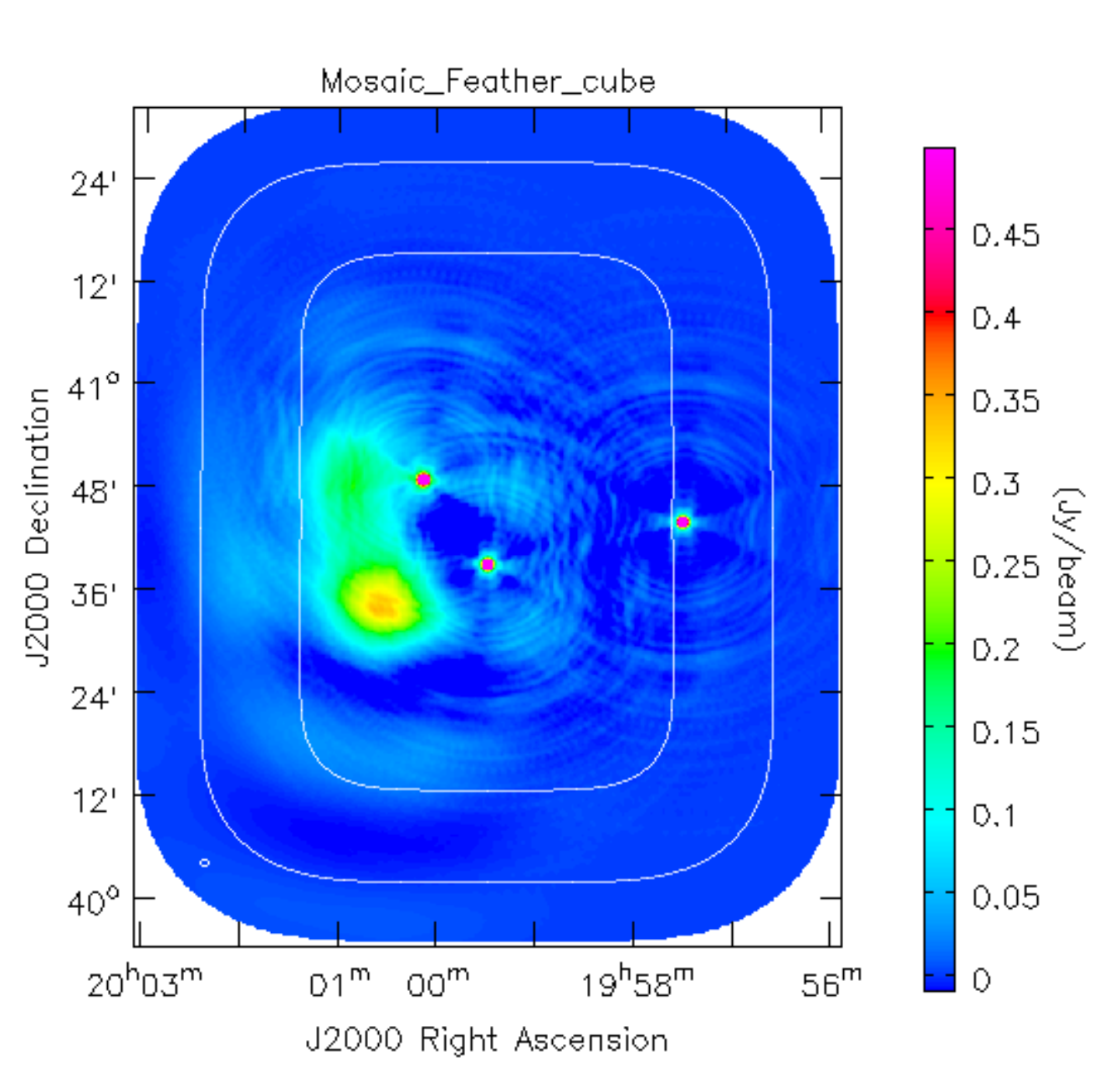}
\includegraphics[width=0.23\textwidth]{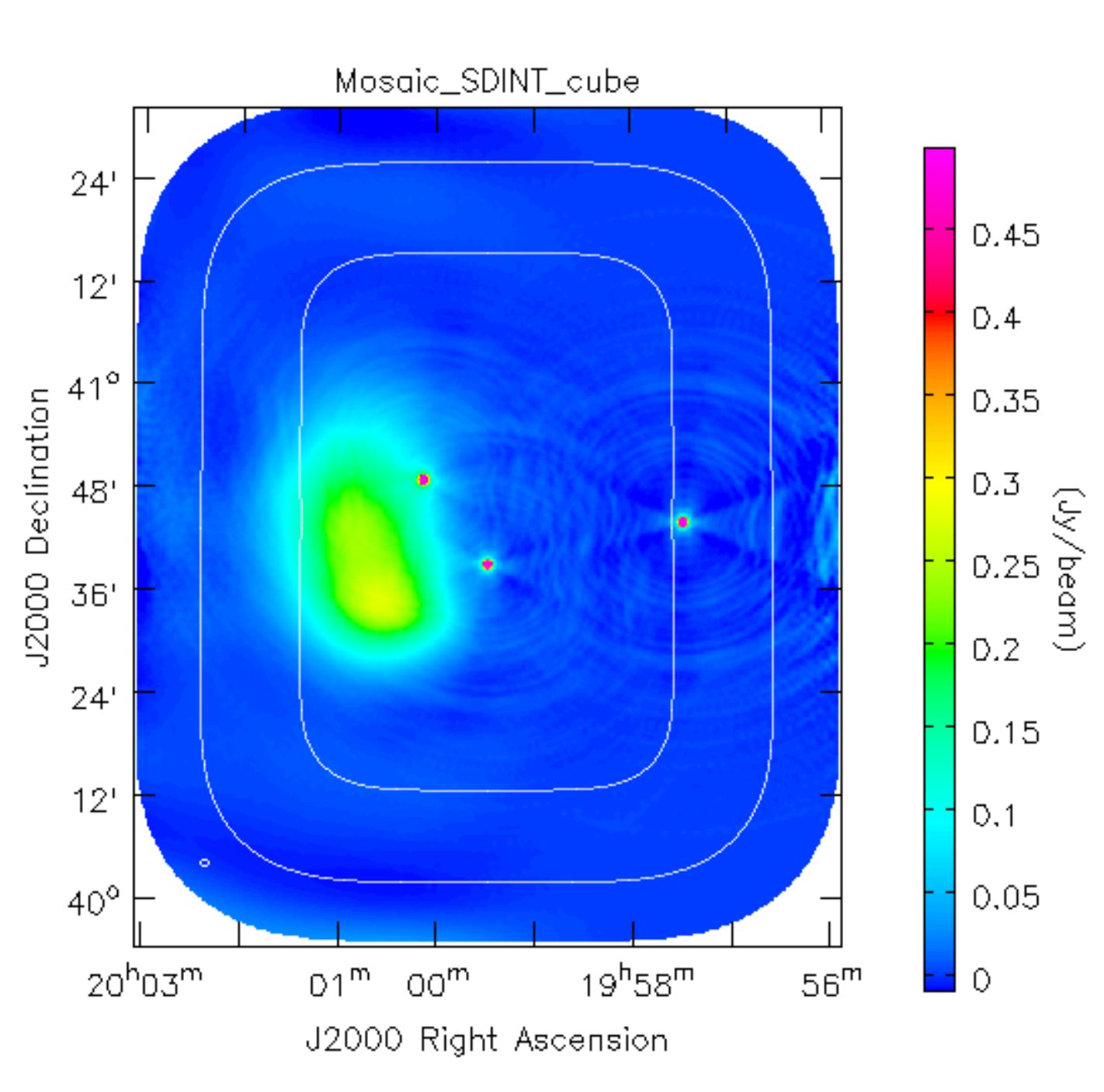}
\includegraphics[width=0.23\textwidth]{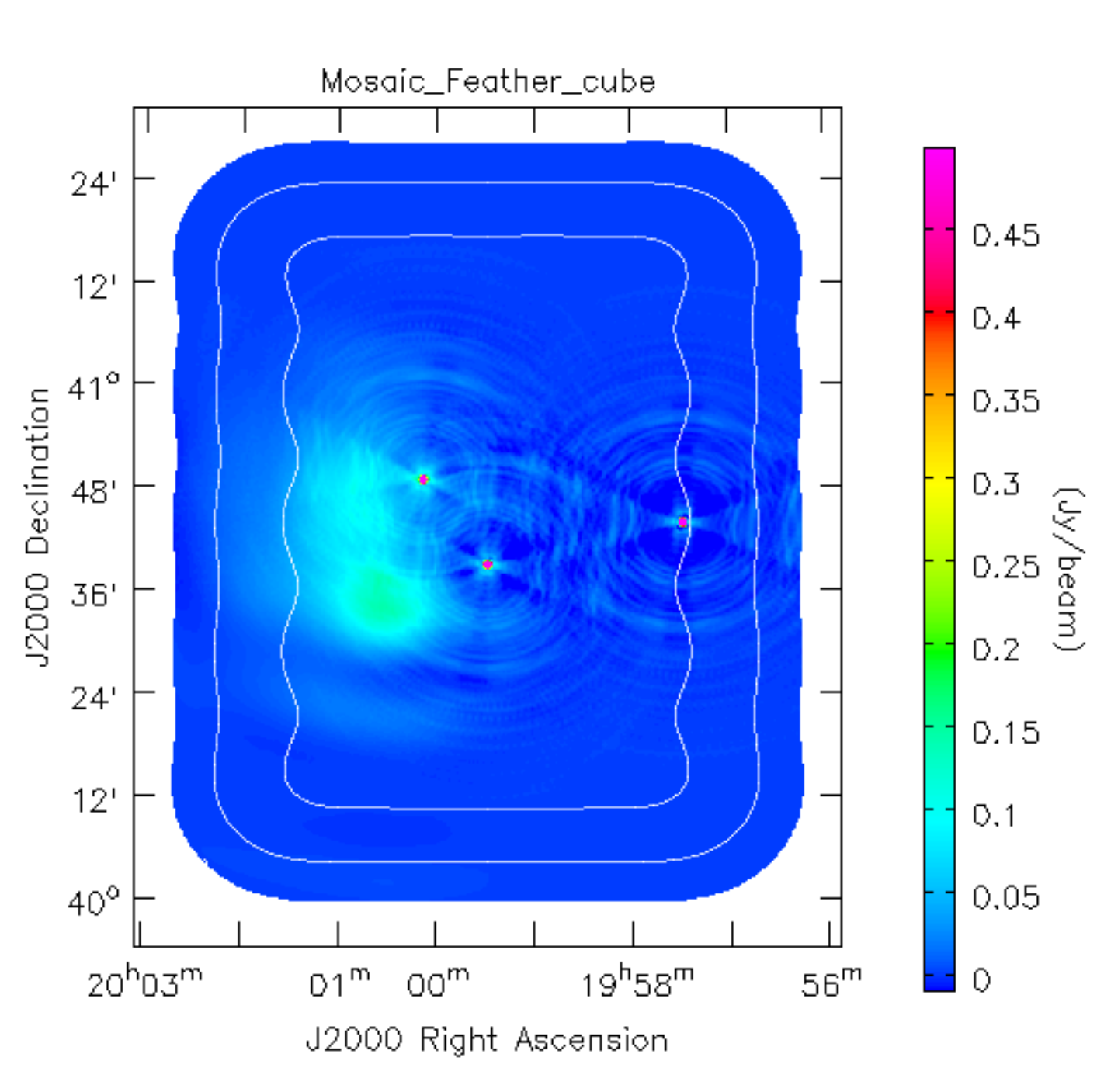}
\includegraphics[width=0.23\textwidth]{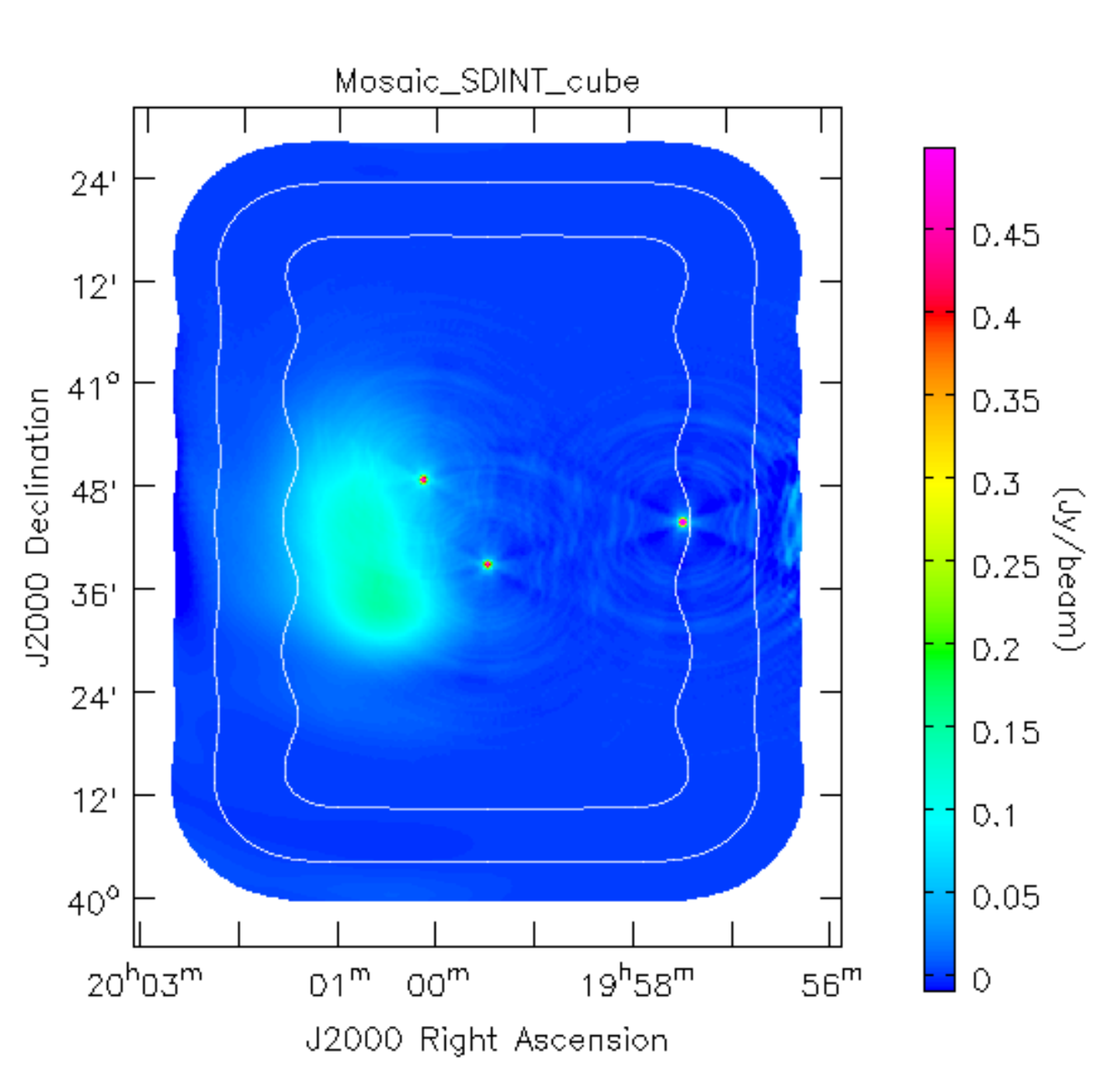}
\caption{Mosaic Spectral Cube from Feathering (LEFT) and Joint SDINT reconstruction (RIGHT)
at 1.0GHz, 1.5GHz and 2.0GHz (top to bottom).}
\label{Fig:mos_cube_feather_sdint}
\end{figure}

\begin{figure}
\centering
\includegraphics[width=0.23\textwidth]{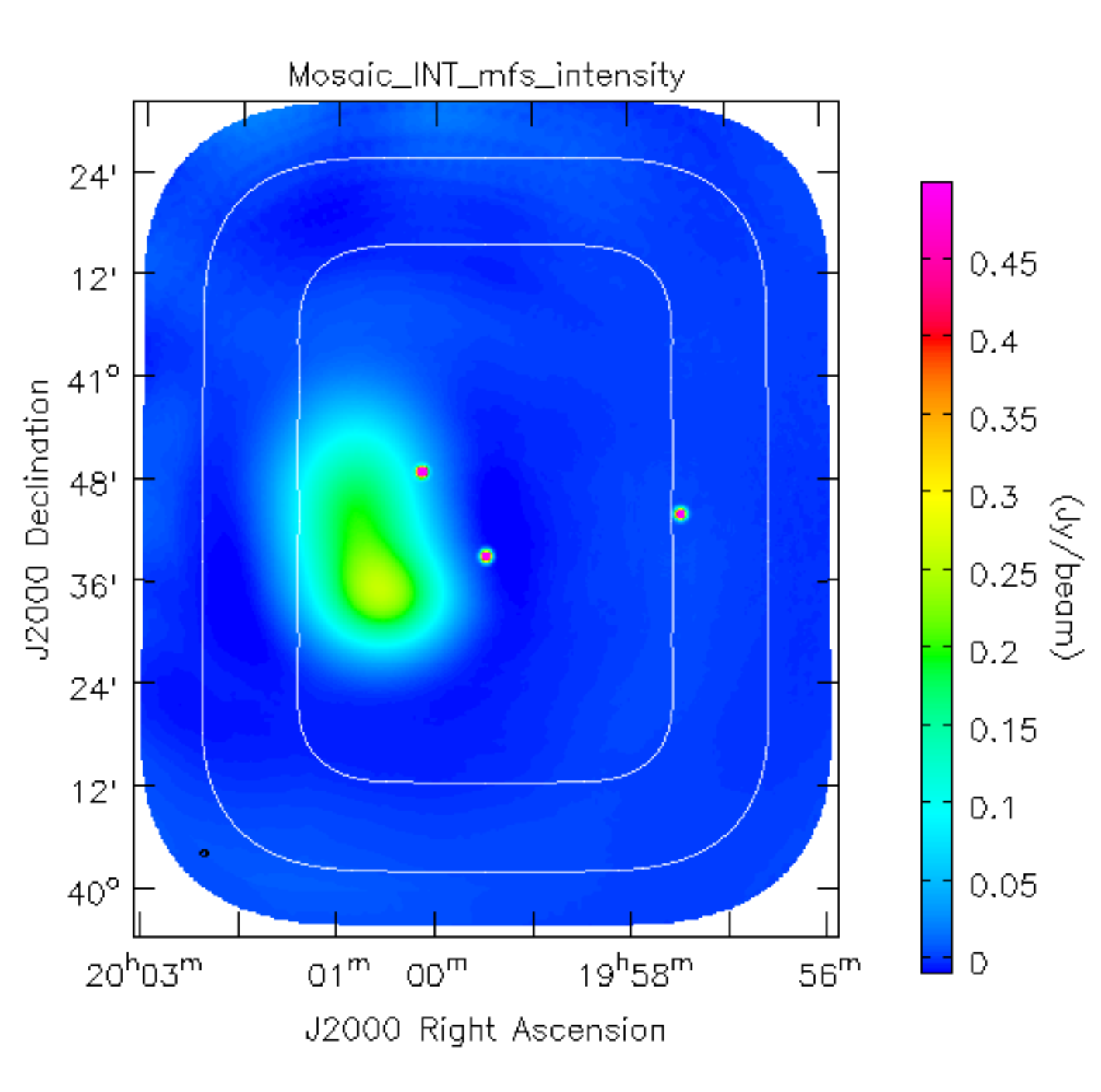}
\includegraphics[width=0.23\textwidth]{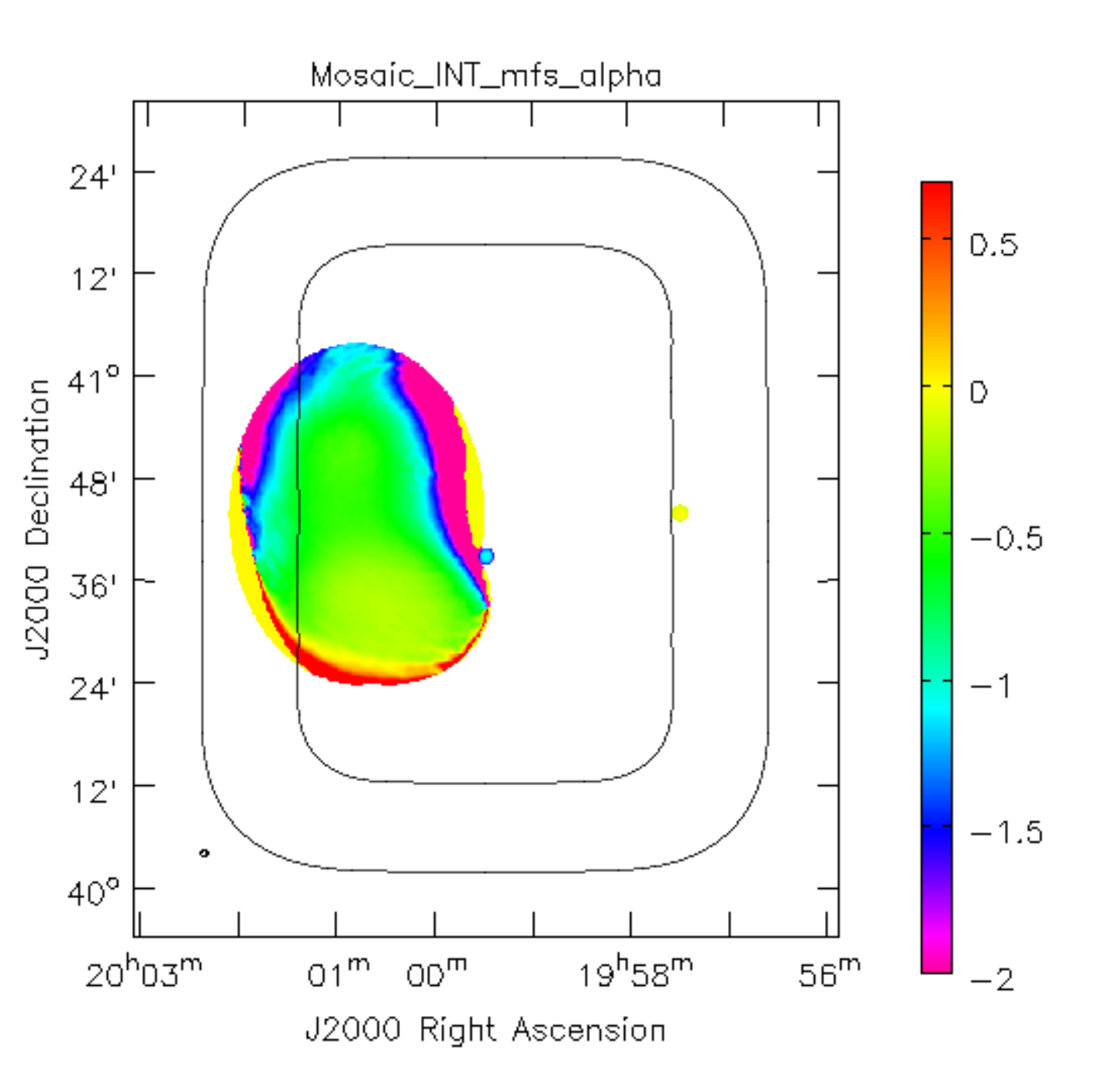}
\caption{Wideband MT-MFS Mosaic Intensity and Spectral index from INT data only }
\label{Fig:mos_mtmfs_intonly}
\includegraphics[width=0.23\textwidth]{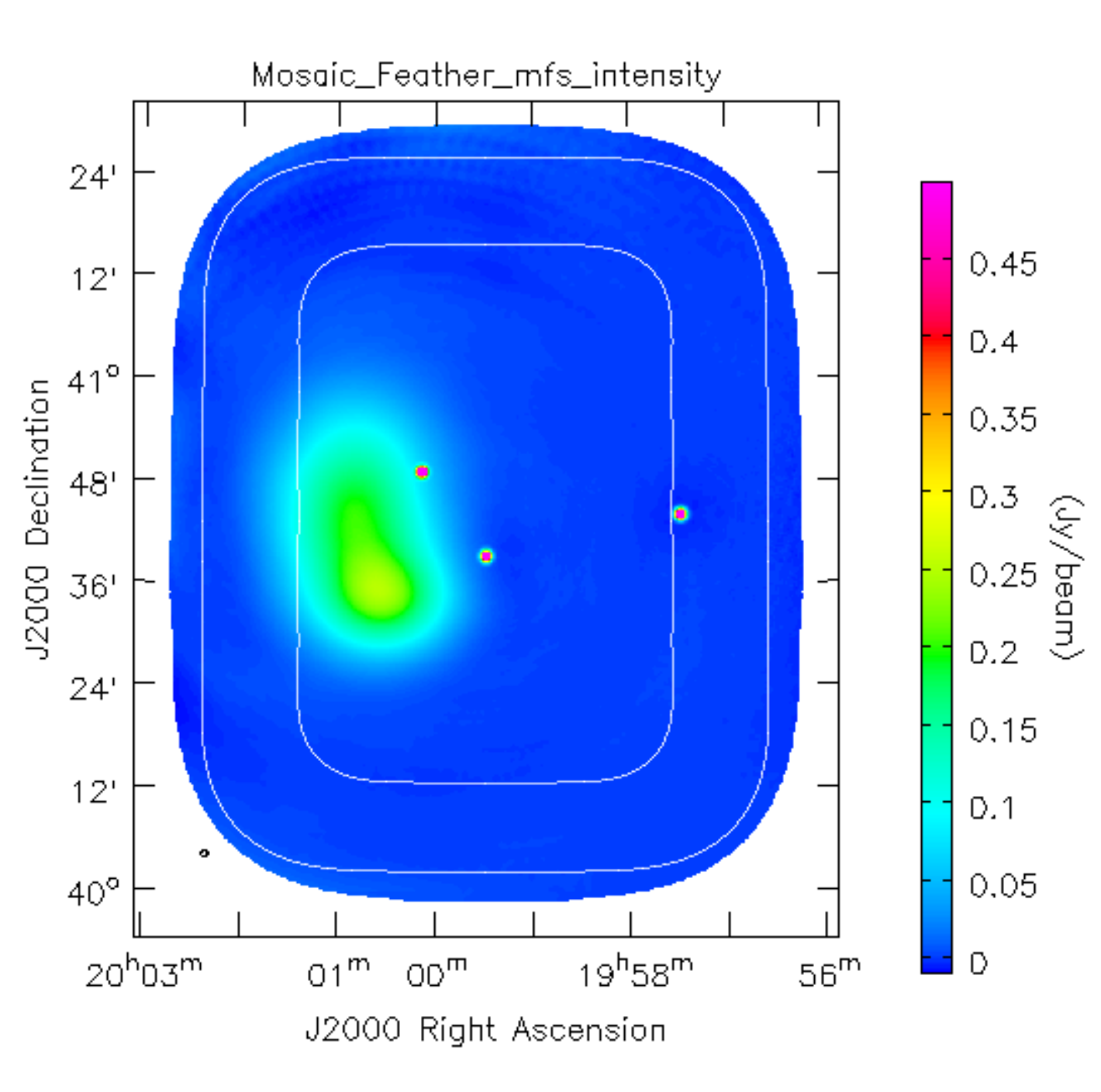}
\includegraphics[width=0.23\textwidth]{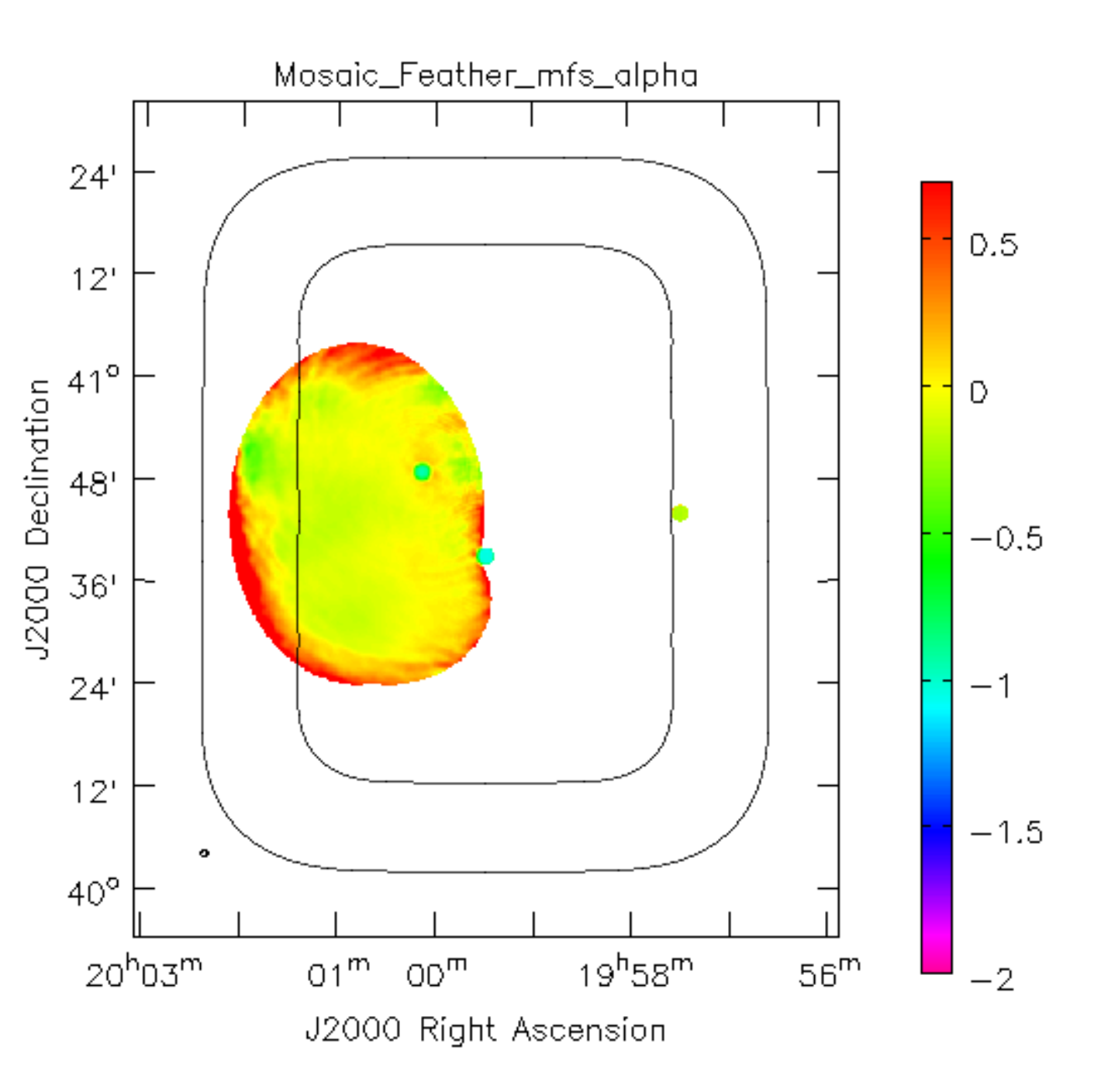}
\caption{Wideband MT-MFS Mosaic Intensity and Spectral index from Feathering}
\label{Fig:mos_mtmfs_feather}
\includegraphics[width=0.23\textwidth]{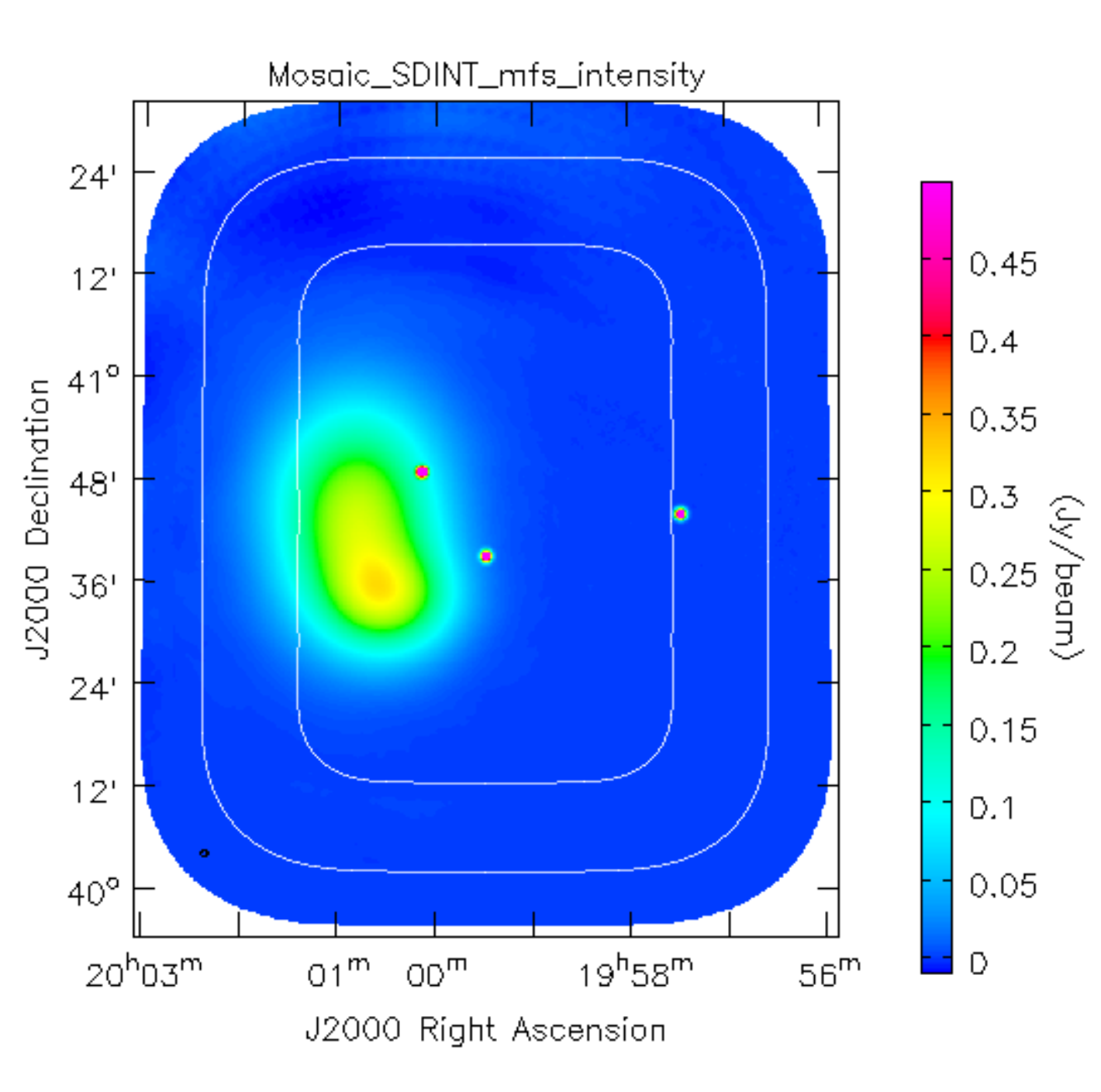}
\includegraphics[width=0.23\textwidth]{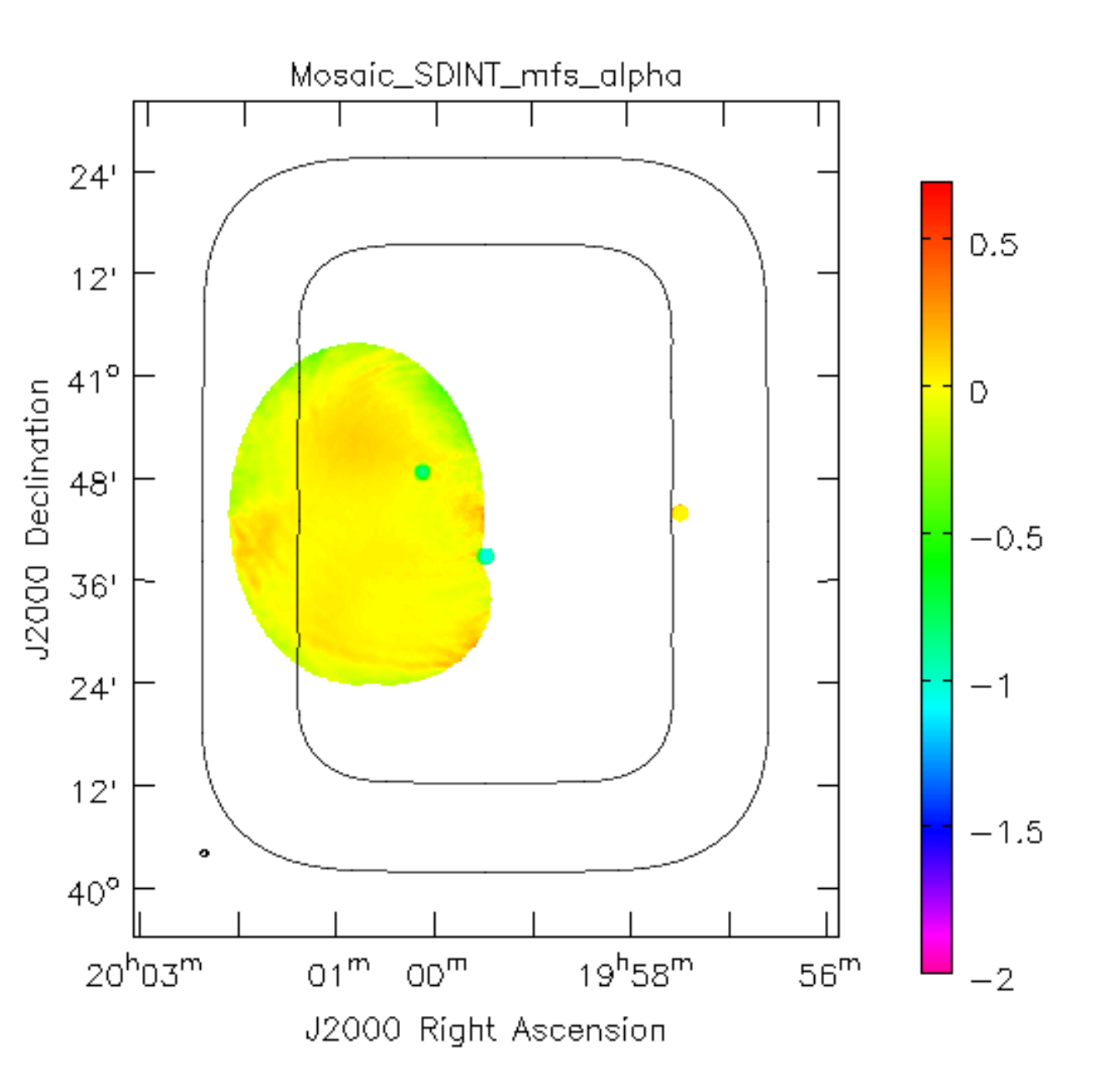}
\caption{Wideband MT-MFS Mosaic Intensity and Spectral index from Joint SDINT reconstruction}
\label{Fig:mos_mtmfs_sdint}
\end{figure}

A series of joint mosaic imaging tests were done using the simulated mosaic 
interferometer data along with the single dish images. 
The A-Projection algorithm offered by the {\tt gridder='mosaic'} option in 
CASA's {\it tclean} was used with a flat-noise normalization and used the same 
frequency dependent primary beam models that were used for the simulation 
of the mosaic interferometer data. 

Figures \ref{Fig:mos_cube_intonly_sdonly} and \ref{Fig:mos_cube_feather_sdint} 
show flat-sky results for mosaic spectral cube imaging using the 
INT-only, SD-only, Feathering and Joint \ALGO approaches. 
Artifacts due to missing short spacing information are clearly visible in the INT-only
images, less prominent in the feathering results and minimized in the joint \ALGO
reconstructions where the on-source structure and flux values are closest to the
expected true intensity distribution (shown in Fig.\ref{Fig:true_intensity}).
In all these figures, interferometry
primary beam contours are drawn at the 0.5 and 0.9 gain levels and they illustrate the
algorithms' performance both within the central region of the mosaic as well as at
the edge (where normalization errors, if present, would most prominently manifest themselves).

Figure \ref{Fig:mos_mtmfs_intonly} shows flat-sky intensity and spectral index maps for the
interferometer-only wideband mosaic reconstruction. The spectral index recovery
at the largest scales is more accurate than the basic single pointing simulation\footnote{Note that the mosaic simulation
contained about 25 times more interferometry visibilities compared to the basic simulation. 
},but 
still too steep by about 0.5 or more. 
Figure \ref{Fig:mos_mtmfs_feather} shows results from feathering the Taylor coefficient
maps from the above interferometer-only reconstruction (after taking them to flat-sky)
with single dish wideband Taylor coefficients derived from the SD-only multi-term deconvolution
run.
Figure \ref{Fig:mos_mtmfs_sdint} shows results from the joint \ALGO method which
produced the most accurate of the results shown here. 

\subsection{Effect of relative noise levels}\label{Sec:compare_algo_noise}

\begin{figure}
\centering
\includegraphics[width=0.23\textwidth]{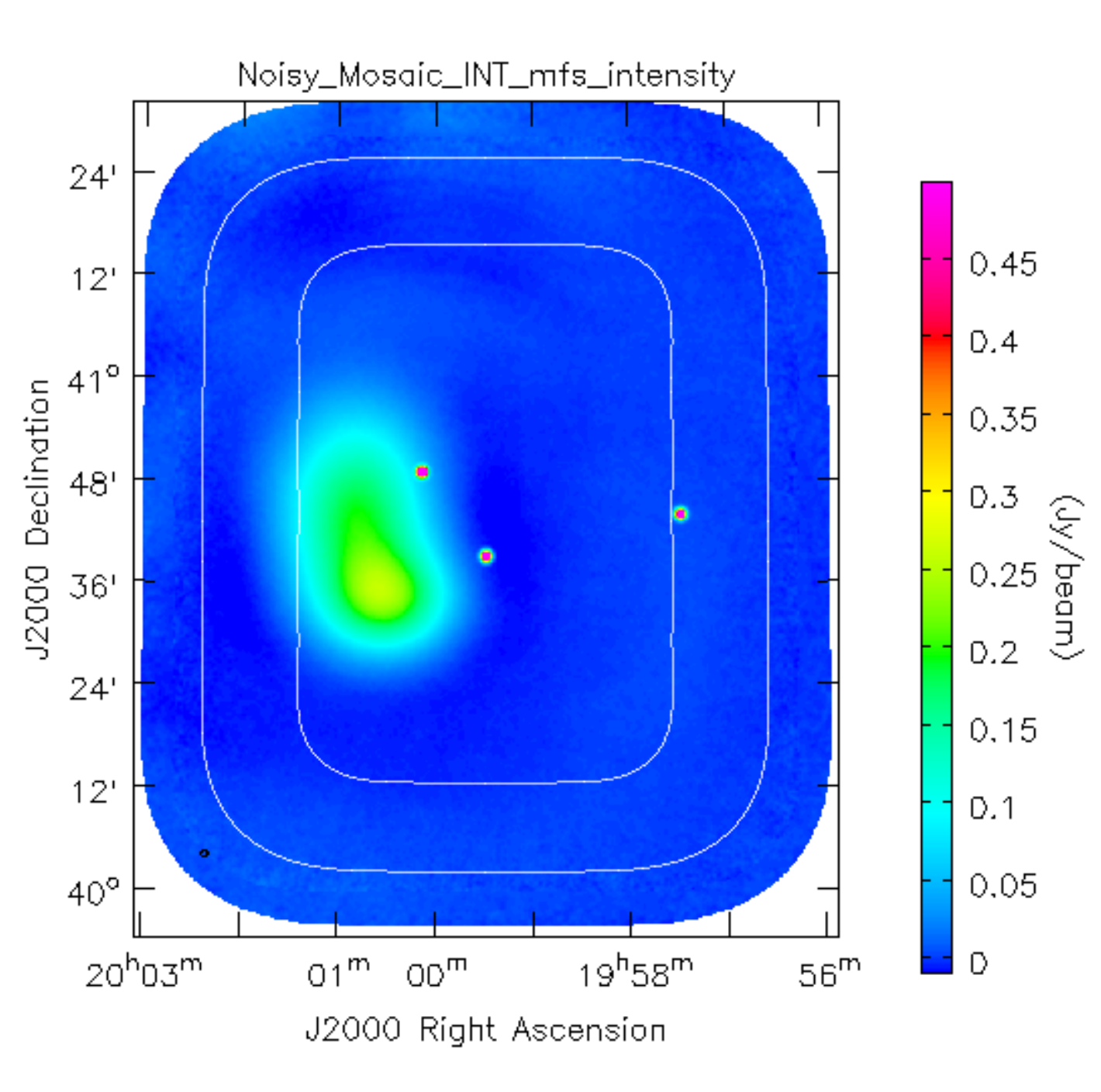}
\includegraphics[width=0.23\textwidth]{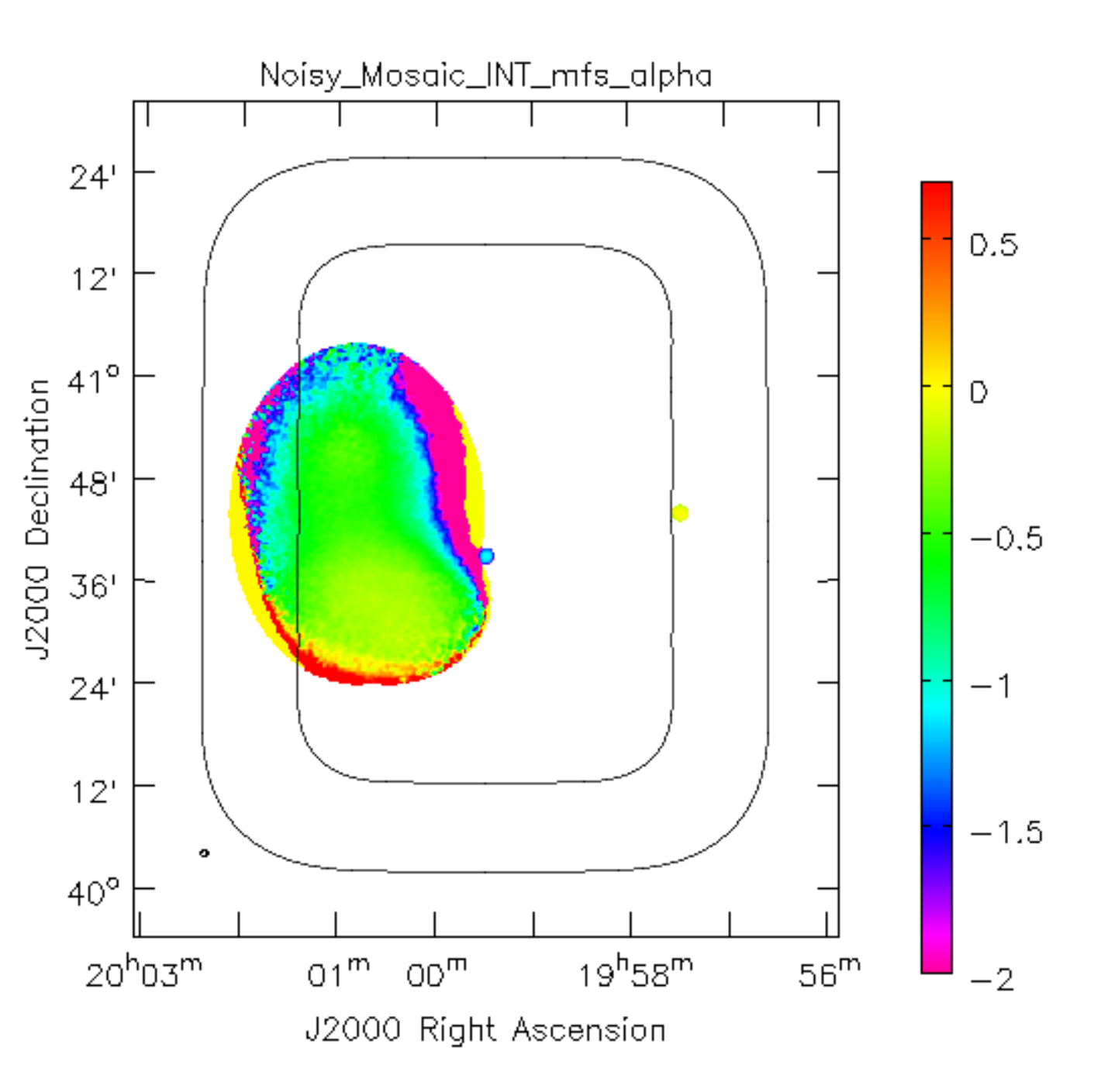}
\caption{With Noise : Wideband MT-MFS Mosaic Intensity and Spectral index from Noisy INT data alone }
\label{Fig:noisy_mos_mtmfs_intonly}
\includegraphics[width=0.23\textwidth]{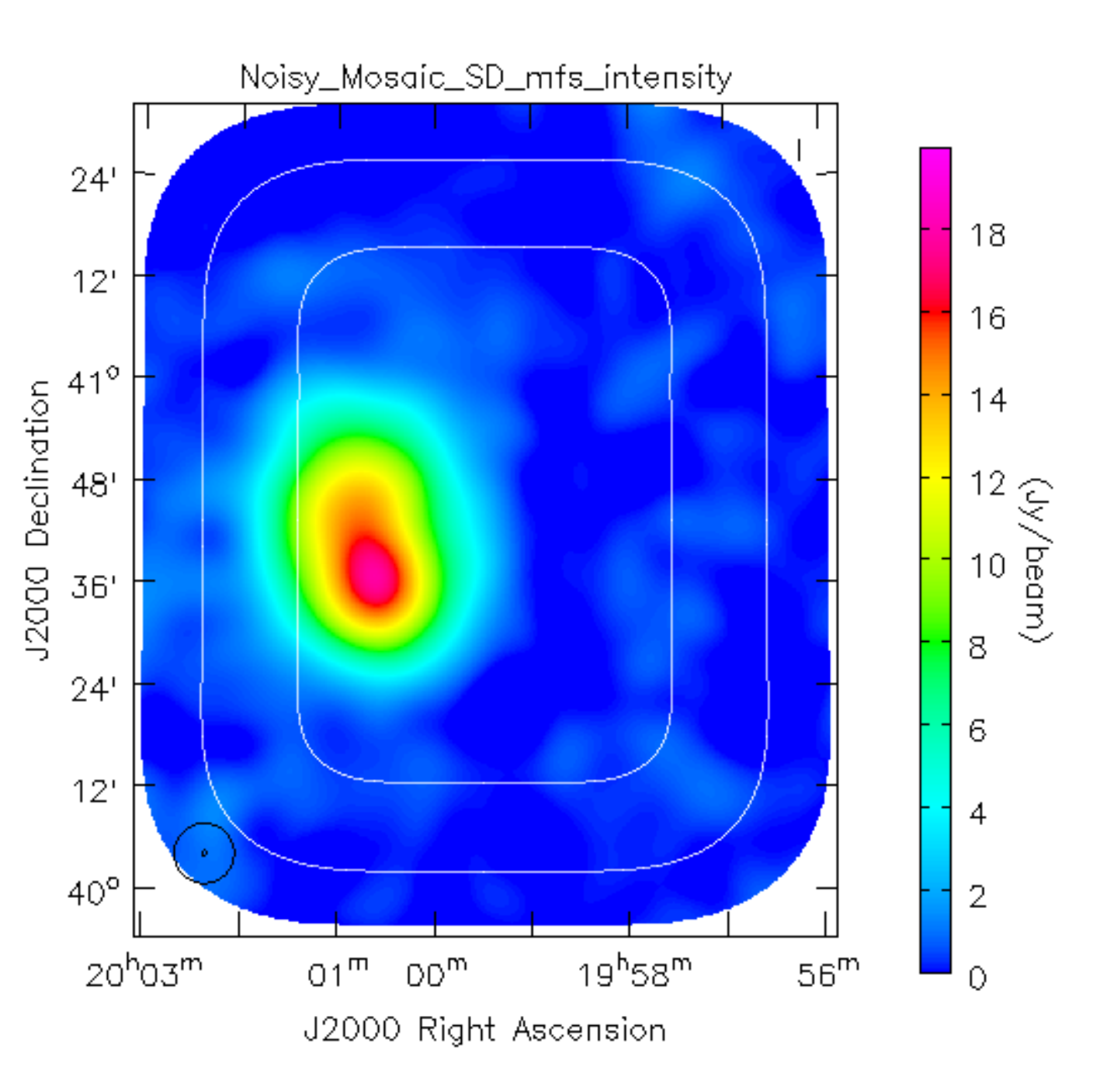}
\includegraphics[width=0.23\textwidth]{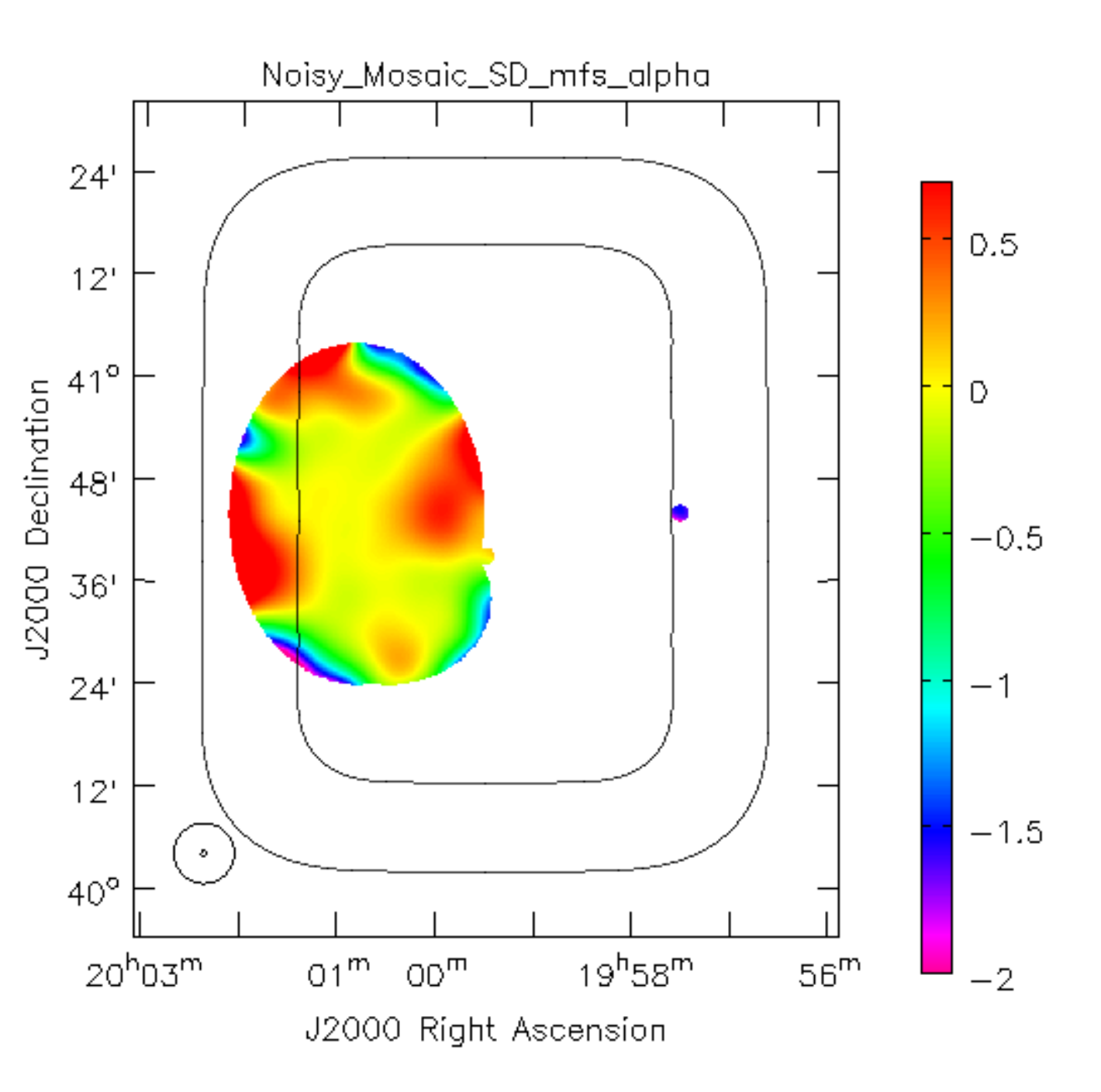}
\caption{With Noise : Wideband MT-MFS Intensity and Spectral index from Noisy SD data alone }
\label{Fig:noisy_mos_mtmfs_sdonly}

\includegraphics[width=0.23\textwidth]{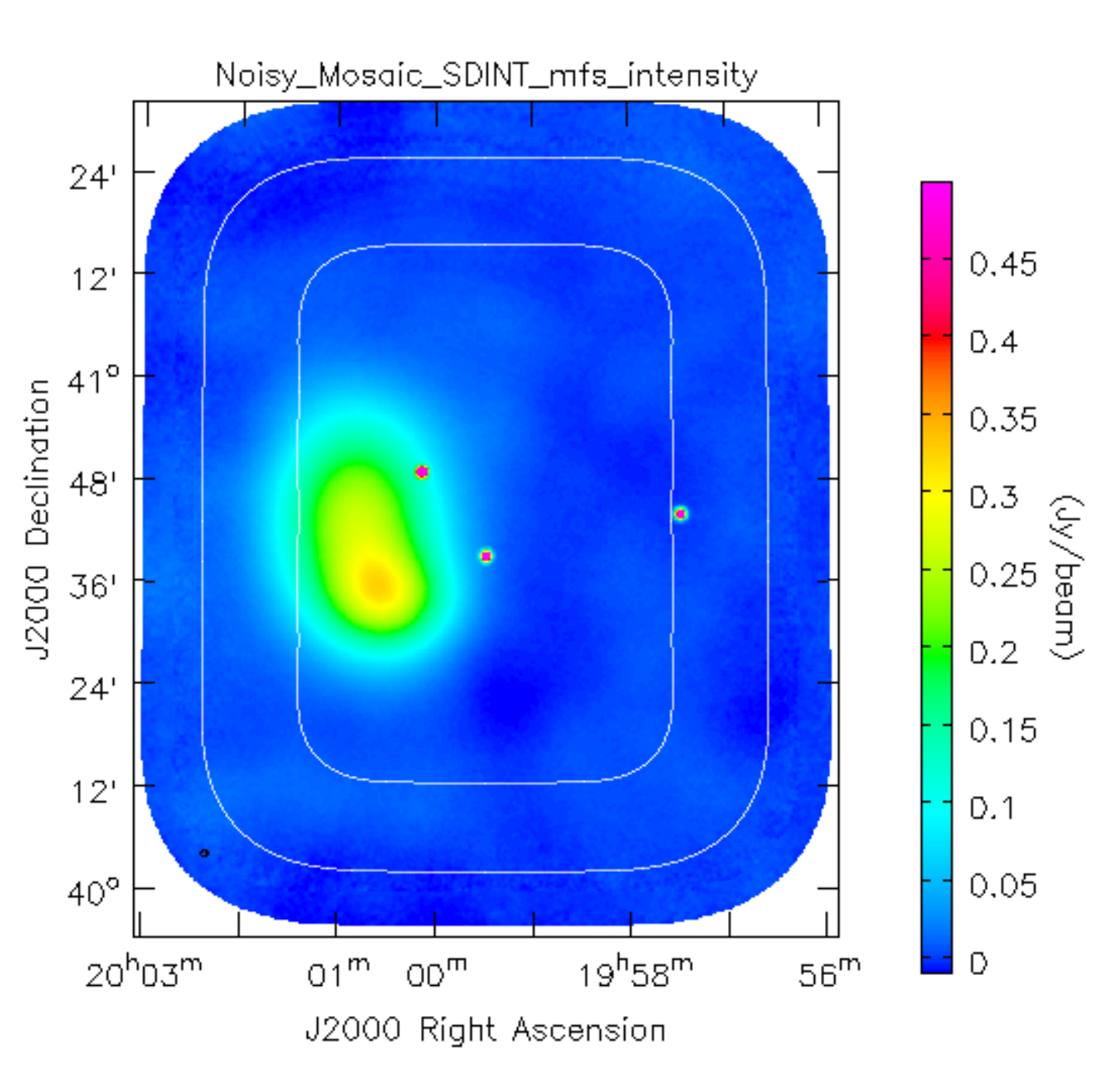}
\includegraphics[width=0.23\textwidth]{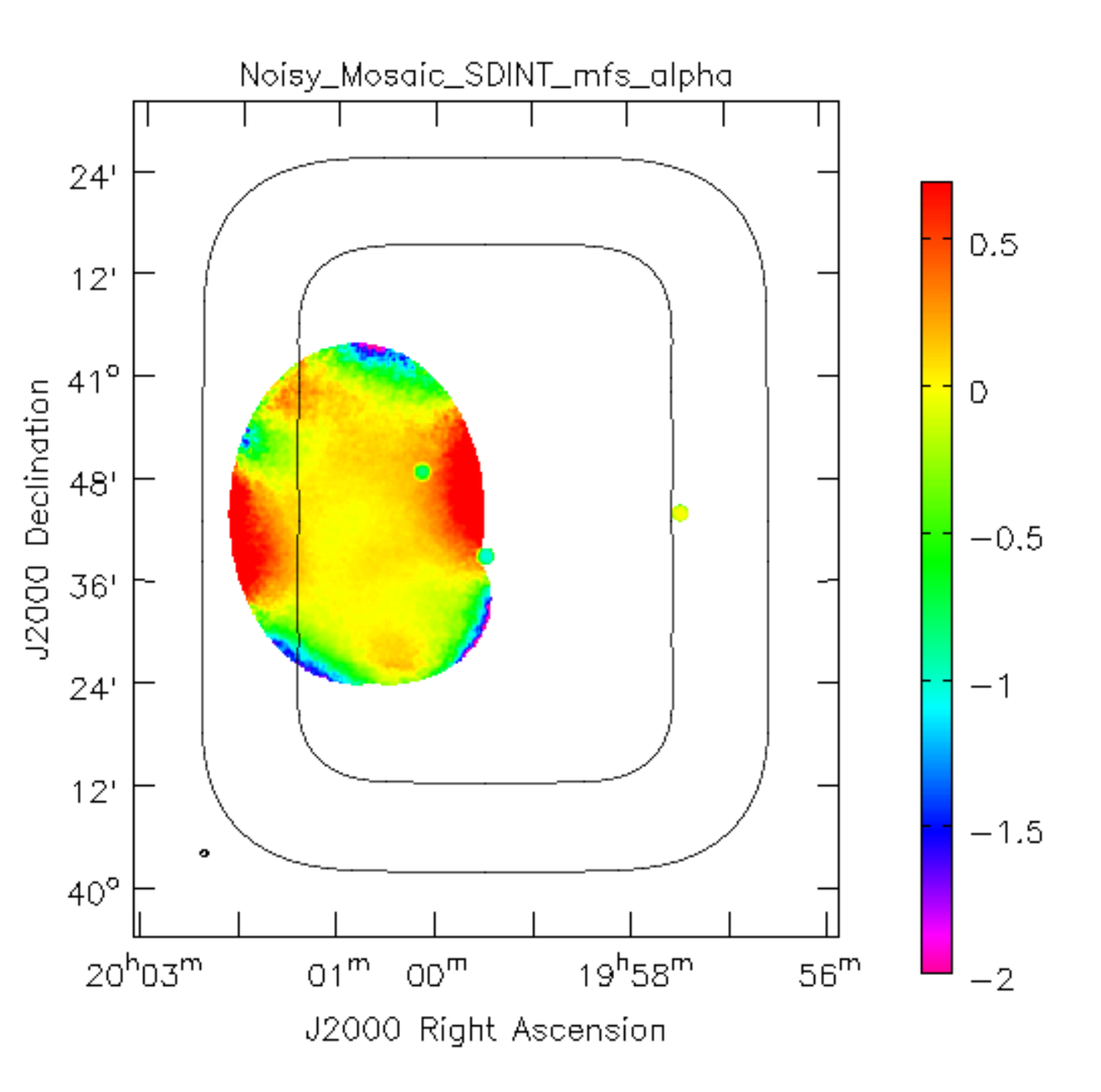}
\caption{With Noise : Wideband MT-MFS Mosaic Intensity and Spectral index from Joint SDINT reconstruction with sdgain=1.0 }
\label{Fig:noisy_mos_mtmfs_sdint}

\includegraphics[width=0.23\textwidth]{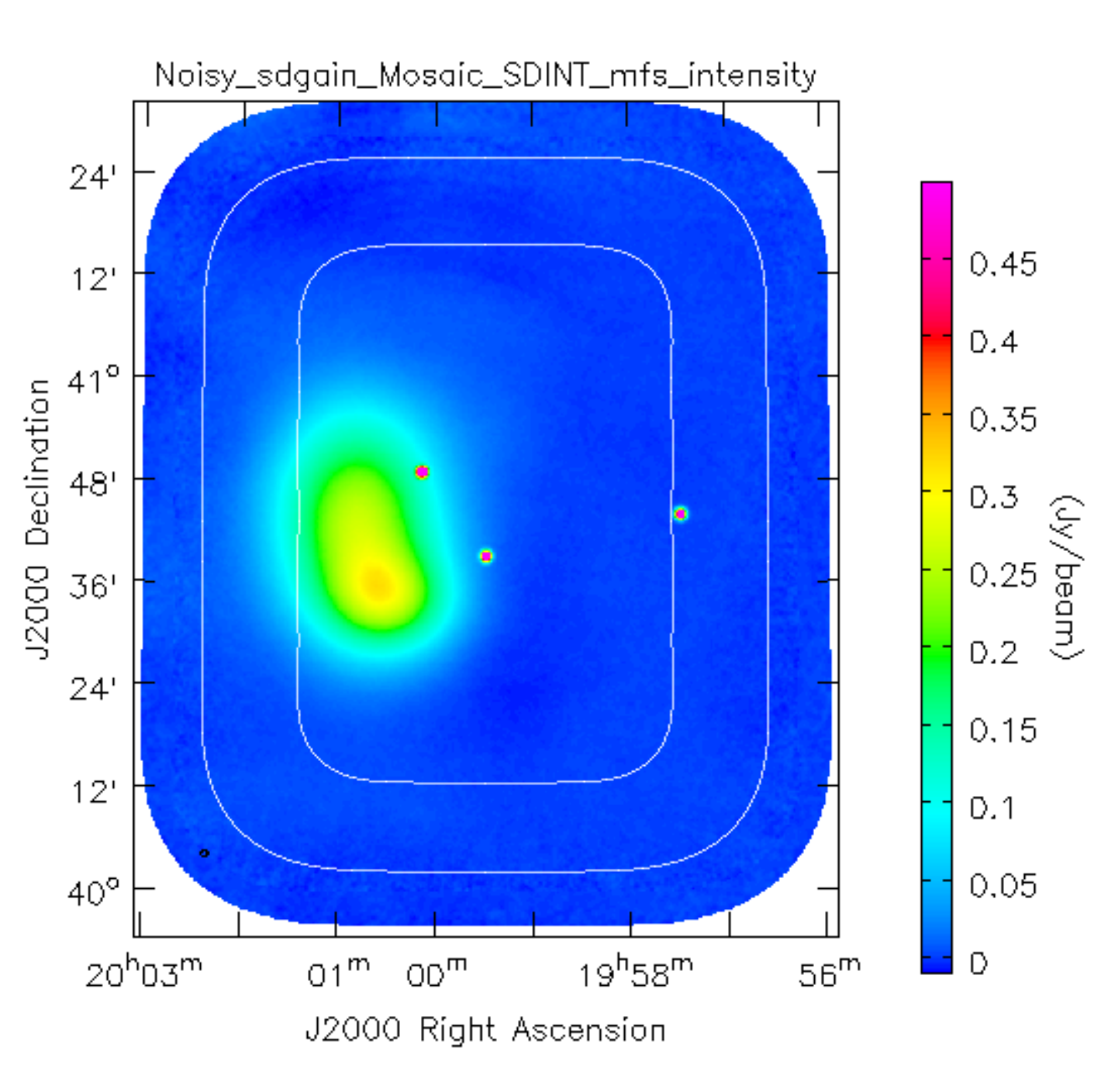}
\includegraphics[width=0.23\textwidth]{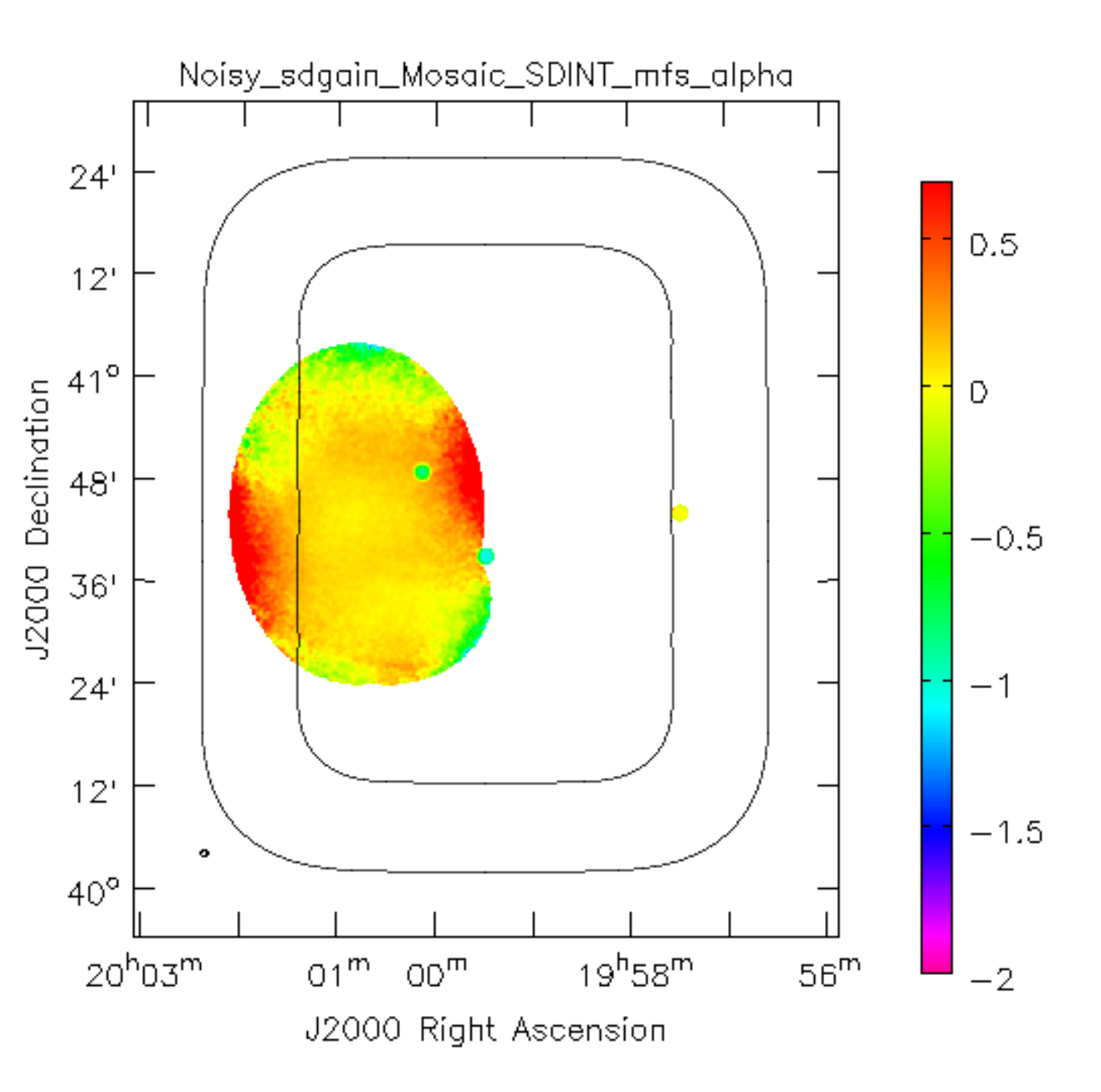}
\caption{With Noise : Wideband MT-MFS Mosaic Intensity and Spectral index from Joint SDINT reconstruction with sdgain=0.2 }
\label{Fig:noisy_mos_mtmfs_sdint_sdgain}

\end{figure}

Noise in a single dish image is often higher than in the interferometric image
and a typical feathering approach will result in a degradation of noise level.
Careful matching of noise levels 
may require suppressing the flux or using complicated spatial-frequency 
weighting functions that attempt to
balance the competing requirements of flux correctness versus noise suppression.
However, as discussed in section \ref{Sec:ScaleWeight} one can in principle 
separate the step of flux scaling from that of applying relative weights within 
a joint reconstruction scheme to suppress noise introduced by  
the single dish data while still preserving flux correctness in the reconstruction.
A basic demonstration of this idea has been explored in this paper with results shown in
Figures \ref{Fig:noisy_mos_mtmfs_intonly} to \ref{Fig:noisy_mos_mtmfs_sdint_sdgain}.

In a test designed specifically to evaluate the use of relative weighting to control single dish noise 
levels, random Gaussian noise of 0.2 Jy was added per 
simulated visibility value to produce a 0.002 Jy/beam image rms 
level for the interferometer data, a peak-to-noise dynamic range of 150 for the 
extended emission with a peak flux of 0.3 Jy/beam and a dynamic range of about 500 for 
the 1 Jy point sources. 
Random Gaussian noise of 0.02 Jy was added per pixel of the simulated single dish image 
prior to smoothing by the SD beam. The resulting single dish image noise level was 0.3 Jy/beam, 
had a peak-to-noise dynamic range of about 50 for the extended structures and with the point 
sources not visible above the noise. This simulates the situation where sky noise dominates 
and produces an image error pattern at scales larger than the pixel size or gridding kernel. 
Such noise is harder to suppress or filter out without affecting the source structure itself and 
was chosen in order to evaluate the harder extreme of the problem. 

Only the wideband multi-term mosaic imaging runs were repeated for these tests with noise. 
Figs.\ref{Fig:noisy_mos_mtmfs_intonly} and \ref{Fig:noisy_mos_mtmfs_sdonly} show 
results with the INT only and SD only data (compare with Figs.\ref{Fig:mtmfs_sdonly}
and \ref{Fig:mtmfs_intonly}). While the interferometer reconstruction is relatively 
unchanged compared to the noiseless simulation, the single dish reconstructions of both
intensity and spectral index are strongly affected by noise at the scale of the single dish beam.
Fig.\ref{Fig:noisy_mos_mtmfs_sdint} shows results from the joint \ALGO reconstruction when
the single dish data were given equal weight compared to the interferometer data 
(i.e. {\tt sdgain = 1.0} in CASA Feather used within the \ALGO algorithm).
The maps show errors that track the noise pattern of the SD-only images
at a level of 0.004 Jy/beam. 
Fig.\ref{Fig:noisy_mos_mtmfs_sdint} shows results when the single dish 
data were downweighted (i.e. {\tt sdgain=0.2}). The intensity
map clearly shows an improvement in the noise level (now at 0.0022 Jy/beam, compared to 
the INT only noise level of 0.002 Jy/beam) with 
on-source intensity and spectral index closer to the truth. The improvement on the spectral index
in this instance was only minor and further reduction of the single dish gain factor began to
reach the edge of the viable range for this simulation and resulted in increased
steepness of the reconstructed spectral index.

In practice, a general rule of thumb for choosing a scale factor for noise suppression is to 
compare the noise levels between an INT only reconstruction and an SDINT 
reconstruction with {\tt sdgain=1.0}, keeping in mind that the noise adds in quadrature
to produce the observed SDINT level.
The derived sdgain factor may then be used only if it is within the viable range driven by
uv-coverage and sky structure for the particular observation at hand.

These results suggest that the  idea of downweighting the single dish data during
combination may yield useful improvements in imaging quality but that it must be done 
with caution especially if there are significant errors at large spatial scales where the interferometer
data too do not provide sufficient constraints (which is precisely what this test probed). 
In situations where the single dish noise is at smaller spatial scales
(such as instrumental noise that is independent per measurement and
smeared only by the imaging gridding kernel) this approach is likely to be more effective.

This view has potentially positive implications on the required time on source for a 
single dish and an interferometer. For example, for the ALMA telescope,
\citet{SDINT_NAASC_2013} derive relative observation times between the ALMA 12m array,
the ACA 7m array and the ALMA 12m Total Power array that are required to match
visibility noise levels in preparation for a single step feathered combination. 
However, for sources with significant flux in extended structures, relaxing this 
requirement
would imply the need for less observing time for the single dish data than
is currently the standard.

In conclusion, as long as the single-dish data are downweighted within the
viable range (from the uv-coverage and structure point of view), the noise
may be suppressed with minimal consequences on the final reconstruction.

\section{Discussion}

Missing short spacing information in interferometric data results in incorrect
reconstructions of the structure and especially the spectrum of the sky brightness 
distributions at large spatial scales. Single dish observations can effectively provide
sufficient constraints to reconstruct these structures and spectra accurately.

Several techniques have been in use for decades, but there is no established standard
especially for wideband imaging. 
In this paper we have evaluated several methods
and proposed the \ALGO algorithm. We have demonstrated that it can accurately solve the
problem of missing short spacings as encountered during a wideband reconstruction and it 
can do this within a generic joint reconstruction algorithmic framework that 
supports spectral cubes or wideband multi-term reconstruction, narrow-field or wide-field
mosaic imaging as well as the option to include only interferometer data or single dish data
or both together. 
In the context of widely differing noise levels between the single dish and interferometer data, we have
also evaluated the idea of using relative weighting between the single dish and interferomter 
data to achieve low joint image rms levels without sacrificing flux accuracy. 

Future work will include real-world applications to VLA and GBT wideband mosaic imaging 
and ALMA's 12m, 7m and 12m total power combinations for spectral cube and continuum 
imaging. 
The option of including a single dish major cycle
that operates directly with ALMA total-power data is within reach within the CASA software 
and it will be useful to evaluate its numerical benefits.
The idea of downweighting single dish data requires careful evaluation on real data
for specific telescope and observing setups (to allow evaluation for specific types of
uv-coverage),
but has the potential of reducing the amount of required single dish observing time.
Finally, various weighting schemes discussed in the literature (e.g. 
in \citet{SDINT_KODA_2011} and \citet{SDINT_KURONO_2009}) will also be evaluated and 
adopted within the \ALGO implementation. 
\newpage
All the above steps are in progress as part of 
commissioning the algorithm for production use and will result in software documentation
and usage guidelines for different applications.

\section{Acknowledgements}
The initial part of this research was conducted under the NRAO  
Summer Student Research Assistantship programme and continued within the 
NRAO Algorithm R\&D group. We acknowledge the use of the CASA software and the 
PySynthesisImager python level prototyping interface to its imaging libraries. 
We would like to thank S.Bhatnagar and K.Golap of NRAO for
 participating in discussions on details of this work.
The National Radio Astronomy Observatory 
is a facility of the National Science Foundation operated under cooperative agreement 
by Associated Universities, Inc. 
\bibliography{wbsdint} 
\bibliographystyle{apj} 

\end{document}